\documentclass[longauth]{aa}

\renewcommand{\eprint}[2][]{%
  \if!#1!#2\else#1\arxivprefixesep\ignorespaces#2\fi
}

\usepackage{graphicx}
\usepackage{natbib}
\usepackage{scalerel}
\usepackage{amsmath}
\usepackage{dcolumn}
\usepackage{bm} 
\usepackage[rightcaption]{sidecap}
\usepackage{siunitx}

\usepackage{multirow}
\usepackage{bigdelim}

\usepackage[version=4]{mhchem}

\usepackage[nolist,nohyperlinks]{acronym}

\usepackage[table]{xcolor}
\usepackage{booktabs}

\usepackage{txfonts}
\usepackage[pdfencoding=auto,psdextra]{hyperref}
\hypersetup{
    colorlinks=true,
    linkcolor=blue,
    filecolor=magenta,      
    urlcolor=blue,
    citecolor=blue
}
\makeatletter
\renewcommand*\aa@pageof{, page \thepage{} of \pageref*{LastPage}}
\makeatother

\usepackage[utf8]{inputenc}
\usepackage[switch, modulo]{lineno}
              
\usepackage{euclid}

\usepackage[nameinlink,capitalise]{cleveref}
\crefname{section}{Sect.}{Sects.}
\Crefname{section}{Section}{Sections}
\crefname{figure}{Fig.}{Figs.}
\Crefname{figure}{Figure}{Figures}
\crefname{equation}{Eq.}{Eqs.}
\Crefname{equation}{Equation}{Equations}
\crefname{table}{Table}{Tables}
\crefname{appendix}{Appendix}{Appendices}
\Crefname{appendix}{Appendix}{Appendices}
\newcommand{\crefp}[1]{Eq.~\ref{#1}}

\newcommand*{\ms}{\si{\micro\second}}

\definecolor{lightgreen}{HTML}{CCFFCC}
\definecolor{lightblue}{HTML}{CCCCFF}
\definecolor{lightyellow}{HTML}{FFFFCC}

\defcitealias{Schirmer-EP29}{ECS2023}

\begin{document}

\acrodef{AA}{azimuth angle}
\acrodef{ACE}{Advanced Composition Explorer}
\acrodef{AIT}{assembly, integration, and testing}
\acrodef{AOCS}{attitude and orbital control system}
\acrodef{AOI}{angle of incidence}
\acrodef{AR}{anti-reflection}
\acrodef{ASIC}{application specific integrated circuit}
\acrodef{BFE}{brighter-fatter effect}
\acrodef{CaLA}{camera-lens assembly}
\acrodef{CCD}{charge-coupled device}
\acrodef{CMU}{compensation mechanism unit}
\acrodef{CoLA}{corrector-lens assembly}
\acrodef{CDS}{Correlated Double Sampling}
\acrodef{CFC}{cryo-flex cable}
\acrodef{CFHT}{Canada-France-Hawaii Telescope}
\acrodef{CFRP}{carbon-fibre reinforced plastic}
\acrodef{CGH}{computer-generated hologram}
\acrodef{CIL}{charge-injection line}
\acrodef{CME}{coronal mass ejection}
\acrodef{CNES}{Centre National d'Etude Spacial}
\acrodef{CPPM}{Centre de Physique des Particules de Marseille}
\acrodef{CPU}{central processing unit}
\acrodef{CR}{cosmic ray}
\acrodef{CTE}{coefficient of thermal expansion}
\acrodef{CME}{coronal mass ejection}
\acrodef{CTI}{charge-transfer inefficiency}
\acrodef{CVZ}{continuous viewing zone}
\acrodef{DCU}{Detector Control Unit}
\acrodef{DES}{Dark Energy Survey}
\acrodef{DGL}{diffuse Galactic light}
\acrodef{DPU}{Data Processing Unit}
\acrodef{DS}{Detector System}
\acrodef{EAF}{Euclid Auxiliary Fields}
\acrodef{EDS}{Euclid Deep Survey}
\acrodef{EDF}{Euclid Deep Fields}
\acrodef{EE}{encircled energy}
\acrodef{EA}{Enterprise Architect}
\acrodef{EBL}{extragalactic background light}
\acrodef{ESA}{European Space Agency}
\acrodef{ESP}{Emission of Solar Protons}
\acrodef{ECSS}{European Cooperation for Space Standardization}
\acrodef{EWS}{Euclid Wide Survey}
\acrodef{EUDF}{Euclid Ultra Deep Field}
\acrodef{EDF-N}{Euclid Deep Field North}
\acrodef{EDF-S}{Euclid Deep Field South}
\acrodef{EMDS}{Euclid Medium Deep Survey}
\acrodef{FDIR}{Fault Detection, Isolation and Recovery}
\acrodef{FGS}{Fine-Guidance Sensor}
\acrodef{FOM}{figure of merit}
\acrodef{FoM3}{folding mirror 3}
\acrodef{FoV}{field of view}
\acrodef{FPA}{focal-plane array}
\acrodef{FWA}{filter-wheel assembly}
\acrodef{FWC}{full-well capacity}
\acrodef{FWHM}{full width at half maximum}
\acrodef{GDPRD}{Ground Data Processing Requirement Document}
\acrodef{GOES}{Geostationary Operational Environmental Satellites}
\acrodef{GCR}{Galactic cosmic ray}
\acrodef{GWA}{grism-wheel assembly}
\acrodef{H2RG}{HAWAII-2RG}
\acrodef{HST}{\textit{Hubble} Space Telescope}
\acrodef{IPC}{interpixel capacitance}
\acrodef{IP2I}{Institut de Physique des 2 Infinis de Lyon}
\acrodef{IPSV}{intrapixel sensitivity variation}
\acrodef{JWST}{{\em James Webb} Space Telescope}
\acrodef{IAD}{ion-assisted deposition}
\acrodef{ICU}{Instrument Control Unit}
\acrodef{IDL}{initial drain level}
\acrodef{IGL}{integrated galaxy light}
\acrodef{IPC}{inter-pixel capacitance}
\acrodef{ISES}{International Space Environmental Services}
\acrodef{ISL}{in-field straylight}
\acrodef{JWST}{\textit{James Webb} Space Telescope}
\acrodef{KiDS}{Kilo-Degree Survey}
\acrodef{LAM}{Laboratoire d'Astrophysique de Marseille}
\acrodef{LED}{light-emitting diode}
\acrodef{LMC}{Large Magellanic Cloud}
\acrodef{LSB}{low surface brightness}
\acrodef{LSF}{large-scale flat}
\acrodef{MACC}{multiple accumulated}
\acrodef{M2M}{secondary-mirror mechanism}
\acrodef{MER}{merge dataset}
\acrodef{MLI}{multi-layer insulation}
\acrodef{MMU}{Mass Memory Unit}
\acrodef{MPE}{Max-Planck-Institut für extraterrestrische Physik}
\acrodef{MOC}{mission operations centre}
\acrodef{MOCD}{Mission Operation Concept Document}
\acrodef{MOS}{metal-oxide semi-conductor}
\acrodef{MPS}{micro-propulsion system}
\acrodef{MRD}{Mission Requirements Document}
\acrodef{MPIA}{Max-Planck-Institut für Astronomie}
\acrodef{NA}{numerical aperture}
\acrodef{NASA}{National Aeronautic and Space Administration}
\acrodef{NDI}{normalised detector irradiance}
\acrodef{NEP}{north Ecliptic pole}
\acrodef{NGP}{north Galactic pole}
\acrodef{NEXUS}{North Ecliptic Pole Extragalactic Unified Survey}
\acrodef{NIEL}{non-ionising energy loss}
\acrodef{JPL}{NASA Jet Propulsion Laboratory}
\acrodef{MZ-CGH}{multi-zonal computer-generated hologram}
\acrodef{NI-CU}{NISP calibration unit}
\acrodef{NI-OA}{near-infrared optical assembly}
\acrodef{NI-GWA}{NISP Grism Wheel Assembly}
\acrodef{NIR}{near-infrared}
\acrodef{NISP}{Near-Infrared Spectrometer and Photometer}
\acrodef{NOAA}{National Oceanic and Atmospheric Administration}
\acrodef{OCM}{orbit-correction manoeuvre}
\acrodef{OSL}{out-of-field straylight}
\acrodef{PARMS}{plasma-assisted reactive magnetron sputtering}
\acrodef{RMS}{root mean square}
\acrodef{PLM}{payload module}
\acrodef{PN}{planetary nebula}
\acrodef{PRNU}{pixel-response non-uniformity}
\acrodef{PTFE}{polytetrafluoroethylene}
\acrodef{PV}{performance verification}
\acrodef{PWM}{pulse-width modulation}
\acrodef{PSF}{point-spread function}
\acrodef{QCM}{quartz crystal microbalance}
\acrodef{QE}{quantum efficiency}
\acrodef{QF}{quality factor}
\acrodef{PDC}{phase-diversity campaign}
\acrodef{PERD}{Payload Elements Requirements Document}
\acrodef{PTC}{photon transfer curve}
\acrodef{RPE}{relative pointing error}
\acrodef{ROE}{readout electronics}
\acrodef{ROI}{region of interest}
\acrodef{ROIC}{readout-integrated circuit}
\acrodef{ROS}{reference observing sequence}
\acrodef{RSD}{reference survey design}
\acrodef{RSU}{VIS rotary shutter unit}
\acrodef{RTN}{random telegraph noise}
\acrodef{SAA}{solar aspect angle}
\acrodef{SCA}{sensor chip array}
\acrodef{SCE}{sensor chip electronic}
\acrodef{SCS}{sensor chip system}
\acrodef{SGS}{science ground segment}
\acrodef{SGPS}{Solar and Galactic Proton Sensor}
\acrodef{SHS}{Shack-Hartmann sensor}
\acrodef{SNR}[S/N]{signal-to-noise ratio}
\acrodef{SolO}{Solar Orbiter}
\acrodef{SRP}{solar radiation pressure}
\acrodef{STIS}{Space Telescope Imaging Spectrograph}
\acrodef{SED}{spectral energy distribution}
\acrodef{SiC}{silicon carbide}
\acrodef{SEP}{South Ecliptic Pole}
\acrodef{SSN}{Sunspot number}
\acrodef{STIX}{Spectrometer/Telescope for Imaging X-rays}
\acrodef{STP}{serial trap pumping}
\acrodef{SolO}{Solar Orbiter}
\acrodef{SOPS}{spacecraft operations}
\acrodef{TER}{thermal-emission recoil}
\acrodef{TP}{trap pumping}
\acrodef{UTR}{up-the-ramp}
\acrodef{SVM}{service module}
\acrodef{VIS}{visible imager}
\acrodef{WD}{white dwarf}
\acrodef{WCS}{world coordinate system}
\acrodef{WFC3}{Wide Field Camera 3}
\acrodef{WFE}{wavefront error}
\acrodef{ZB}{zodiacal background}
\acrodef{ZP}{zero point}

\title{Euclid Data Release 1 (DR1) } 
\subtitle{Spacecraft performance in the L2 environment}

\newcommand{\orcid}[1]{} 
\author{Euclid Collaboration: M.~Schirmer\orcid{0000-0003-2568-9994}\thanks{\email{schirmer@mpia.de}}\inst{\ref{aff1}}
\and K.~Paterson\orcid{0000-0001-8340-3486}\inst{\ref{aff1}}
\and B.~Altieri\orcid{0000-0003-3936-0284}\inst{\ref{aff2}}
\and H.~Aussel\orcid{0000-0002-1371-5705}\inst{\ref{aff3}}
\and S.~Awan\inst{\ref{aff4}}
\and P.~Battaglia\orcid{0000-0002-7337-5909}\inst{\ref{aff5}}
\and G.~Buenadicha\inst{\ref{aff2}}
\and C.~Burigana\orcid{0000-0002-3005-5796}\inst{\ref{aff6},\ref{aff7}}
\and P.~Casenove\orcid{0009-0006-6736-1670}\inst{\ref{aff8}}
\and J.~Dinis\orcid{0000-0001-5075-1601}\inst{\ref{aff9},\ref{aff10}}
\and J.~Mart\'{i}n-Fleitas\orcid{0000-0002-8594-569X}\inst{\ref{aff11}}
\and P.-A.~Frugier\inst{\ref{aff3}}
\and W.~Gillard\orcid{0000-0003-4744-9748}\inst{\ref{aff12}}
\and A.~Gregorio\orcid{0000-0003-4028-8785}\inst{\ref{aff13},\ref{aff14},\ref{aff15}}
\and S.~Hemmati\orcid{0000-0003-2226-5395}\inst{\ref{aff16}}
\and O.~Herent\inst{\ref{aff17}}
\and J.~Hoar\inst{\ref{aff2}}
\and H.~Hoekstra\orcid{0000-0002-0641-3231}\inst{\ref{aff18}}
\and W.~Holmes\orcid{0009-0007-8554-4646}\inst{\ref{aff19}}
\and P.~Hudelot\inst{\ref{aff17}}
\and K.~Jahnke\orcid{0000-0003-3804-2137}\inst{\ref{aff1}}
\and R.~Kohley\inst{\ref{aff2}}
\and K.~Kuijken\orcid{0000-0002-3827-0175}\inst{\ref{aff18}}
\and R.~Laureijs\inst{\ref{aff20}}
\and G.~Libet\inst{\ref{aff8}}
\and P.~Liebing\orcid{0009-0005-8043-7150}\inst{\ref{aff4}}
\and H.~J.~McCracken\orcid{0000-0002-9489-7765}\inst{\ref{aff17}}
\and E.~Medinaceli\orcid{0000-0002-4040-7783}\inst{\ref{aff5}}
\and A.~Mora\orcid{0000-0002-1922-8529}\inst{\ref{aff21}}
\and P.~W.~Morris\orcid{0000-0002-5186-4381}\inst{\ref{aff22}}
\and S.~Mottet\inst{\ref{aff17}}
\and R.~Nakajima\orcid{0009-0009-1213-7040}\inst{\ref{aff23}}
\and D.~Neumann\orcid{0009-0006-3926-6756}\inst{\ref{aff18}}
\and G.~Polenta\orcid{0000-0003-4067-9196}\inst{\ref{aff24}}
\and M.~Portaluppi\inst{\ref{aff25}}
\and G.~D.~Racca\orcid{0000-0002-9883-8981}\inst{\ref{aff18},\ref{aff25}}
\and F.~Schmidt\inst{\ref{aff26}}
\and M.~Schmidt\inst{\ref{aff27}}
\and C.~Sirignano\orcid{0000-0002-0995-7146}\inst{\ref{aff28},\ref{aff29}}
\and J.~Skottfelt\orcid{0000-0003-1310-8283}\inst{\ref{aff30}}
\and F.~Soldano\inst{\ref{aff17}}
\and I.~Tereno\orcid{0000-0002-4537-6218}\inst{\ref{aff9},\ref{aff10}}
\and T.~Trombetti\orcid{0000-0001-5166-2467}\inst{\ref{aff6},\ref{aff7}}
\and R.~Vavrek\inst{\ref{aff2}}
\and L.~M.~Gaspar~Venancio\orcid{0000-0002-9634-2982}\inst{\ref{aff25}}
\and A.~Zacchei\orcid{0000-0003-0396-1192}\inst{\ref{aff14},\ref{aff31}}
\and S.~Andreon\orcid{0000-0002-2041-8784}\inst{\ref{aff32}}
\and N.~Auricchio\orcid{0000-0003-4444-8651}\inst{\ref{aff5}}
\and C.~Baccigalupi\orcid{0000-0002-8211-1630}\inst{\ref{aff31},\ref{aff14},\ref{aff15},\ref{aff33}}
\and M.~Baldi\orcid{0000-0003-4145-1943}\inst{\ref{aff34},\ref{aff5},\ref{aff35}}
\and A.~Balestra\orcid{0000-0002-6967-261X}\inst{\ref{aff36}}
\and S.~Bardelli\orcid{0000-0002-8900-0298}\inst{\ref{aff5}}
\and R.~Bender\orcid{0000-0001-7179-0626}\inst{\ref{aff37},\ref{aff38}}
\and A.~Biviano\orcid{0000-0002-0857-0732}\inst{\ref{aff14},\ref{aff31}}
\and R.~P.~Blake\inst{\ref{aff39}}
\and E.~Branchini\orcid{0000-0002-0808-6908}\inst{\ref{aff40},\ref{aff41},\ref{aff32}}
\and M.~Brescia\orcid{0000-0001-9506-5680}\inst{\ref{aff42},\ref{aff43}}
\and J.~Brinchmann\orcid{0000-0003-4359-8797}\inst{\ref{aff44},\ref{aff45},\ref{aff46}}
\and S.~Camera\orcid{0000-0003-3399-3574}\inst{\ref{aff47},\ref{aff48},\ref{aff49}}
\and G.~Ca\~nas-Herrera\orcid{0000-0003-2796-2149}\inst{\ref{aff18}}
\and V.~Capobianco\orcid{0000-0002-3309-7692}\inst{\ref{aff49}}
\and C.~Carbone\orcid{0000-0003-0125-3563}\inst{\ref{aff50}}
\and J.~Carretero\orcid{0000-0002-3130-0204}\inst{\ref{aff51},\ref{aff52}}
\and F.~J.~Castander\orcid{0000-0001-7316-4573}\inst{\ref{aff53},\ref{aff54}}
\and M.~Castellano\orcid{0000-0001-9875-8263}\inst{\ref{aff55}}
\and G.~Castignani\orcid{0000-0001-6831-0687}\inst{\ref{aff5}}
\and S.~Cavuoti\orcid{0000-0002-3787-4196}\inst{\ref{aff43},\ref{aff56}}
\and P.-Y.~Chabaud\orcid{0009-0002-7741-4722}\inst{\ref{aff57}}
\and A.~Cimatti\inst{\ref{aff58}}
\and C.~Colodro-Conde\inst{\ref{aff59}}
\and G.~Congedo\orcid{0000-0003-2508-0046}\inst{\ref{aff39}}
\and C.~J.~Conselice\orcid{0000-0003-1949-7638}\inst{\ref{aff60}}
\and L.~Conversi\orcid{0000-0002-6710-8476}\inst{\ref{aff61},\ref{aff2}}
\and Y.~Copin\orcid{0000-0002-5317-7518}\inst{\ref{aff62}}
\and A.~Costille\inst{\ref{aff57}}
\and F.~Courbin\orcid{0000-0003-0758-6510}\inst{\ref{aff63},\ref{aff64},\ref{aff65}}
\and H.~M.~Courtois\orcid{0000-0003-0509-1776}\inst{\ref{aff66}}
\and M.~Cropper\orcid{0000-0003-4571-9468}\inst{\ref{aff4}}
\and A.~Da~Silva\orcid{0000-0002-6385-1609}\inst{\ref{aff9},\ref{aff10}}
\and H.~Degaudenzi\orcid{0000-0002-5887-6799}\inst{\ref{aff67}}
\and G.~De~Lucia\orcid{0000-0002-6220-9104}\inst{\ref{aff14}}
\and H.~Dole\orcid{0000-0002-9767-3839}\inst{\ref{aff68},\ref{aff69}}
\and F.~Dubath\orcid{0000-0002-6533-2810}\inst{\ref{aff67}}
\and X.~Dupac\inst{\ref{aff2}}
\and S.~Dusini\orcid{0000-0002-1128-0664}\inst{\ref{aff29}}
\and S.~Escoffier\orcid{0000-0002-2847-7498}\inst{\ref{aff12}}
\and M.~Fabricius\orcid{0000-0002-7025-6058}\inst{\ref{aff37},\ref{aff38}}
\and M.~Farina\orcid{0000-0002-3089-7846}\inst{\ref{aff70}}
\and R.~Farinelli\inst{\ref{aff5}}
\and S.~Farrens\orcid{0000-0002-9594-9387}\inst{\ref{aff3}}
\and F.~Faustini\orcid{0000-0001-6274-5145}\inst{\ref{aff55}}
\and S.~Ferriol\inst{\ref{aff62}}
\and S.~Fotopoulou\orcid{0000-0002-9686-254X}\inst{\ref{aff71}}
\and N.~Fourmanoit\orcid{0009-0005-6816-6925}\inst{\ref{aff12}}
\and M.~Frailis\orcid{0000-0002-7400-2135}\inst{\ref{aff14}}
\and E.~Franceschi\orcid{0000-0002-0585-6591}\inst{\ref{aff5}}
\and M.~Fumana\orcid{0000-0001-6787-5950}\inst{\ref{aff50}}
\and L.~Gabarra\orcid{0000-0002-8486-8856}\inst{\ref{aff72}}
\and S.~Galeotta\orcid{0000-0002-3748-5115}\inst{\ref{aff14}}
\and K.~George\orcid{0000-0002-1734-8455}\inst{\ref{aff73}}
\and C.~Giocoli\orcid{0000-0002-9590-7961}\inst{\ref{aff5},\ref{aff35}}
\and P.~G\'omez-Alvarez\orcid{0000-0002-8594-5358}\inst{\ref{aff74},\ref{aff2}}
\and J.~Gracia-Carpio\orcid{0000-0003-4689-3134}\inst{\ref{aff37}}
\and B.~R.~Granett\orcid{0000-0003-2694-9284}\inst{\ref{aff32}}
\and A.~Grazian\orcid{0000-0002-5688-0663}\inst{\ref{aff36}}
\and F.~Grupp\inst{\ref{aff37},\ref{aff38}}
\and L.~Guzzo\orcid{0000-0001-8264-5192}\inst{\ref{aff75},\ref{aff32},\ref{aff76}}
\and W.~G.~Hartley\orcid{0000-0002-4563-5101}\inst{\ref{aff67}}
\and S.~V.~H.~Haugan\orcid{0000-0001-9648-7260}\inst{\ref{aff77}}
\and I.~M.~Hook\orcid{0000-0002-2960-978X}\inst{\ref{aff78}}
\and F.~Hormuth\inst{\ref{aff79}}
\and A.~Hornstrup\orcid{0000-0002-3363-0936}\inst{\ref{aff80},\ref{aff81}}
\and M.~Jhabvala\inst{\ref{aff82}}
\and S.~Kermiche\orcid{0000-0002-0302-5735}\inst{\ref{aff12}}
\and A.~Kiessling\orcid{0000-0002-2590-1273}\inst{\ref{aff19}}
\and B.~Kubik\orcid{0009-0006-5823-4880}\inst{\ref{aff62}}
\and M.~K\"ummel\orcid{0000-0003-2791-2117}\inst{\ref{aff38}}
\and M.~Kunz\orcid{0000-0002-3052-7394}\inst{\ref{aff83}}
\and H.~Kurki-Suonio\orcid{0000-0002-4618-3063}\inst{\ref{aff84},\ref{aff85}}
\and A.~M.~C.~Le~Brun\orcid{0000-0002-0936-4594}\inst{\ref{aff86}}
\and V.~Le~Brun\orcid{0000-0002-5027-1939}\inst{\ref{aff57}}
\and S.~Ligori\orcid{0000-0003-4172-4606}\inst{\ref{aff49}}
\and P.~B.~Lilje\orcid{0000-0003-4324-7794}\inst{\ref{aff77}}
\and V.~Lindholm\orcid{0000-0003-2317-5471}\inst{\ref{aff84},\ref{aff85}}
\and I.~Lloro\orcid{0000-0001-5966-1434}\inst{\ref{aff87}}
\and M.~Magliocchetti\orcid{0000-0001-9158-4838}\inst{\ref{aff70}}
\and G.~Mainetti\orcid{0000-0003-2384-2377}\inst{\ref{aff88}}
\and O.~Mansutti\orcid{0000-0001-5758-4658}\inst{\ref{aff14}}
\and S.~Marcin\inst{\ref{aff89}}
\and O.~Marggraf\orcid{0000-0001-7242-3852}\inst{\ref{aff23}}
\and M.~Martinelli\orcid{0000-0002-6943-7732}\inst{\ref{aff55},\ref{aff90}}
\and N.~Martinet\orcid{0000-0003-2786-7790}\inst{\ref{aff57}}
\and F.~Marulli\orcid{0000-0002-8850-0303}\inst{\ref{aff91},\ref{aff5},\ref{aff35}}
\and R.~J.~Massey\orcid{0000-0002-6085-3780}\inst{\ref{aff92}}
\and N.~Mauri\orcid{0000-0001-8196-1548}\inst{\ref{aff58},\ref{aff35}}
\and S.~Mei\orcid{0000-0002-2849-559X}\inst{\ref{aff93},\ref{aff94}}
\and M.~Melchior\inst{\ref{aff95}}
\and M.~Meneghetti\orcid{0000-0003-1225-7084}\inst{\ref{aff5},\ref{aff35}}
\and E.~Merlin\orcid{0000-0001-6870-8900}\inst{\ref{aff36}}
\and G.~Meylan\orcid{0000-0001-6503-0209}\inst{\ref{aff96}}
\and P.~Monaco\orcid{0000-0003-2083-7564}\inst{\ref{aff13},\ref{aff14},\ref{aff15},\ref{aff31}}
\and M.~Moresco\orcid{0000-0002-7616-7136}\inst{\ref{aff91},\ref{aff5}}
\and C.~Moretti\orcid{0000-0003-3314-8936}\inst{\ref{aff14},\ref{aff31},\ref{aff15}}
\and L.~Moscardini\orcid{0000-0002-3473-6716}\inst{\ref{aff91},\ref{aff5},\ref{aff35}}
\and C.~Neissner\orcid{0000-0001-8524-4968}\inst{\ref{aff97},\ref{aff52}}
\and R.~C.~Nichol\orcid{0000-0003-0939-6518}\inst{\ref{aff98}}
\and S.-M.~Niemi\orcid{0009-0005-0247-0086}\inst{\ref{aff25}}
\and C.~Padilla\orcid{0000-0001-7951-0166}\inst{\ref{aff97}}
\and S.~Paltani\orcid{0000-0002-8108-9179}\inst{\ref{aff67}}
\and F.~Pasian\orcid{0000-0002-4869-3227}\inst{\ref{aff14}}
\and W.~J.~Percival\orcid{0000-0002-0644-5727}\inst{\ref{aff99},\ref{aff100},\ref{aff101}}
\and V.~Pettorino\orcid{0000-0002-4203-9320}\inst{\ref{aff25}}
\and S.~Pires\orcid{0000-0002-0249-2104}\inst{\ref{aff3}}
\and M.~Poncet\inst{\ref{aff8}}
\and L.~A.~Popa\inst{\ref{aff102}}
\and F.~Raison\orcid{0000-0002-7819-6918}\inst{\ref{aff37}}
\and A.~Refregier\inst{\ref{aff103}}
\and A.~Renzi\orcid{0000-0001-9856-1970}\inst{\ref{aff28},\ref{aff29},\ref{aff5}}
\and J.~Rhodes\orcid{0000-0002-4485-8549}\inst{\ref{aff19}}
\and G.~Riccio\inst{\ref{aff43}}
\and I.~Risso\orcid{0000-0003-2525-7761}\inst{\ref{aff40},\ref{aff41},\ref{aff32}}
\and F.~Rizzo\orcid{0000-0002-9407-585X}\inst{\ref{aff14}}
\and E.~Romelli\orcid{0000-0003-3069-9222}\inst{\ref{aff14}}
\and M.~Roncarelli\orcid{0000-0001-9587-7822}\inst{\ref{aff5}}
\and C.~Rosset\orcid{0000-0003-0286-2192}\inst{\ref{aff93}}
\and B.~Rusholme\orcid{0000-0001-7648-4142}\inst{\ref{aff16}}
\and R.~Saglia\orcid{0000-0003-0378-7032}\inst{\ref{aff38},\ref{aff37}}
\and Z.~Sakr\orcid{0000-0002-4823-3757}\inst{\ref{aff104},\ref{aff105},\ref{aff106}}
\and A.~G.~S\'anchez\orcid{0000-0003-1198-831X}\inst{\ref{aff37}}
\and D.~Sapone\orcid{0000-0001-7089-4503}\inst{\ref{aff107}}
\and B.~Sartoris\orcid{0000-0003-1337-5269}\inst{\ref{aff108}}
\and P.~Schneider\orcid{0000-0001-8561-2679}\inst{\ref{aff23}}
\and T.~Schrabback\orcid{0000-0002-6987-7834}\inst{\ref{aff109}}
\and M.~Scodeggio\inst{\ref{aff50}}
\and A.~Secroun\orcid{0000-0003-0505-3710}\inst{\ref{aff12}}
\and E.~Sihvola\orcid{0000-0003-1804-7715}\inst{\ref{aff84},\ref{aff85}}
\and P.~Simon\inst{\ref{aff23}}
\and G.~Sirri\orcid{0000-0003-2626-2853}\inst{\ref{aff35}}
\and L.~Stanco\orcid{0000-0002-9706-5104}\inst{\ref{aff29}}
\and P.~Tallada-Cresp\'{i}\orcid{0000-0002-1336-8328}\inst{\ref{aff51},\ref{aff52}}
\and A.~N.~Taylor\inst{\ref{aff39}}
\and H.~I.~Teplitz\orcid{0000-0002-7064-5424}\inst{\ref{aff110}}
\and N.~Tessore\orcid{0000-0002-9696-7931}\inst{\ref{aff4}}
\and S.~Toft\orcid{0000-0003-3631-7176}\inst{\ref{aff111},\ref{aff112}}
\and R.~Toledo-Moreo\orcid{0000-0002-2997-4859}\inst{\ref{aff113},\ref{aff114}}
\and F.~Torradeflot\orcid{0000-0003-1160-1517}\inst{\ref{aff52},\ref{aff51}}
\and I.~Tutusaus\orcid{0000-0002-3199-0399}\inst{\ref{aff53},\ref{aff54},\ref{aff105}}
\and J.~Valiviita\orcid{0000-0001-6225-3693}\inst{\ref{aff84},\ref{aff85}}
\and T.~Vassallo\orcid{0000-0001-6512-6358}\inst{\ref{aff14},\ref{aff73}}
\and A.~Veropalumbo\orcid{0000-0003-2387-1194}\inst{\ref{aff32},\ref{aff41},\ref{aff40}}
\and D.~Vibert\orcid{0009-0008-0607-631X}\inst{\ref{aff57}}
\and Y.~Wang\orcid{0000-0002-4749-2984}\inst{\ref{aff16}}
\and J.~Weller\orcid{0000-0002-8282-2010}\inst{\ref{aff38},\ref{aff37}}
\and G.~Zamorani\orcid{0000-0002-2318-301X}\inst{\ref{aff5}}
\and F.~M.~Zerbi\orcid{0000-0002-9996-973X}\inst{\ref{aff32}}
\and E.~Zucca\orcid{0000-0002-5845-8132}\inst{\ref{aff5}}
\and I.~T.~Andika\orcid{0000-0001-6102-9526}\inst{\ref{aff38}}
\and M.~Ballardini\orcid{0000-0003-4481-3559}\inst{\ref{aff115},\ref{aff116},\ref{aff5}}
\and P.~Bergamini\orcid{0000-0003-1383-9414}\inst{\ref{aff5}}
\and S.~Borgani\orcid{0000-0001-6151-6439}\inst{\ref{aff13},\ref{aff31},\ref{aff14},\ref{aff15},\ref{aff117}}
\and E.~Bozzo\orcid{0000-0002-8201-1525}\inst{\ref{aff67}}
\and R.~Cabanac\orcid{0000-0001-6679-2600}\inst{\ref{aff105}}
\and M.~Calabrese\orcid{0000-0002-2637-2422}\inst{\ref{aff118},\ref{aff50}}
\and F.~Caro\orcid{0009-0003-1053-0507}\inst{\ref{aff55}}
\and T.~Castro\orcid{0000-0002-6292-3228}\inst{\ref{aff119},\ref{aff14},\ref{aff31}}
\and S.~Conseil\orcid{0000-0002-3657-4191}\inst{\ref{aff120},\ref{aff62}}
\and D.~Di~Ferdinando\inst{\ref{aff35}}
\and J.~A.~Escartin~Vigo\inst{\ref{aff37}}
\and G.~Fabbian\orcid{0000-0002-3255-4695}\inst{\ref{aff68}}
\and T.~Flanet\inst{\ref{aff17}}
\and J.~Garc\'ia-Bellido\orcid{0000-0002-9370-8360}\inst{\ref{aff104}}
\and T.~Gasparetto\orcid{0000-0002-7913-4866}\inst{\ref{aff55}}
\and C.~M.~Gutierrez\orcid{0000-0001-7854-783X}\inst{\ref{aff59},\ref{aff121}}
\and A.~Hall\orcid{0000-0002-3139-8651}\inst{\ref{aff39}}
\and Y.~Kang\orcid{0009-0000-8588-7250}\inst{\ref{aff109}}
\and G.~Leroy\orcid{0009-0004-2523-4425}\inst{\ref{aff122},\ref{aff92}}
\and A.~Loureiro\orcid{0000-0002-4371-0876}\inst{\ref{aff123},\ref{aff124}}
\and J.~Macias-Perez\orcid{0000-0002-5385-2763}\inst{\ref{aff125}}
\and R.~Maoli\orcid{0000-0002-6065-3025}\inst{\ref{aff126},\ref{aff55}}
\and R.~B.~Metcalf\orcid{0000-0003-3167-2574}\inst{\ref{aff91},\ref{aff5}}
\and S.~Nadathur\orcid{0000-0001-9070-3102}\inst{\ref{aff127}}
\and L.~Patrizii\inst{\ref{aff35}}
\and M.~P\"ontinen\orcid{0000-0001-5442-2530}\inst{\ref{aff84}}
\and E.~Sarpa\orcid{0000-0002-1256-655X}\inst{\ref{aff14}}
\and V.~Scottez\orcid{0009-0008-3864-940X}\inst{\ref{aff128},\ref{aff129}}
\and M.~Sereno\orcid{0000-0003-0302-0325}\inst{\ref{aff5},\ref{aff35}}
\and K.~Tanidis\orcid{0000-0001-9843-5130}\inst{\ref{aff130}}
\and M.~Tenti\orcid{0000-0002-4254-5901}\inst{\ref{aff35}}
\and A.~Troja\orcid{0000-0003-0239-4595}\inst{\ref{aff14}}
\and M.~Tucci\inst{\ref{aff67}}
\and M.~Viel\orcid{0000-0002-2642-5707}\inst{\ref{aff31},\ref{aff14},\ref{aff33},\ref{aff15},\ref{aff117}}
\and M.~Wiesmann\orcid{0009-0000-8199-5860}\inst{\ref{aff77}}
\and Abdurro'uf\orcid{0000-0002-5258-8761}\inst{\ref{aff131}}
\and J.~A.~Acevedo~Barroso\orcid{0000-0002-9654-1711}\inst{\ref{aff19}}
\and Y.~Akrami\orcid{0000-0002-2407-7956}\inst{\ref{aff104},\ref{aff132}}
\and G.~Alguero\inst{\ref{aff125}}
\and G.~Angora\orcid{0000-0002-0316-6562}\inst{\ref{aff43},\ref{aff115}}
\and S.~Anselmi\orcid{0000-0002-3579-9583}\inst{\ref{aff29},\ref{aff28},\ref{aff133}}
\and M.~Archidiacono\orcid{0000-0003-4952-9012}\inst{\ref{aff75},\ref{aff76}}
\and G.~Aric\`o\orcid{0000-0002-2802-2928}\inst{\ref{aff35}}
\and F.~Atrio-Barandela\orcid{0000-0002-2130-2513}\inst{\ref{aff134}}
\and M.~Aubert\orcid{0009-0002-7667-8814}\inst{\ref{aff62}}
\and E.~Aubourg\orcid{0000-0002-5592-023X}\inst{\ref{aff93},\ref{aff135}}
\and M.~Baes\orcid{0000-0002-3930-2757}\inst{\ref{aff136}}
\and N.~Bartolo\inst{\ref{aff28},\ref{aff29},\ref{aff36}}
\and L.~Bazzanini\orcid{0000-0003-0727-0137}\inst{\ref{aff115},\ref{aff5}}
\and J.~R.~Bermejo-Climent\inst{\ref{aff137},\ref{aff138}}
\and D.~Bertacca\orcid{0000-0002-2490-7139}\inst{\ref{aff28},\ref{aff36},\ref{aff29}}
\and M.~Bethermin\orcid{0000-0002-3915-2015}\inst{\ref{aff139}}
\and F.~Beutler\orcid{0000-0003-0467-5438}\inst{\ref{aff39}}
\and A.~Blanchard\orcid{0000-0001-8555-9003}\inst{\ref{aff105}}
\and M.~Bonici\orcid{0000-0002-8430-126X}\inst{\ref{aff99},\ref{aff50}}
\and F.~Bouch\`e\orcid{0000-0002-4663-1786}\inst{\ref{aff140},\ref{aff56}}
\and M.~L.~Brown\orcid{0000-0002-0370-8077}\inst{\ref{aff60}}
\and S.~Bruton\orcid{0000-0002-6503-5218}\inst{\ref{aff22}}
\and A.~Calabro\orcid{0000-0003-2536-1614}\inst{\ref{aff55}}
\and C.~S.~Carvalho\inst{\ref{aff141}}
\and N.~Chartab\orcid{0000-0003-3691-937X}\inst{\ref{aff16}}
\and F.~Cogato\orcid{0000-0003-4632-6113}\inst{\ref{aff5}}
\and T.~E.~Collett\orcid{0000-0001-5564-3140}\inst{\ref{aff127}}
\and A.~R.~Cooray\orcid{0000-0002-3892-0190}\inst{\ref{aff142}}
\and P.~Corcho-Caballero\orcid{0000-0001-6327-7080}\inst{\ref{aff20}}
\and M.~Correnti\orcid{0000-0001-6464-3257}\inst{\ref{aff55},\ref{aff24}}
\and B.~Csizi\orcid{0000-0003-3227-6581}\inst{\ref{aff109}}
\and O.~Cucciati\orcid{0000-0002-9336-7551}\inst{\ref{aff5}}
\and H.~Dannerbauer\orcid{0000-0001-7147-3575}\inst{\ref{aff59},\ref{aff121}}
\and J.~E.~Davies\orcid{0000-0002-5079-9098}\inst{\ref{aff1}}
\and T.~de~Boer\orcid{0000-0001-5486-2747}\inst{\ref{aff143}}
\and G.~Desprez\orcid{0000-0001-8325-1742}\inst{\ref{aff20}}
\and A.~D\'iaz-S\'anchez\orcid{0000-0003-0748-4768}\inst{\ref{aff144}}
\and S.~Di~Domizio\orcid{0000-0003-2863-5895}\inst{\ref{aff40},\ref{aff41}}
\and J.~M.~Diego\orcid{0000-0001-9065-3926}\inst{\ref{aff145}}
\and P.~Dimauro\orcid{0000-0001-7399-2854}\inst{\ref{aff55},\ref{aff146}}
\and P.-A.~Duc\orcid{0000-0003-3343-6284}\inst{\ref{aff139}}
\and V.~Duret\orcid{0009-0009-0383-4960}\inst{\ref{aff147}}
\and Y.~Fang\orcid{0000-0002-0334-6950}\inst{\ref{aff38}}
\and A.~Farina\orcid{0009-0000-3420-929X}\inst{\ref{aff32},\ref{aff41}}
\and A.~M.~N.~Ferguson\orcid{0000-0001-7934-1278}\inst{\ref{aff39}}
\and A.~Finoguenov\orcid{0000-0002-4606-5403}\inst{\ref{aff84}}
\and A.~Franco\orcid{0000-0002-4761-366X}\inst{\ref{aff148},\ref{aff149},\ref{aff150}}
\and N.~Frusciante\orcid{0000-0002-7375-1230}\inst{\ref{aff151},\ref{aff152}}
\and A.~Fumagalli\orcid{0009-0004-0300-2535}\inst{\ref{aff14}}
\and Y.~Fu\orcid{0000-0002-0759-0504}\inst{\ref{aff18},\ref{aff20}}
\and K.~Ganga\orcid{0000-0001-8159-8208}\inst{\ref{aff93}}
\and R.~Gavazzi\orcid{0000-0002-5540-6935}\inst{\ref{aff57},\ref{aff17}}
\and E.~Gaztanaga\orcid{0000-0001-9632-0815}\inst{\ref{aff53},\ref{aff54},\ref{aff127}}
\and Z.~Ghaffari\orcid{0000-0002-6467-8078}\inst{\ref{aff14},\ref{aff31}}
\and F.~Giacomini\orcid{0000-0002-3129-2814}\inst{\ref{aff35}}
\and F.~Gianotti\orcid{0000-0003-4666-119X}\inst{\ref{aff5}}
\and G.~Gozaliasl\orcid{0000-0002-0236-919X}\inst{\ref{aff153},\ref{aff84}}
\and M.~Guidi\orcid{0000-0001-9408-1101}\inst{\ref{aff34},\ref{aff5}}
\and N.~A.~Hatch\orcid{0000-0001-5600-0534}\inst{\ref{aff154}}
\and C.~Hern\'andez-Monteagudo\orcid{0000-0001-5471-9166}\inst{\ref{aff121},\ref{aff59}}
\and T.~Heuberger\orcid{0009-0009-3474-8467}\inst{\ref{aff89}}
\and J.~Hjorth\orcid{0000-0002-4571-2306}\inst{\ref{aff155}}
\and M.~Jauzac\orcid{0000-0003-1974-8732}\inst{\ref{aff122},\ref{aff92},\ref{aff156},\ref{aff157},\ref{aff105}}
\and S.~Joudaki\orcid{0000-0001-8820-673X}\inst{\ref{aff51}}
\and J.~J.~E.~Kajava\orcid{0000-0002-3010-8333}\inst{\ref{aff158},\ref{aff159},\ref{aff160}}
\and D.~Karagiannis\orcid{0000-0002-4927-0816}\inst{\ref{aff161},\ref{aff115},\ref{aff162}}
\and J.~Kim\orcid{0000-0003-2776-2761}\inst{\ref{aff72}}
\and C.~C.~Kirkpatrick\inst{\ref{aff163}}
\and A.~Kov\'acs\orcid{0000-0002-5825-579X}\inst{\ref{aff138},\ref{aff137}}
\and I.~Kova{\v{c}}i{\'{c}}\orcid{0000-0001-6751-3263}\inst{\ref{aff136}}
\and K.~Koyama\orcid{0000-0001-6727-6915}\inst{\ref{aff127}}
\and S.~Kruk\orcid{0000-0001-8010-8879}\inst{\ref{aff2}}
\and M.~C.~Lam\orcid{0000-0002-9347-2298}\inst{\ref{aff39}}
\and F.~Leclercq\orcid{0000-0002-9339-1404}\inst{\ref{aff17}}
\and S.~Lee\orcid{0000-0002-8289-740X}\inst{\ref{aff164},\ref{aff19}}
\and J.~Le~Graet\orcid{0000-0001-6523-7971}\inst{\ref{aff12}}
\and J.~Lesgourgues\orcid{0000-0001-7627-353X}\inst{\ref{aff165}}
\and T.~I.~Liaudat\orcid{0000-0002-9104-314X}\inst{\ref{aff135}}
\and L.~Linke\orcid{0000-0002-2622-8113}\inst{\ref{aff109}}
\and A.~Manj\'on-Garc\'ia\orcid{0000-0002-7413-8825}\inst{\ref{aff144}}
\and F.~Mannucci\orcid{0000-0002-4803-2381}\inst{\ref{aff166}}
\and M.~Marinucci\orcid{0000-0003-1159-3756}\inst{\ref{aff167}}
\and F.~R.~Marleau\orcid{0000-0002-1442-2947}\inst{\ref{aff109}}
\and C.~J.~A.~P.~Martins\orcid{0000-0002-4886-9261}\inst{\ref{aff168},\ref{aff45}}
\and M.~Migliaccio\inst{\ref{aff169},\ref{aff170}}
\and M.~Miluzio\inst{\ref{aff2},\ref{aff171}}
\and G.~Morgante\inst{\ref{aff5}}
\and K.~Naidoo\orcid{0000-0002-9182-1802}\inst{\ref{aff127},\ref{aff1}}
\and P.~Natoli\orcid{0000-0003-0126-9100}\inst{\ref{aff115},\ref{aff116}}
\and A.~Navarro-Alsina\orcid{0000-0002-3173-2592}\inst{\ref{aff23}}
\and A.~Nersesian\orcid{0000-0001-6843-409X}\inst{\ref{aff172},\ref{aff136}}
\and S.~Nesseris\orcid{0000-0002-0567-0324}\inst{\ref{aff104}}
\and L.~Nicastro\orcid{0000-0001-8534-6788}\inst{\ref{aff5}}
\and F.~Oppizzi\orcid{0000-0003-3904-8370}\inst{\ref{aff41},\ref{aff29},\ref{aff28}}
\and F.~Pace\orcid{0000-0001-8039-0480}\inst{\ref{aff47},\ref{aff48},\ref{aff49},\ref{aff10}}
\and M.~J.~Page\orcid{0000-0002-6689-6271}\inst{\ref{aff4}}
\and D.~Paoletti\orcid{0000-0003-4761-6147}\inst{\ref{aff5},\ref{aff7}}
\and F.~Passalacqua\orcid{0000-0002-8606-4093}\inst{\ref{aff29}}
\and C.~Pattison\orcid{0000-0003-3272-2617}\inst{\ref{aff127}}
\and A.~Pisani\orcid{0000-0002-6146-4437}\inst{\ref{aff12}}
\and L.~Pizzuti\orcid{0000-0001-5654-7580}\inst{\ref{aff173}}
\and D.~Potter\orcid{0000-0002-0757-5195}\inst{\ref{aff174}}
\and M.~Poulain\orcid{0000-0002-7664-4510}\inst{\ref{aff175}}
\and G.~W.~Pratt\inst{\ref{aff3}}
\and S.~Quai\orcid{0000-0002-0449-8163}\inst{\ref{aff91},\ref{aff5}}
\and M.~Radovich\orcid{0000-0002-3585-866X}\inst{\ref{aff36}}
\and G.~Rodighiero\orcid{0000-0002-9415-2296}\inst{\ref{aff28},\ref{aff36}}
\and W.~Roster\orcid{0000-0002-9149-6528}\inst{\ref{aff37}}
\and M.~Ruffenach\orcid{0000-0001-8772-7330}\inst{\ref{aff8}}
\and M.~Sahl\'en\orcid{0000-0003-0973-4804}\inst{\ref{aff176}}
\and D.~B.~Sanders\orcid{0000-0002-1233-9998}\inst{\ref{aff143}}
\and A.~Schneider\orcid{0000-0001-7055-8104}\inst{\ref{aff174}}
\and D.~Sciotti\orcid{0009-0008-4519-2620}\inst{\ref{aff55},\ref{aff90}}
\and E.~Sellentin\orcid{0009-0002-2655-3458}\inst{\ref{aff177},\ref{aff18}}
\and S.~Serjeant\orcid{0000-0002-0517-7943}\inst{\ref{aff178}}
\and A.~Shulevski\orcid{0000-0002-1827-0469}\inst{\ref{aff179},\ref{aff20},\ref{aff180}}
\and L.~C.~Smith\orcid{0000-0002-3259-2771}\inst{\ref{aff181}}
\and I.~Szapudi\orcid{0000-0003-2274-0301}\inst{\ref{aff143}}
\and M.~Talia\orcid{0000-0003-4352-2063}\inst{\ref{aff91},\ref{aff5}}
\and G.~Testera\orcid{0000-0003-2970-766X}\inst{\ref{aff41}}
\and M.~Tewes\orcid{0000-0002-1155-8689}\inst{\ref{aff23}}
\and R.~Teyssier\orcid{0000-0001-7689-0933}\inst{\ref{aff182}}
\and S.~Tosi\orcid{0000-0002-7275-9193}\inst{\ref{aff40},\ref{aff32},\ref{aff41}}
\and C.~Uhlemann\orcid{0000-0001-7831-1579}\inst{\ref{aff183}}
\and C.~Valieri\inst{\ref{aff35}}
\and A.~Venhola\orcid{0000-0001-6071-4564}\inst{\ref{aff175}}
\and D.~Vergani\orcid{0000-0003-0898-2216}\inst{\ref{aff5}}
\and E.~Vilenius\orcid{0000-0002-6184-7681}\inst{\ref{aff4}}
\and S.~Vinciguerra\orcid{0009-0005-4018-3184}\inst{\ref{aff57}}
\and M.~von~Wietersheim-Kramsta\orcid{0000-0003-4986-5091}\inst{\ref{aff92},\ref{aff122}}
\and A.~H.~Wright\orcid{0000-0001-7363-7932}\inst{\ref{aff184}}
\and H.~W.~Yeung\orcid{0000-0002-4993-9014}\inst{\ref{aff39}}
\and D.~Scott\orcid{0000-0002-6878-9840}\inst{\ref{aff185}}}
										   
\institute{Max-Planck-Institut f\"ur Astronomie, K\"onigstuhl 17, 69117 Heidelberg, Germany\label{aff1}
\and
ESAC/ESA, Camino Bajo del Castillo, s/n., Urb. Villafranca del Castillo, 28692 Villanueva de la Ca\~nada, Madrid, Spain\label{aff2}
\and
Universit\'e Paris-Saclay, Universit\'e Paris Cit\'e, CEA, CNRS, AIM, 91191, Gif-sur-Yvette, France\label{aff3}
\and
Mullard Space Science Laboratory, University College London, Holmbury St Mary, Dorking, Surrey RH5 6NT, UK\label{aff4}
\and
INAF-Osservatorio di Astrofisica e Scienza dello Spazio di Bologna, Via Piero Gobetti 93/3, 40129 Bologna, Italy\label{aff5}
\and
INAF, Istituto di Radioastronomia, Via Piero Gobetti 101, 40129 Bologna, Italy\label{aff6}
\and
INFN-Bologna, Via Irnerio 46, 40126 Bologna, Italy\label{aff7}
\and
Centre National d'Etudes Spatiales -- Centre spatial de Toulouse, 18 avenue Edouard Belin, 31401 Toulouse Cedex 9, France\label{aff8}
\and
Departamento de F\'isica, Faculdade de Ci\^encias, Universidade de Lisboa, Edif\'icio C8, Campo Grande, PT1749-016 Lisboa, Portugal\label{aff9}
\and
Instituto de Astrof\'isica e Ci\^encias do Espa\c{c}o, Faculdade de Ci\^encias, Universidade de Lisboa, Campo Grande, 1749-016 Lisboa, Portugal\label{aff10}
\and
Aurora Technology for European Space Agency (ESA), Camino bajo del Castillo, s/n, Urbanizacion Villafranca del Castillo, Villanueva de la Ca\~nada, 28692 Madrid, Spain\label{aff11}
\and
Aix-Marseille Universit\'e, CNRS/IN2P3, CPPM, Marseille, France\label{aff12}
\and
Dipartimento di Fisica - Sezione di Astronomia, Universit\`a di Trieste, Via Tiepolo 11, 34131 Trieste, Italy\label{aff13}
\and
INAF-Osservatorio Astronomico di Trieste, Via G. B. Tiepolo 11, 34143 Trieste, Italy\label{aff14}
\and
INFN, Sezione di Trieste, Via Valerio 2, 34127 Trieste TS, Italy\label{aff15}
\and
Caltech/IPAC, 1200 E. California Blvd., Pasadena, CA 91125, USA\label{aff16}
\and
Institut d'Astrophysique de Paris, UMR 7095, CNRS, and Sorbonne Universit\'e, 98 bis boulevard Arago, 75014 Paris, France\label{aff17}
\and
Leiden Observatory, Leiden University, Einsteinweg 55, 2333 CC Leiden, The Netherlands\label{aff18}
\and
Jet Propulsion Laboratory, California Institute of Technology, 4800 Oak Grove Drive, Pasadena, CA, 91109, USA\label{aff19}
\and
Kapteyn Astronomical Institute, University of Groningen, PO Box 800, 9700 AV Groningen, The Netherlands\label{aff20}
\and
Telespazio UK S.L. for European Space Agency (ESA), Camino bajo del Castillo, s/n, Urbanizacion Villafranca del Castillo, Villanueva de la Ca\~nada, 28692 Madrid, Spain\label{aff21}
\and
California Institute of Technology, 1200 E California Blvd, Pasadena, CA 91125, USA\label{aff22}
\and
Universit\"at Bonn, Argelander-Institut f\"ur Astronomie, Auf dem H\"ugel 71, 53121 Bonn, Germany\label{aff23}
\and
Space Science Data Center, Italian Space Agency, via del Politecnico snc, 00133 Roma, Italy\label{aff24}
\and
European Space Agency/ESTEC, Keplerlaan 1, 2201 AZ Noordwijk, The Netherlands\label{aff25}
\and
Telespazio Germany GmbH, Europapl. 5, 64293 Darmstadt, Germany\label{aff26}
\and
European Space Agency/ESOC, Robert-Bosch-Str. 5, 64293 Darmstadt, Germany\label{aff27}
\and
Dipartimento di Fisica e Astronomia "G. Galilei", Universit\`a di Padova, Via Marzolo 8, 35131 Padova, Italy\label{aff28}
\and
INFN-Padova, Via Marzolo 8, 35131 Padova, Italy\label{aff29}
\and
Centre for Electronic Imaging, Open University, Walton Hall, Milton Keynes, MK7~6AA, UK\label{aff30}
\and
IFPU, Institute for Fundamental Physics of the Universe, via Beirut 2, 34151 Trieste, Italy\label{aff31}
\and
INAF-Osservatorio Astronomico di Brera, Via Brera 28, 20122 Milano, Italy\label{aff32}
\and
SISSA, International School for Advanced Studies, Via Bonomea 265, 34136 Trieste TS, Italy\label{aff33}
\and
Dipartimento di Fisica e Astronomia, Universit\`a di Bologna, Via Gobetti 93/2, 40129 Bologna, Italy\label{aff34}
\and
INFN-Sezione di Bologna, Viale Berti Pichat 6/2, 40127 Bologna, Italy\label{aff35}
\and
INAF-Osservatorio Astronomico di Padova, Via dell'Osservatorio 5, 35122 Padova, Italy\label{aff36}
\and
Max Planck Institute for Extraterrestrial Physics, Giessenbachstr. 1, 85748 Garching, Germany\label{aff37}
\and
Universit\"ats-Sternwarte M\"unchen, Fakult\"at f\"ur Physik, Ludwig-Maximilians-Universit\"at M\"unchen, Scheinerstr.~1, 81679 M\"unchen, Germany\label{aff38}
\and
Institute for Astronomy, University of Edinburgh, Royal Observatory, Blackford Hill, Edinburgh EH9 3HJ, UK\label{aff39}
\and
Dipartimento di Fisica, Universit\`a di Genova, Via Dodecaneso 33, 16146, Genova, Italy\label{aff40}
\and
INFN-Sezione di Genova, Via Dodecaneso 33, 16146, Genova, Italy\label{aff41}
\and
Department of Physics "E. Pancini", University Federico II, Via Cinthia 6, 80126, Napoli, Italy\label{aff42}
\and
INAF-Osservatorio Astronomico di Capodimonte, Via Moiariello 16, 80131 Napoli, Italy\label{aff43}
\and
European Southern Observatory, Karl-Schwarzschild-Str.~2, 85748 Garching, Germany\label{aff44}
\and
Instituto de Astrof\'isica e Ci\^encias do Espa\c{c}o, Universidade do Porto, CAUP, Rua das Estrelas, PT4150-762 Porto, Portugal\label{aff45}
\and
Faculdade de Ci\^encias da Universidade do Porto, Rua do Campo de Alegre, 4150-007 Porto, Portugal\label{aff46}
\and
Dipartimento di Fisica, Universit\`a degli Studi di Torino, Via P. Giuria 1, 10125 Torino, Italy\label{aff47}
\and
INFN-Sezione di Torino, Via P. Giuria 1, 10125 Torino, Italy\label{aff48}
\and
INAF-Osservatorio Astrofisico di Torino, Via Osservatorio 20, 10025 Pino Torinese (TO), Italy\label{aff49}
\and
INAF-IASF Milano, Via Alfonso Corti 12, 20133 Milano, Italy\label{aff50}
\and
Centro de Investigaciones Energ\'eticas, Medioambientales y Tecnol\'ogicas (CIEMAT), Avenida Complutense 40, 28040 Madrid, Spain\label{aff51}
\and
Port d'Informaci\'{o} Cient\'{i}fica, Campus UAB, C. Albareda s/n, 08193 Bellaterra (Barcelona), Spain\label{aff52}
\and
Institute of Space Sciences (ICE, CSIC), Campus UAB, Carrer de Can Magrans, s/n, 08193 Barcelona, Spain\label{aff53}
\and
Institut d'Estudis Espacials de Catalunya (IEEC),  Edifici RDIT, Campus UPC, 08860 Castelldefels, Barcelona, Spain\label{aff54}
\and
INAF-Osservatorio Astronomico di Roma, Via Frascati 33, 00078 Monteporzio Catone, Italy\label{aff55}
\and
INFN -- Sezione di Napoli, Via Cinthia 6, 80126, Napoli, Italy\label{aff56}
\and
Aix-Marseille Universit\'e, CNRS, CNES, LAM, Marseille, France\label{aff57}
\and
Dipartimento di Fisica e Astronomia "Augusto Righi" - Alma Mater Studiorum Universit\`a di Bologna, Viale Berti Pichat 6/2, 40127 Bologna, Italy\label{aff58}
\and
Instituto de Astrof\'{\i}sica de Canarias, E-38205 La Laguna, Tenerife, Spain\label{aff59}
\and
Jodrell Bank Centre for Astrophysics, Department of Physics and Astronomy, University of Manchester, Oxford Road, Manchester M13 9PL, UK\label{aff60}
\and
European Space Agency/ESRIN, Largo Galileo Galilei 1, 00044 Frascati, Roma, Italy\label{aff61}
\and
Universit\'e Lyon 1, CNRS/IN2P3, IP2I Lyon, Villeurbanne, France\label{aff62}
\and
Institut de Ci\`{e}ncies del Cosmos (ICCUB), Universitat de Barcelona (IEEC-UB), Mart\'{i} i Franqu\`{e}s 1, 08028 Barcelona, Spain\label{aff63}
\and
Instituci\'o Catalana de Recerca i Estudis Avan\c{c}ats (ICREA), Passeig de Llu\'{\i}s Companys 23, 08010 Barcelona, Spain\label{aff64}
\and
Institut de Ciencies de l'Espai (IEEC-CSIC), Campus UAB, Carrer de Can Magrans, s/n Cerdanyola del Vall\'es, 08193 Barcelona, Spain\label{aff65}
\and
Universit\'e Lyon 1, CNRS/IN2P3, IUF, IP2I Lyon, Villeurbanne, France \label{aff66}
\and
Department of Astronomy, University of Geneva, ch. d'Ecogia 16, 1290 Versoix, Switzerland\label{aff67}
\and
Universit\'e Paris-Saclay, CNRS, Institut d'astrophysique spatiale, 91405, Orsay, France\label{aff68}
\and
Institut universitaire de France (IUF), 1 rue Descartes, 75231 PARIS CEDEX 05, France\label{aff69}
\and
INAF-Istituto di Astrofisica e Planetologia Spaziali, via del Fosso del Cavaliere, 100, 00100 Roma, Italy\label{aff70}
\and
School of Physics, HH Wills Physics Laboratory, University of Bristol, Tyndall Avenue, Bristol, BS8 1TL, UK\label{aff71}
\and
Department of Physics, University of Oxford, Keble Road, Oxford OX1 3RH, UK\label{aff72}
\and
University Observatory, LMU Faculty of Physics, Scheinerstr.~1, 81679 Munich, Germany\label{aff73}
\and
FRACTAL S.L.N.E., calle Tulip\'an 2, Portal 13 1A, 28231, Las Rozas de Madrid, Spain\label{aff74}
\and
Dipartimento di Fisica "Aldo Pontremoli", Universit\`a degli Studi di Milano, Via Celoria 16, 20133 Milano, Italy\label{aff75}
\and
INFN-Sezione di Milano, Via Celoria 16, 20133 Milano, Italy\label{aff76}
\and
Institute of Theoretical Astrophysics, University of Oslo, P.O. Box 1029 Blindern, 0315 Oslo, Norway\label{aff77}
\and
School of Physics and Astronomy, Lancaster University, Lancaster, LA1 4YB, UK\label{aff78}
\and
Felix Hormuth Engineering, Goethestr. 17, 69181 Leimen, Germany\label{aff79}
\and
Technical University of Denmark, Elektrovej 327, 2800 Kgs. Lyngby, Denmark\label{aff80}
\and
Cosmic Dawn Center (DAWN), Denmark\label{aff81}
\and
NASA Goddard Space Flight Center, Greenbelt, MD 20771, USA\label{aff82}
\and
Universit\'e de Gen\`eve, D\'epartement de Physique Th\'eorique and Centre for Astroparticle Physics, 24 quai Ernest-Ansermet, CH-1211 Gen\`eve 4, Switzerland\label{aff83}
\and
Department of Physics, P.O. Box 64, University of Helsinki, 00014 Helsinki, Finland\label{aff84}
\and
Helsinki Institute of Physics, Gustaf H{\"a}llstr{\"o}min katu 2, University of Helsinki, 00014 Helsinki, Finland\label{aff85}
\and
Laboratoire d'etude de l'Univers et des phenomenes eXtremes, Observatoire de Paris, Universit\'e PSL, Sorbonne Universit\'e, CNRS, 92190 Meudon, France\label{aff86}
\and
SKAO, Jodrell Bank, Lower Withington, Macclesfield SK11 9FT, UK\label{aff87}
\and
Centre de Calcul de l'IN2P3/CNRS, 21 avenue Pierre de Coubertin 69627 Villeurbanne Cedex, France\label{aff88}
\and
University of Applied Sciences and Arts of Northwestern Switzerland, School of Computer Science, 5210 Windisch, Switzerland\label{aff89}
\and
INFN-Sezione di Roma, Piazzale Aldo Moro, 2 - c/o Dipartimento di Fisica, Edificio G. Marconi, 00185 Roma, Italy\label{aff90}
\and
Dipartimento di Fisica e Astronomia "Augusto Righi" - Alma Mater Studiorum Universit\`a di Bologna, via Piero Gobetti 93/2, 40129 Bologna, Italy\label{aff91}
\and
Department of Physics, Institute for Computational Cosmology, Durham University, South Road, Durham, DH1 3LE, UK\label{aff92}
\and
Universit\'e Paris Cit\'e, CNRS, Astroparticule et Cosmologie, 75013 Paris, France\label{aff93}
\and
CNRS-UCB International Research Laboratory, Centre Pierre Bin\'etruy, IRL2007, CPB-IN2P3, Berkeley, USA\label{aff94}
\and
University of Applied Sciences and Arts of Northwestern Switzerland, School of Engineering, 5210 Windisch, Switzerland\label{aff95}
\and
Institute of Physics, Laboratory of Astrophysics, Ecole Polytechnique F\'ed\'erale de Lausanne (EPFL), Observatoire de Sauverny, 1290 Versoix, Switzerland\label{aff96}
\and
Institut de F\'{i}sica d'Altes Energies (IFAE), The Barcelona Institute of Science and Technology, Campus UAB, 08193 Bellaterra (Barcelona), Spain\label{aff97}
\and
School of Mathematics and Physics, University of Surrey, Guildford, Surrey, GU2 7XH, UK\label{aff98}
\and
Waterloo Centre for Astrophysics, University of Waterloo, Waterloo, Ontario N2L 3G1, Canada\label{aff99}
\and
Department of Physics and Astronomy, University of Waterloo, Waterloo, Ontario N2L 3G1, Canada\label{aff100}
\and
Perimeter Institute for Theoretical Physics, Waterloo, Ontario N2L 2Y5, Canada\label{aff101}
\and
Institute of Space Science, Str. Atomistilor, nr. 409 M\u{a}gurele, Ilfov, 077125, Romania\label{aff102}
\and
Institute for Particle Physics and Astrophysics, Dept. of Physics, ETH Zurich, Wolfgang-Pauli-Strasse 27, 8093 Zurich, Switzerland\label{aff103}
\and
Instituto de F\'isica Te\'orica UAM-CSIC, Campus de Cantoblanco, 28049 Madrid, Spain\label{aff104}
\and
Institut de Recherche en Astrophysique et Plan\'etologie (IRAP), Universit\'e de Toulouse, CNRS, UPS, CNES, 14 Av. Edouard Belin, 31400 Toulouse, France\label{aff105}
\and
Universit\'e St Joseph; Faculty of Sciences, Beirut, Lebanon\label{aff106}
\and
Departamento de F\'isica, FCFM, Universidad de Chile, Blanco Encalada 2008, Santiago, Chile\label{aff107}
\and
Universit\'e C\^{o}te d'Azur, Observatoire de la C\^{o}te d'Azur, CNRS, Laboratoire Lagrange, Bd de l'Observatoire, CS 34229, 06304 Nice cedex 4, France\label{aff108}
\and
Universit\"at Innsbruck, Institut f\"ur Astro- und Teilchenphysik, Technikerstr. 25/8, 6020 Innsbruck, Austria\label{aff109}
\and
Infrared Processing and Analysis Center, California Institute of Technology, Pasadena, CA 91125, USA\label{aff110}
\and
Cosmic Dawn Center (DAWN)\label{aff111}
\and
Niels Bohr Institute, University of Copenhagen, Jagtvej 128, 2200 Copenhagen, Denmark\label{aff112}
\and
Universidad Polit\'ecnica de Cartagena, Departamento de Electr\'onica y Tecnolog\'ia de Computadoras,  Plaza del Hospital 1, 30202 Cartagena, Spain\label{aff113}
\and
European University of Technology EUt+, European Union\label{aff114}
\and
Dipartimento di Fisica e Scienze della Terra, Universit\`a degli Studi di Ferrara, Via Giuseppe Saragat 1, 44122 Ferrara, Italy\label{aff115}
\and
Istituto Nazionale di Fisica Nucleare, Sezione di Ferrara, Via Giuseppe Saragat 1, 44122 Ferrara, Italy\label{aff116}
\and
ICSC - Centro Nazionale di Ricerca in High Performance Computing, Big Data e Quantum Computing, Via Magnanelli 2, Bologna, Italy\label{aff117}
\and
Astronomical Observatory of the Autonomous Region of the Aosta Valley (OAVdA), Loc. Lignan 39, I-11020, Nus (Aosta Valley), Italy\label{aff118}
\and
Department of Mathematical Physics, Institute of Physics, University of S\~ao Paulo, R. do Mat\~ao 1371, 05508-090, S\~ao Paulo, SP, Brazil\label{aff119}
\and
Centre de Recherche Astrophysique de Lyon, UMR5574, CNRS, Universit\'e Claude Bernard Lyon 1, ENS de Lyon, 69230, Saint-Genis-Laval, France\label{aff120}
\and
Universidad de La Laguna, Dpto. Astrof\'\i sica, E-38206 La Laguna, Tenerife, Spain\label{aff121}
\and
Department of Physics, Centre for Extragalactic Astronomy, Durham University, South Road, Durham, DH1 3LE, UK\label{aff122}
\and
Oskar Klein Centre for Cosmoparticle Physics, Department of Physics, Stockholm University, Stockholm, SE-106 91, Sweden\label{aff123}
\and
Astrophysics Group, Blackett Laboratory, Imperial College London, London SW7 2AZ, UK\label{aff124}
\and
Univ. Grenoble Alpes, CNRS, Grenoble INP, LPSC-IN2P3, 53, Avenue des Martyrs, 38000 Grenoble, France\label{aff125}
\and
Dipartimento di Fisica, Sapienza Universit\`a di Roma, Piazzale Aldo Moro 2, 00185 Roma, Italy\label{aff126}
\and
Institute of Cosmology and Gravitation, University of Portsmouth, Portsmouth PO1 3FX, UK\label{aff127}
\and
Institut d'Astrophysique de Paris, 98bis Boulevard Arago, 75014, Paris, France\label{aff128}
\and
ICL, Junia, Universit\'e Catholique de Lille, LITL, 59000 Lille, France\label{aff129}
\and
Center for Astrophysics and Cosmology, University of Nova Gorica, Nova Gorica, Slovenia\label{aff130}
\and
Department of Astronomy, Indiana University, 727 East Third Street, Bloomington, IN 47405, USA\label{aff131}
\and
CERCA/ISO, Department of Physics, Case Western Reserve University, 10900 Euclid Avenue, Cleveland, OH 44106, USA\label{aff132}
\and
Laboratoire Univers et Th\'eorie, Observatoire de Paris, Universit\'e PSL, Universit\'e Paris Cit\'e, CNRS, 92190 Meudon, France\label{aff133}
\and
Departamento de F{\'\i}sica Fundamental. Universidad de Salamanca. Plaza de la Merced s/n. 37008 Salamanca, Spain\label{aff134}
\and
IRFU, CEA, Universit\'e Paris-Saclay 91191 Gif-sur-Yvette Cedex, France\label{aff135}
\and
Universiteit Gent, Department of Physics and Astronomy, Proeftuinstraat 86 N3, 9000 Ghent, Belgium
\label{aff136}
\and
Konkoly Observatory, HUN-REN CSFK, MTA Centre of Excellence, Budapest, Konkoly Thege Mikl\'os {\'u}t 15-17. H-1121, Hungary\label{aff137}
\and
MTA-CSFK Lend\"ulet Large-Scale Structure Research Group, Konkoly-Thege Mikl\'os \'ut 15-17, H-1121 Budapest, Hungary\label{aff138}
\and
Universit\'e de Strasbourg, CNRS, Observatoire astronomique de Strasbourg, UMR 7550, 67000 Strasbourg, France\label{aff139}
\and
Scuola Superiore Meridionale, Via Mezzocannone 4, 80138, Napoli, Italy\label{aff140}
\and
Instituto de Astrof\'isica e Ci\^encias do Espa\c{c}o, Faculdade de Ci\^encias, Universidade de Lisboa, Tapada da Ajuda, 1349-018 Lisboa, Portugal\label{aff141}
\and
Department of Physics \& Astronomy, University of California Irvine, Irvine CA 92697, USA\label{aff142}
\and
Institute for Astronomy, University of Hawaii, 2680 Woodlawn Drive, Honolulu, HI 96822, USA\label{aff143}
\and
Departamento F\'isica Aplicada, Universidad Polit\'ecnica de Cartagena, Campus Muralla del Mar, 30202 Cartagena, Murcia, Spain\label{aff144}
\and
Instituto de F\'isica de Cantabria, Edificio Juan Jord\'a, Avenida de los Castros, 39005 Santander, Spain\label{aff145}
\and
Observatorio Nacional, Rua General Jose Cristino, 77-Bairro Imperial de Sao Cristovao, Rio de Janeiro, 20921-400, Brazil\label{aff146}
\and
Univ. Grenoble Alpes, CNRS, IPAG, 38000 Grenoble, France\label{aff147}
\and
Department of Mathematics and Physics E. De Giorgi, University of Salento, Via per Arnesano, CP-I93, 73100, Lecce, Italy\label{aff148}
\and
INFN, Sezione di Lecce, Via per Arnesano, CP-193, 73100, Lecce, Italy\label{aff149}
\and
INAF-Sezione di Lecce, c/o Dipartimento Matematica e Fisica, Via per Arnesano, 73100, Lecce, Italy\label{aff150}
\and
Dipartimento di Fisica "E. Pancini", Universita degli Studi di Napoli Federico II, Via Cinthia 6, 80126, Napoli, Italy\label{aff151}
\and
INFN Sezione di Napoli, Universita degli Studi di Napoli "Federico II", Compl. Univ. di Monte S. Angelo, Edificio G, Via Cinthia, I-80126, Napoli, Italy\label{aff152}
\and
Department of Computer Science, Aalto University, PO Box 15400, Espoo, FI-00 076, Finland\label{aff153}
\and
School of Physics and Astronomy, University of Nottingham, University Park, Nottingham NG7 2RD, UK\label{aff154}
\and
DARK, Niels Bohr Institute, University of Copenhagen, Jagtvej 155, 2200 Copenhagen, Denmark\label{aff155}
\and
Astrophysics Research Centre, University of KwaZulu-Natal, Westville Campus, Durban 4041, South Africa\label{aff156}
\and
School of Mathematics, Statistics \& Computer Science, University of KwaZulu-Natal, Westville Campus, Durban 4041, South Africa\label{aff157}
\and
Department of Physics and Astronomy, Vesilinnantie 5, University of Turku, 20014 Turku, Finland\label{aff158}
\and
Finnish Centre for Astronomy with ESO (FINCA), Quantum, Vesilinnantie 5, University of Turku, 20014 Turku, Finland\label{aff159}
\and
Serco for European Space Agency (ESA), Camino bajo del Castillo, s/n, Urbanizacion Villafranca del Castillo, Villanueva de la Ca\~nada, 28692 Madrid, Spain\label{aff160}
\and
Van Swinderen Institute, University of Groningen, Nijenborgh 3, 9747 AG Groningen, The Netherlands\label{aff161}
\and
Department of Physics and Astronomy, University of the Western Cape, Bellville, Cape Town, 7535, South Africa\label{aff162}
\and
Department of Physics and Helsinki Institute of Physics, Gustaf H\"allstr\"omin katu 2, University of Helsinki, 00014 Helsinki, Finland\label{aff163}
\and
Ohio University, Physics \& Astronomy Department,1 Ohio University, Athens, OH 45701, USA\label{aff164}
\and
Institute for Theoretical Particle Physics and Cosmology (TTK), RWTH Aachen University, 52056 Aachen, Germany\label{aff165}
\and
INAF-Osservatorio Astrofisico di Arcetri, Largo E. Fermi 5, 50125, Firenze, Italy\label{aff166}
\and
Institute for Theoretical Physics, ETH Zurich, Wolfgang-Pauli-Str. 27, 8093 Zurich, Switzerland\label{aff167}
\and
Centro de Astrof\'{\i}sica da Universidade do Porto, Rua das Estrelas, 4150-762 Porto, Portugal\label{aff168}
\and
Dipartimento di Fisica, Universit\`a di Roma Tor Vergata, Via della Ricerca Scientifica 1, Roma, Italy\label{aff169}
\and
INFN, Sezione di Roma 2, Via della Ricerca Scientifica 1, Roma, Italy\label{aff170}
\and
HE Space for European Space Agency (ESA), Camino bajo del Castillo, s/n, Urbanizacion Villafranca del Castillo, Villanueva de la Ca\~nada, 28692 Madrid, Spain\label{aff171}
\and
STAR Institute, University of Li{\`e}ge, Quartier Agora, All\'ee du six Ao\^ut 19c, 4000 Li\`ege, Belgium\label{aff172}
\and
Dipartimento di Fisica ``G. Occhialini", Universit\`a degli Studi di Milano Bicocca, Piazza della Scienza 3, 20126 Milano, Italy\label{aff173}
\and
Department of Astrophysics, University of Zurich, Winterthurerstrasse 190, 8057 Zurich, Switzerland\label{aff174}
\and
Space physics and astronomy research unit, University of Oulu, Pentti Kaiteran katu 1, FI-90014 Oulu, Finland\label{aff175}
\and
Theoretical astrophysics, Department of Physics and Astronomy, Uppsala University, Box 524, 751 20 Uppsala, Sweden\label{aff176}
\and
Mathematical Institute, University of Leiden, Einsteinweg 55, 2333 CA Leiden, The Netherlands\label{aff177}
\and
School of Physical Sciences, The Open University, Milton Keynes, MK7 6AA, UK\label{aff178}
\and
ASTRON, the Netherlands Institute for Radio Astronomy, Postbus 2, 7990 AA, Dwingeloo, The Netherlands\label{aff179}
\and
Center for Advanced Interdisciplinary Research, Ss. Cyril and Methodius University in Skopje, Macedonia\label{aff180}
\and
Institute of Astronomy, University of Cambridge, Madingley Road, Cambridge CB3 0HA, UK\label{aff181}
\and
Department of Astrophysical Sciences, Peyton Hall, Princeton University, Princeton, NJ 08544, USA\label{aff182}
\and
Fakult\"at f\"ur Physik, Universit\"at Bielefeld, Postfach 100131, 33501 Bielefeld, Germany\label{aff183}
\and
Ruhr University Bochum, Faculty of Physics and Astronomy, Astronomical Institute (AIRUB), German Centre for Cosmological Lensing (GCCL), 44780 Bochum, Germany\label{aff184}
\and
Department of Physics and Astronomy, University of British Columbia, Vancouver, BC V6T 1Z1, Canada\label{aff185}}    
   
\abstract{
The \Euclid mission was designed around the need for thermo-optical and mechanical stability, straylight suppression, low outgassing, and precise guiding. After three years in space, a wealth of data is available to assess the performance in these areas. We review the L2 halo-orbit design and consumables budget. Solar straylight gradually increases by 14\% per year and is consistent with meteoroid puncturing, but remains low overall. Data loss from high solar activity has declined notably since the maximum of the 25th Solar cycle passed. The thermodynamical response of the optical elements to a thermal step impulse follows sigmoidal growth curves, entering the asymptotic equilibration regime after 21--46\,h. Water-ice contamination has affected data from both instruments and requires continued monitoring; we review detection and decontamination methods, and \Euclid's contamination history. High-precision measurements of the zodiacal background reveal, at $8\,\sigma$ significance, the subtle modulation caused by the changing perspective on the dust distribution due to the L2 halo-orbit libration. These data also show that conservative pre-launch estimates overestimated the out-of-field straylight by factors of 4 (\HE band) to 27 (\IE band). Telemetry data demonstrate excellent guiding, and enable highly significant detections of solar radiation pressure and of thermal-emission recoil from \Euclid's unilluminated dark side. Lastly, we present a list of 34 meteoroid impacts, discuss their mechanical and optical signatures, and derive a minimum detectable impactor mass of $5\times10^{-7}$\,g.
}

%

\titlerunning{Spacecraft performance at L2}
\authorrunning{Euclid Collaboration: M.~Schirmer et al.: DR1}

   \maketitle
   \nolinenumbers

\section{Introduction}
\label{sec:intro}

\Euclid's mission specifications include extraordinary optical stability and image quality, to achieve the best possible shear measurements from space \citep{EuclidSkyOverview}. These manifested in numerous key characteristics, among others the following that affect calibration of mission data: (i) a silicon-carbide structure for all relevant opto-mechanical parts of the scientific \ac{PLM};  (ii) an eclipse-free halo orbit around the Sun--Earth L2 point, offering maximum thermal stability; (iii) a survey design minimising thermal disturbances; (iv) a spacecraft design that minimises outgassing and molecular contamination of optical surfaces; and (v) minimal levels of optical in- and out-of-field stray light.

In this paper we present a science-centred perspective of \Euclid's technical performance after three years in orbit. While this is already half the nominal mission's duration, DR1 \citep{DR1cite} just begins to deliver on \Euclid's promise. In \cref{sec:operationalstatus} we demonstrate that the spacecraft is in excellent operational condition, delivering a maximum of scientifically useful data with operational time exceeding 99\%. In \cref{sec:straylightxrays} we review \Euclid's sensitivity to solar stray light and X-ray emission from solar flares. In \cref{sec:thermalstability} we model the thermal response and equilibration timescales of opto-mechanical elements to a thermal step impulse. \Cref{sec:contaminationanddecontamination} covers molecular contamination by water ice, which was expected and explicitly folded into the calibration plan; although its effect on the optical and photometric performance remained highly uncertain until in-flight data became available. We present detection methods and our decontamination strategies, and in \cref{sec:inflight_contamination_history} the spacecraft's actual in-flight contamination history. In \cref{sec:zodi} we use monthly observations of the self-calibration field to compare the background's time-variability against \ac{ZB} model predictions. In \cref{sec:guiding_review} we review the guiding performance in terms of \ac{RPE}, and use the corrective torques from the cold-gas thrusters to determine \ac{SRP} and \ac{TER}. Finally, we discuss the mechanical and optical signatures of meteoroid impacts in \cref{sec:impacts}, and conclude in \cref{sec:conclusions}.

\section{Reliability, orbit, and consumables\label{sec:operationalstatus}}
Here we review \Euclid's status and how it is managed by the \ac{ESA}, specifically the \ac{MOC} in Darmstadt, Germany. Some of the technical information presented here is taken -- and updated where necessary -- from the consolidated report on mission analysis \citep[mostly orbital maintenance;][]{renk2019} and the September 2026 monthly status report from the flight-control team \citep{EuclidFCT2026}, which are unclassified by \ac{ESA}, albeit not in the public domain.
 
\subsection{Spacecraft reliability}
\Euclid is in very good health. Indications for ageing of mechanisms and electronics -- for example in terms of torques, temperatures, voltages, charge-transfer inefficiency \citep[CTI; see][]{Skottfelt2024,DR1-TP002}, and power consumption -- are either absent, or well within expectations, and do not limit the potential lifetime of the mission as set by its consumables (\cref{sec:consumables}).

Between February 2024, the beginning of routine operations, and June 2026, on average 99.13\% of all pre-planned scientific and telemetry files were retrieved on ground. The data return is nearly perfect, despite the first years in orbit coinciding with the maximum of the 25th solar cycle.

\subsection{Halo orbit\label{sec:orbit}}
\begin{figure*}[t]
\centering
\includegraphics[angle=0,width=1.0\hsize]{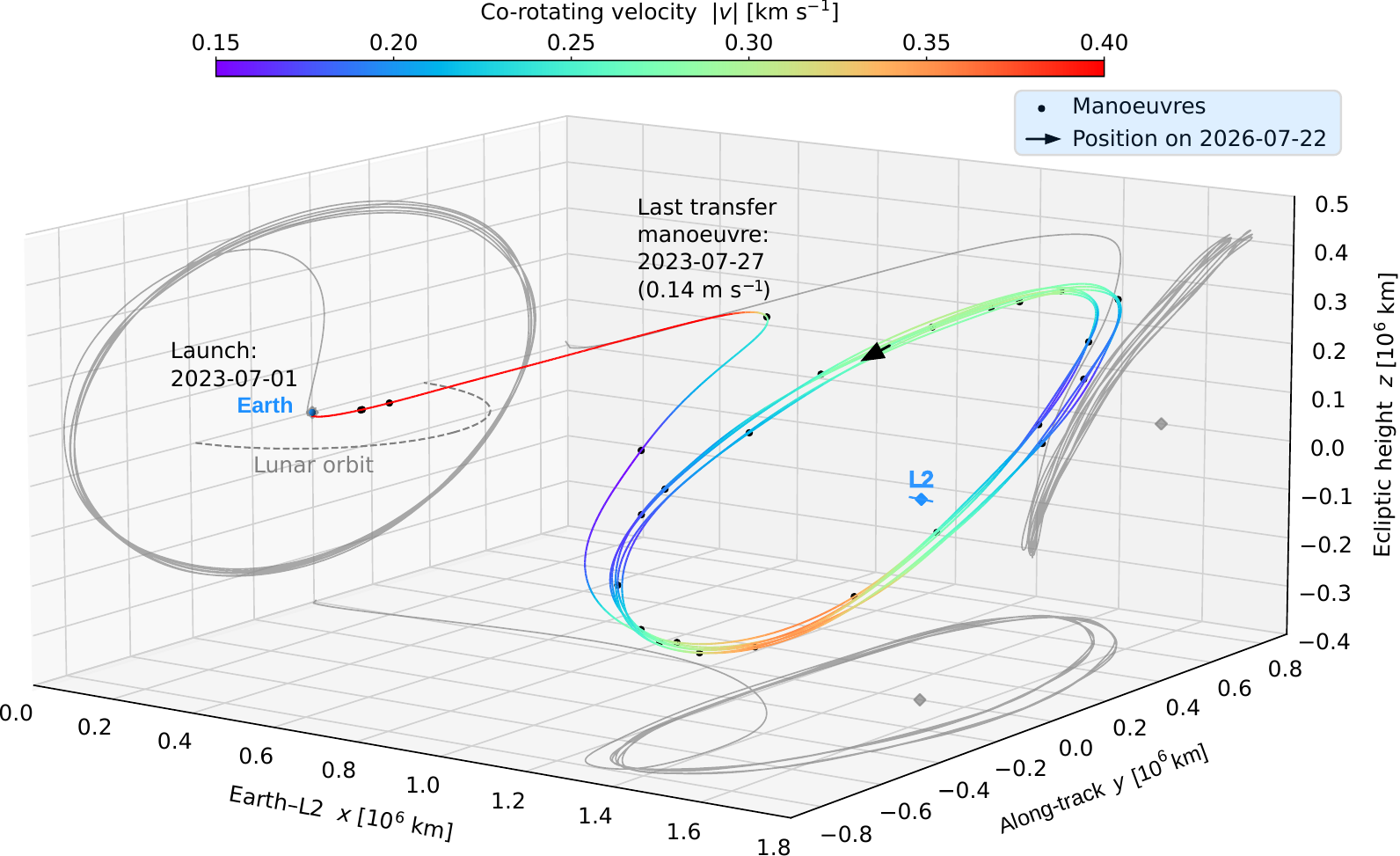}
\caption{\Euclid's eclipse-free trajectory in the Sun--Earth co-rotating frame. Equal scaling is used along all axes, preserving the correct spatial proportions, including those of the wall projections (grey curves). The orbit is colour-coded by the spacecraft's velocity in the co-rotating frame. The black dots mark transfer and orbit-correction manoeuvres; the last transfer manoeuvre corresponds to the second orange bar in \cref{fig:ocm_deltav}. Because of Earth's eccentric orbit, the Earth--L2 distance varies by 50\,200\,km -- indicated by the short line segment across the L2 marker. As a result, the halo loops with a period of 180 days gradually shift by a corresponding amount along the Earth--L2 axis. The \ac{JWST} is in a  similar orbit, mirrored on the ecliptic plane; seen from Earth, \Euclid orbits clockwise, and \ac{JWST} counter-clockwise.}
\label{fig:euclid_orbit}
\end{figure*}

\Euclid is in a wide, eclipse-free halo orbit around the second Sun--Earth Lagrange point L2, avoiding shadow passages (\cref{fig:euclid_orbit}). It offers maximum thermal stability, and has a period of approximately 180 days set by the orbit's geometry and the mass ratio between the Earth and the Sun. \Euclid's orbital velocity in the Sun--Earth co-rotating frame varies periodically between 0.16 and 0.34\,km\,s$^{-1}$. 

The characteristic warped and tilted shape of the orbit is a result of the nonlinearity of the restricted three-body problem in which \Euclid is considered massless. While the linearised equations of motion near L2 decouple into independent oscillators in the $x$--$y$ ecliptic plane and perpendicular ($z$) to it, the full gravitational potential depends on the three-dimensional distances of \Euclid to the Earth and the Sun, therefore mixing all three coordinates whenever the out-of-plane amplitude $A_z$ is large, as required for shadow avoidance. In the third-order approximation of \cite{richardson1980}, the orbit's frequency is coupled to the amplitudes as
\begin{equation}
    \omega^2 = 1 + l_1 A_x^2 + l_2 A_z^2\,,
\end{equation}
where $l_1$ and $l_2$ parametrise the effective potential. This coupling introduces higher harmonics that warp the orbit away from a simple tilted ellipse. 

\subsection{Station keeping\label{sec:stationkeeping}}
\Euclid's L2 halo orbit is dynamically unstable \citep{richardson1980}. It is sensitive to gravitational and non-gravitational perturbations, which grow exponentially with a time constant of around 23 days, leading to an escape from orbit after approximately 90 days if uncorrected. If all perturbing accelerations were perfectly known, together with \Euclid's orbit state vector ${\bf S}({\bf x},{\bf v},t)$, then a velocity change $\Delta v$ of a few millimetres per second every 180 days would be sufficient to stay in orbit. Actual corrections, however, are orders of magnitude larger.

\subsubsection{Non-gravitational orbit perturbations\label{sec:orbitalperturbations}}
Five non-gravitational orbit perturbations are of relevance for \Euclid. 
First and foremost is \ac{SRP}, which causes an acceleration on the surface normal, 
\begin{equation}
    a_{\rm SRP}\approx\frac{S\,A\,(\alpha+\rho)}{m\,c}\,{\cos^2(\theta)}\,.
    \label{eq:radiationpressure}
\end{equation}
Here, $S=1360$\,W\,m$^{-2}$ is the solar constant at Earth's heliocentric distance, $c$ the speed of light, $A/m\approx0.008$\,m$^2$\,kg$^{-1}$ is \Euclid's area-to-mass ratio, $\rho=0.1$ the sunshield's reflectivity, $\alpha=0.9$ its absorptivity, and $\theta$ the angle of incidence. With these values, $a\approx4.0\times10^{-8}$\,m\,s$^{-2}$, with a cumulative $\Delta v=0.10$\,m\,s$^{-1}$ after one month. The acceleration varies periodically with an amplitude of about 3.5\% due to heliocentric distance variation, and up to 3\% from $\theta\in[-3^\circ,20^\circ]$ when \Euclid steps to a new survey field every 74\,min. \Ac{SRP} also changes gradually over time because of space-weathering effects from dust pitting and particle irradiation \citep[][]{miller2011,engelhart2019}. 

Second is \ac{TER} from warm surfaces, predominantly the sunshield, but also from radiators and \ac{SVM} panels on \Euclid's unilluminated sides.   \ac{TER} scales with $T^4$ and is therefore particularly sensitive to small relative changes in heliocentric distance and temperature. For example, radiation-induced darkening of the solar-cells' coverglass increases the sunshields' temperature, and thus the \ac{TER} acting on it \citep[][see also \cref{sec:timeseries_fittingresults}]{distefano2023,Aihara2026}.

Third is the \ac{AOCS} operating in non-science mode, because the thruster system is not perfectly balanced and noisy. This adds a stochastic perturbation, for example, when offloading reaction wheels near saturation, with a pre-launch estimate of $6\times10^{-8}$\,m\,s$^{-2}$ at the 1\,$\sigma$ level. Propellant and cold-gas leaks can also be relevant. 

Fourth is molecular outgassing \citep[e.g.,][]{schlaeppi2010,Schirmer-EP29}. The \Euclid spacecraft attitude, by design, results in an extremely stable thermal state. The perturbation from outgassing is expected to be smaller than \ac{SRP}, and is slowly decreasing over years until the volatile reservoirs are depleted. However, outgassing rates can change suddenly, for example when previously shadowed structures get exposed to solar or planetary radiation. The acceleration from such an event exceeded that of \ac{SRP} by one order of magnitude for OSIRIS-Rex \citep{sandford2020}, and by two orders of magnitude for \textit{BepiColombo} following a Venus flyby \citep{defilippis2025}. 

At the lowest level, several factors come into play, such as antenna recoil, stochastic momentum transfer from dust and meteoroid impacts (see also \cref{sec:impacts}), and Solar wind. In general, these are several orders of magnitude below \ac{SRP}.

\subsubsection{Floquet method to null the escape component\label{sec:floquet}}
\Euclid uses hydrazine monopropellant and ten pairs of 20\,N-thrusters \citep{delanoe2016} to execute \acp{OCM}. \ac{ESA} uses the Floquet method \citep{simo1986,simo1987} for station keeping: ${\bf S}({\bf x},{\bf v},t_0)$ is estimated from range finding and Doppler measurements (\cref{tab:orbitdetermination}) at time $t_0$, and a correction impulse $\Delta {\bf v}_{\rm corr}$ is computed for an \ac{OCM} at a later time $t_{\rm c}$ that removes only the current escape component from ${\bf S}$. It does not attempt to keep the spacecraft on a launch-date-specific reference orbit.\footnote{\Euclid's resulting orbit is still close to its reference orbit.} This nulling method propagates the spacecraft along a continuously evolving non-escaping trajectory (\cref{fig:euclid_orbit}), and is approximately twice as fuel-efficient as pursuing a reference orbit. Determination and execution of $\Delta {\bf v}_{\rm corr}$, which are intrinsically uncertain (\cref{tab:orbitdetermination}), accumulate orbit perturbations that are eventually nulled by a future \ac{OCM}.

\begin{figure*}[t]
\centering
\includegraphics[angle=0,width=0.48\hsize]{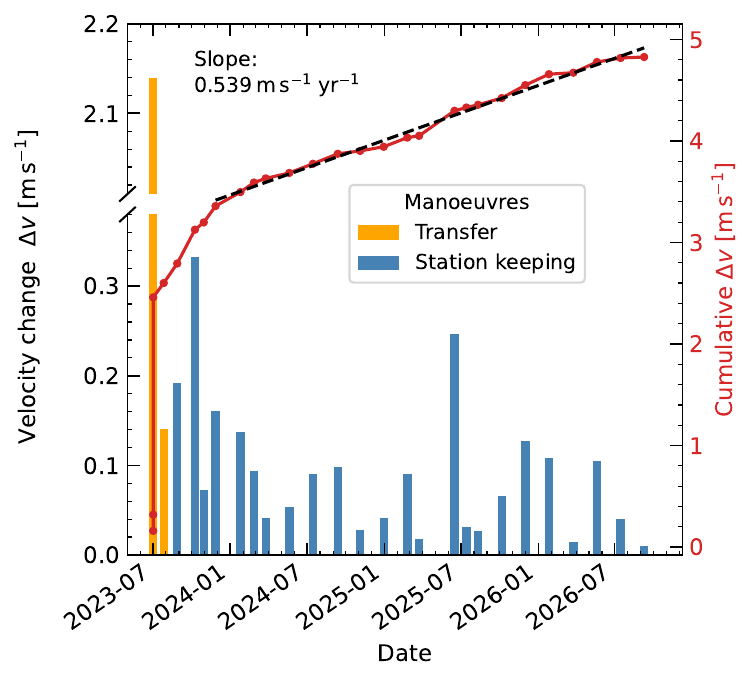}
\includegraphics[angle=0,width=0.48\hsize]{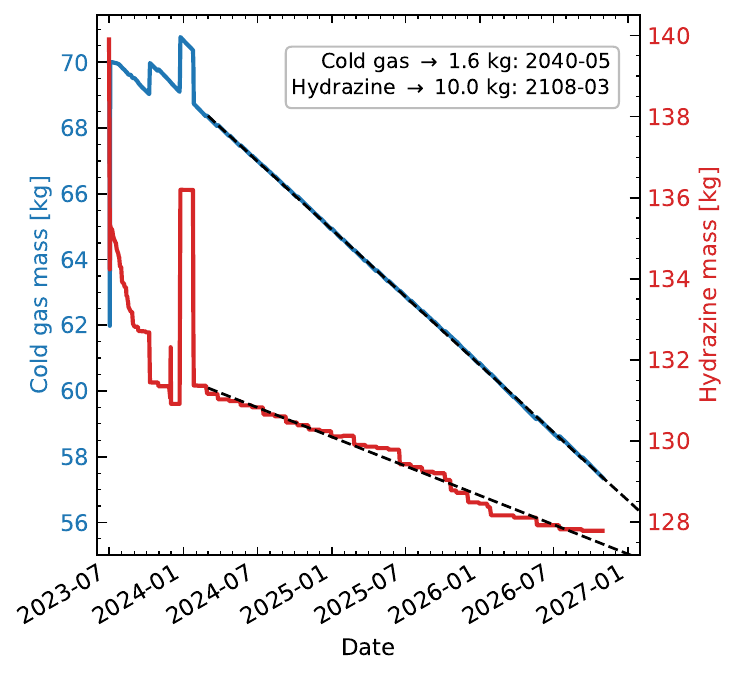}
\caption{\textit{Left}: Manoeuvres since launch (see also \cref{fig:euclid_orbit}). The red curve shows the cumulative velocity change; it increases by 0.539\,m\,s$^{-1}$\,yr$^{-1}$ (black dashed line) since November 2023, demonstrating a well-controlled orbit. \textit{Right}: Residual masses for cold gas (micro-propulsion, parameter APPT1022) and hydrazine (station keeping, parameter APPT1291). Depletion dates are linear extrapolations of the consumptions since the beginning of nominal survey operations in February 2024, presuming unusable residual masses of 1.6\,kg and 10.0\,kg for cold gas and hydrazine, respectively. The large vertical jumps in 2023 are due to a software update and the upload (and subsequent fix) of an erroneous configuration file.}
\label{fig:ocm_deltav}
\end{figure*}

\begin{table}[h]
\centering
\footnotesize
\caption{General ESA station-keeping references.}
\label{tab:orbitdetermination}
\begin{tabular}{llll}
\hline
\hline\noalign{\vskip 1pt}
\multicolumn{4}{l}{Determination of state vector ${\bf S}({\bf x},{\bf v},t)$}\\
\hline\noalign{\vskip 1pt}
Type & Frequency & Precision & Accuracy (bias)\\
\hline\noalign{\vskip 1pt}
Range & Once per pass & 20\,m & 20\,m\\
Doppler & Every 10\,min & $0.3$\,mm\,s$^{-1}$ & $<10^{-3}$\,mm\,s$^{-1}$\\
 & during pass & & \\
\hline\noalign{\vskip 1pt}
\multicolumn{4}{l}{Correction-manoeuvre execution ($3\,\sigma$ requirements)}\\
\hline\noalign{\vskip 1pt}
\multicolumn{2}{l}{$\Delta {\bf v}_{\rm corr}$ magnitude ($>0.2$\,m\,s$^{-1}$)} & 3\% & - \\
\multicolumn{2}{l}{$\Delta {\bf v}_{\rm corr}$ direction ($>0.2$\,m\,s$^{-1}$)} & \ang{0.75} & -\\
\multicolumn{2}{l}{$\Delta {\bf v}_{\rm corr}$ direction ($<0.2$\,m\,s$^{-1}$)} & $<0.006$ \,m s$^{-1}$ & -\\
\hline
\end{tabular}
\tablefoot{Required precision and accuracy on range and velocity measurements for standard tracking of a spacecraft in orbit around a libration point. Actual performance may differ. A pass occurs daily and has a duration of 4\,h for the data downlink.}
\end{table}

\subsubsection{Cadence of orbit corrections\label{sec:sopscadence}}
The precomputed survey includes six-hour long \ac{SOPS} windows for maintenance and station keeping \citep[see section 7.6 in][]{Scaramella-EP1}. Whether an \ac{OCM} is needed during \ac{SOPS} becomes known only a few days in advance from the measured state vector (\cref{sec:floquet}). 

\ac{OCM} cadence is a trade-off. More frequent nulling means less overall fuel consumption because perturbations have less time for exponential growth. Science operations, however, are not possible during \ac{SOPS} and thus limit their cadence. At the beginning of the mission, \ac{SOPS} windows were placed in the survey once every four weeks. This was reduced to once every eight weeks from 2026-01-26 onward because of favourable performance.

The \ac{OCM} history is displayed in \cref{fig:ocm_deltav}. The orbit stabilised after four months with a long-term cumulative $\langle\Delta v\rangle= 0.554$\,m\,s$^{-1}$ per year. This is seven times better than estimated from orbit simulations using a residual acceleration of $3\times10^{-8}$\,m\,s$^{-2}$, similar to our radiation-pressure estimate. 

\subsection{Consumables\label{sec:consumables}}
Hydrazine monopropellant is used for transfer manoeuvres and \acp{OCM}, with an initial mass of 137\,kg. A conservative load of hydrazine was needed to compensate for a potential under- or over-performance of the launch system and margins during L2 transfer, provisioning for a $\Delta v$ of approximately 50\,m\,s$^{-1}$. Enough propellant must remain at the end of the mission to provide $\Delta v=10$\,m\,s$^{-1}$ for disposal into a heliocentric orbit according to \ac{ESA}'s space-debris mitigation policy \citep{ESA2008,renk2019}. With the current usage profile, hydrazine will last until 2105. 

A \ac{MPS} with an initial load of 70\,kg nitrogen as cold gas is used for spacecraft stabilisation after slews and during science exposures, when it predominantly counters the continuous radiation-pressure torques 
(\cref{sec:guiding_review}). Using a linear extrapolation of the consumption during routine operations, and a residual unusable mass of 1.6\,kg, the cold gas will be depleted by 2040 (right panel in \cref{fig:ocm_deltav}).

\begin{figure}[h]
    \centering
    \includegraphics[angle=0,width=0.7\hsize]{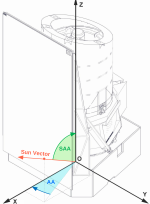}
    \caption{Spacecraft coordinate system and attitude angles. An interactive, 3D model is at \url{https://scifleet.esa.int/model/euclid/}.}
    \label{fig:PLM_cropped}
\end{figure}

\begin{figure*}[t]
\centering
\includegraphics[angle=0,width=1.0\hsize]{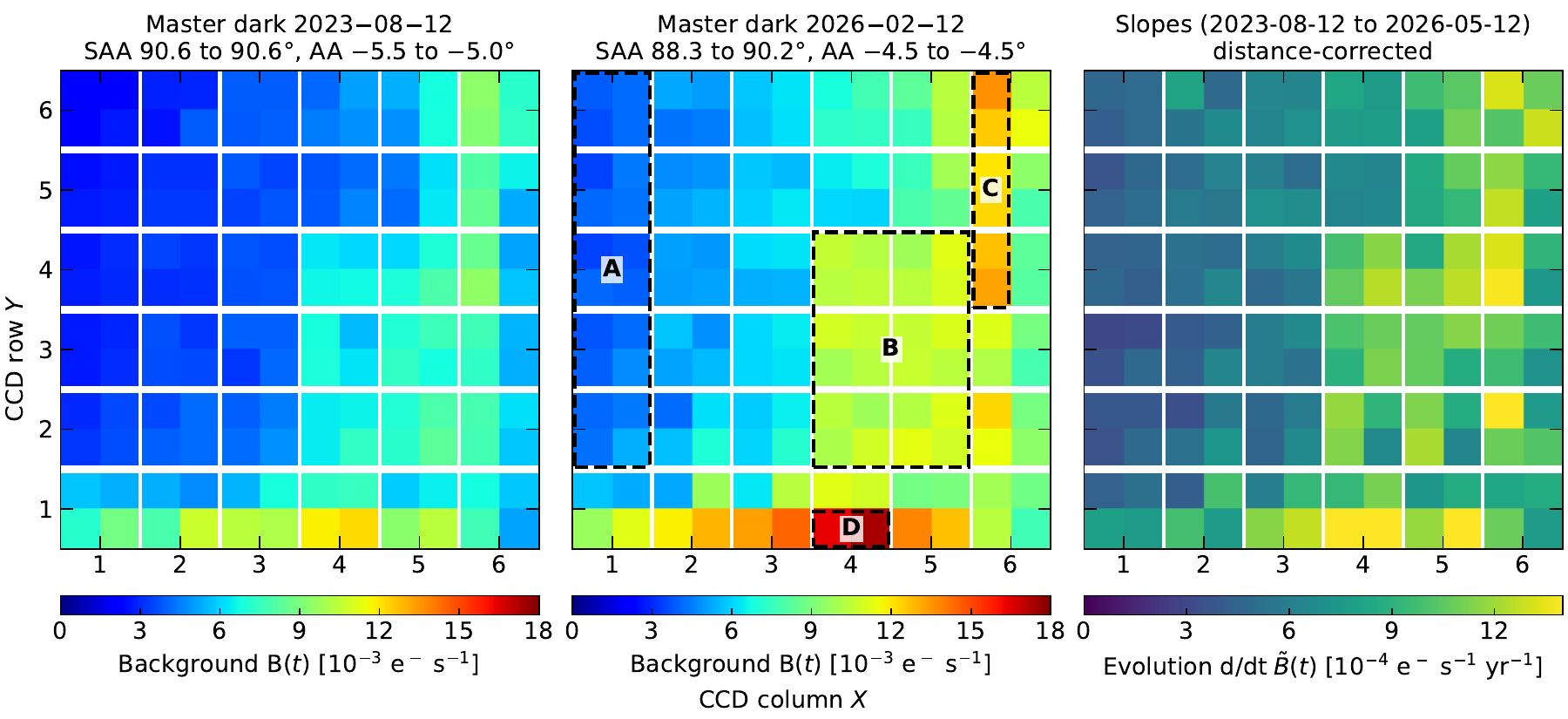}
\caption{Evolution of the 2D VIS background in the shutter-closed configuration, measured at detector-quadrant level. \textit{Left}: Master dark from August 2023. \textit{Middle}: Master dark from February 2026 with regions A to D.  \textit{Right}: Best-fit slopes from a linear regression of the time-series data. See also \cref{fig:timeseries_groups_med}.}
\label{fig:dark_evolution_maps}
\end{figure*}

\begin{figure*}[t]
\centering
\includegraphics[angle=0,width=0.48\hsize]{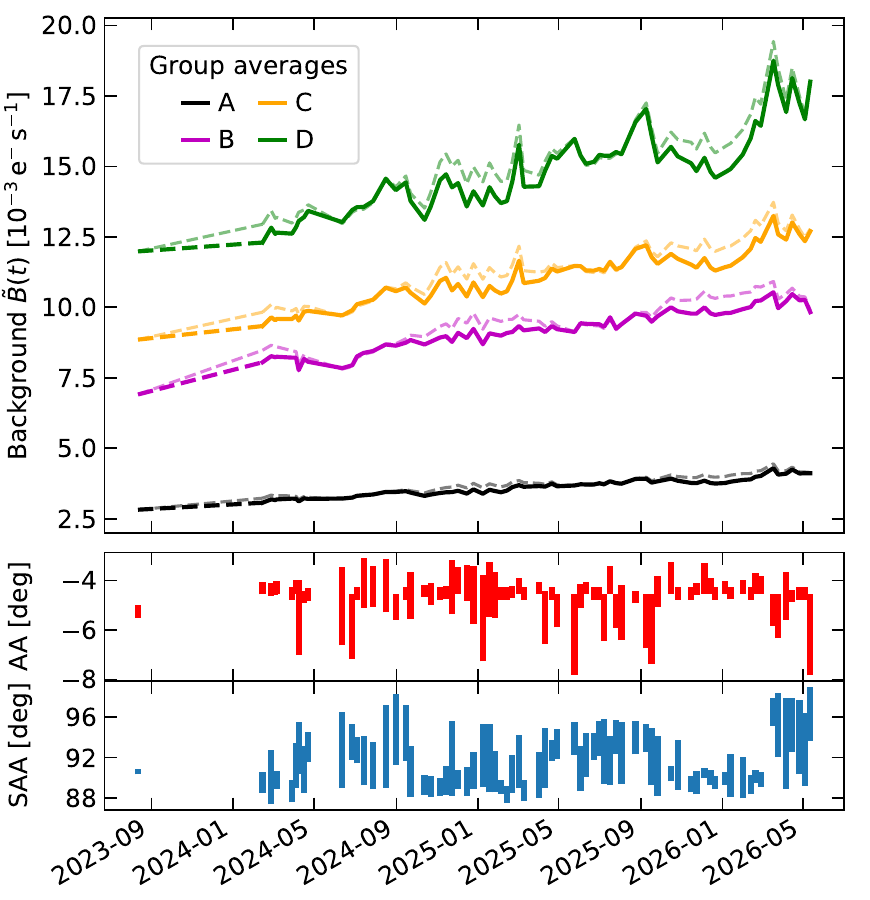}
\includegraphics[angle=0,width=0.48\hsize]{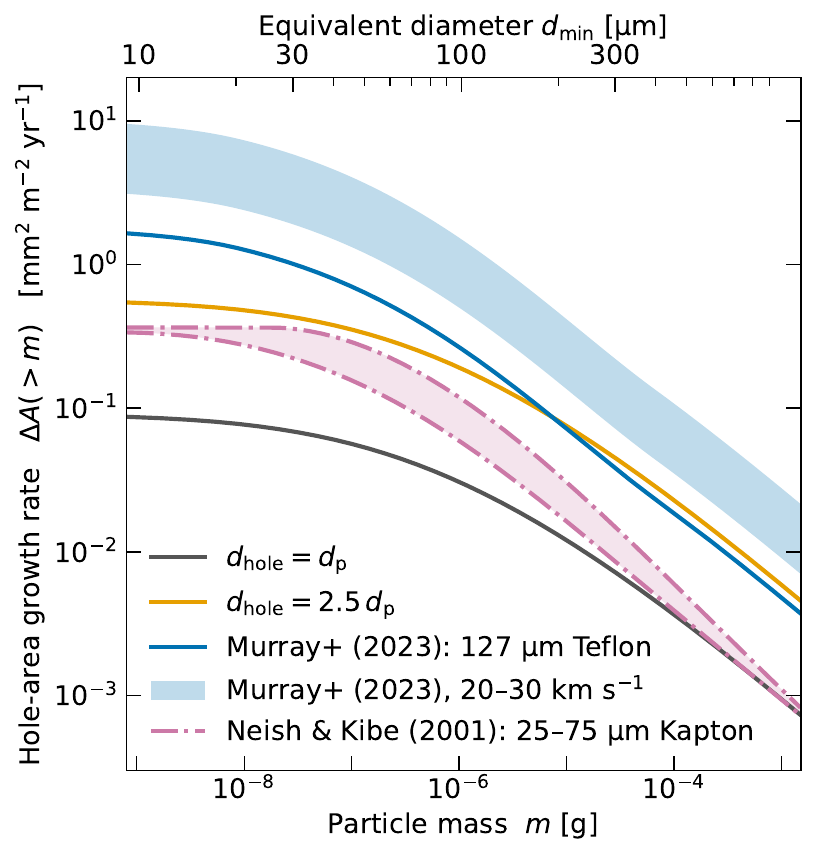}
\caption{\textit{Upper left}: Time series of the VIS background in groups A to D (\cref{fig:dark_evolution_maps}). The dashed lines show the raw signal, and the  solid lines include heliocentric-distance correction. \textit{Lower left}: Spacecraft attitude, which governs the amount of stray light. Groups C and D are susceptible to higher \ac{SAA}. \textit{Right}: \ac{MLI} hole-area growth rate from impacts by particles above a certain minimum mass. The curves integrate the interplanetary particle flux model of \cite{Gruen1985}, using five different models to translate particle mass to hole diameter.}
\label{fig:timeseries_groups_med}
\end{figure*}

\section{Susceptibility of VIS to solar stray light and X-rays \label{sec:straylightxrays}}
\subsection{Solar stray light\label{sec:straylight}}
Sunlight can reach the VIS instrument \citep{EuclidSkyVIS} by scattering off an illuminated thruster nozzle toward the thermally insulated, reflective backside of the sunshield. From there, photons enter the instrument cavity through small openings and, potentially, holes in its \ac{MLI}; a solid encapsulation of the lift-to-L2 cavity would have exceeded the capacity of the originally foreseen Soyuz launcher. These parasitic light paths were largely neutralised by systematically offsetting the \acl{AA} (\acs{AA}, \cref{fig:PLM_cropped})\acused{AA}, moving the nozzle into the shadow. The background decreased by factors 100--1000 to around 10\%--15\% of \acf{ZB} level, at the expense of a 5\% area loss due to increased field overlaps. Since the remaining stray-light components depend on \ac{AA} and \ac{SAA}, further entry paths must exist that have remained unclear, despite thorough investigation. Whether parasitic light paths were always present, or were created by space debris or meteoroid punctures (\cref{sec:impacts}) after launch and before commissioning, is unknown \citep[see also sections 5.4 and 5.5 in][]{EuclidSkyOverview}.

\subsubsection{Measurements}
Given the similarity of stray-light patterns in the open- and closed-shutter configuration, most parasitic light paths bypass the shutter blade regardless of its position, and reach the detectors even when the shutter is closed. Therefore, VIS darks are well suited to detect stray-light evolution, such as from an increasing total hole area due to meteoroid puncturing. Master darks are built on a regular basis from 60--70 single darks accumulated during 7--14 days of survey operation; the single exposures are bias-corrected and gain-normalised prior to stacking. The pre-flight detector characterisation -- having an estimated sensitivity of $10^{-4}$\,e$^-$\,s$^{-1}$ -- could not detect a measurable thermal dark current near the operating temperature of 153\,K; the measured signal in the master darks is at least a factor of 40 higher than this upper limit. Therefore, the master darks are considered to be pure straylight maps. 

The median background flux per detector quadrant forms a time series $B(t)$. Quadrant groups A to D (middle panel in \cref{fig:dark_evolution_maps}) select regions with increasing absolute stray-light levels, but within each group the signal is very similar. We correct the straylight in master dark $i$ for its seasonal dependence on heliocentric distance, 
\begin{equation}
    \tilde B_i = B_i \,\left(\frac{d_{\rm h,i}}{d_{\rm h,0}}\right)^2\,.
    \label{eq:btilde}
\end{equation}
Here, $d_{\rm h,0}$ normalises the straylight to the heliocentric distance at the time of the first master dark, August 2023.

The slopes of the individual time series $\tilde B(t)$ were obtained using iterative linear regression, and are shown in the right panel of \cref{fig:dark_evolution_maps}. In general, slopes are higher in areas of higher absolute straylight. The group-averaged time series are shown in \cref{fig:timeseries_groups_med}, and their slopes given in \cref{tab:dark_group_slopes}. The heliocentric distance correction has the expected, slight flattening effect on the trends. The relative increase of stray light between August 2023 and May 2026 was the same across the four groups, 
on average 38.1\% or 13.9\%\,yr$^{-1}$. 

\subsubsection{Puncturing by meteoroids}
Apart from the short-term dependence on \ac{AA} and \ac{SAA}, the long-term absolute and relative increase of stray light appears stable and uniform at group level. A plausible explanation is the continuous puncturing of the \ac{MLI} by the dust in the interplanetary particle field, relative to which \Euclid moves with a velocity of $\sim30$\,km\,s$^{-1}$. 

The \ac{MLI} stack used around the instrument cavity consists of, from outside to inside: (i) 25\,\micron\ Kapton with vapour-deposited aluminium (VDA) on both sides, perforated; (ii) 19 layers of 7.6\,\micron\ thickness made from a stiffer polyimide than Kapton, both sides with VDA, perforated; and (iii) 40\,\micron\ black Kapton (carbon-charged) with VDA facing to the 19-layer stack, not perforated. The total polyimide thickness is 209\,\micron, with a total surface mass density of around 303\,g\,m$^{-2}$.

Hydrocode simulations show that soda-lime glass spheres of 50--200\,\micron\ diameter penetrate \ac{MLI} with a surface mass density of up to $315$\,g\,m$^{-2}$ at $20$\,km\,s$^{-1}$ \citep{Ito2024}, and that an \ac{MLI} blanket with 20 layers Kapton is perforated at $7$--$150$\,km\,s$^{-1}$ \citep{Iyer2015}. We note that their surface mass density likely includes the netting between layers, because the total polyimide mass of their heaviest \ac{MLI} stack adds up to only 238\,g\,m$^-2$ (compared to the listed 315\,g\,m$^{-2}$). It is therefore plausible that particles with diameter $65\,\micron = 50\,\micron\times303/238=65$ at 20--30\,km\,s$^{-1}$ penetrate the \ac{MLI} skirt; here, we upscaled the smallest diameter used by \cite{Ito2024} by the 1.3 times higher surface mass density of the \Euclid MLI, and the mass of this particle is $2.9\times10^{-7}$\,g for a standard silicate density of 2\,g\,cm$^{-3}$. 

The hole diameter depends on the layer thickness and the impact speed \citep{Neish2001,Murray2023}; holes in multi-layer stacks appear larger than in a single layer \citep{Murray2023}, perhaps because of blow-back from the inner layers. These relations are calibrated at $\approx5$\,km\,s$^{-1}$ for Kapton \citep{Neish2001,Burchell2026} and below $\approx9$\,km\,s$^{-1}$ for \ac{MLI} \citep{Neish2001,Murray2023}; the speed dependence remains to be determined \citep{Burchell2026}. 

To check the plausibility of the resulting damage, we estimated its growth rate with various models: hole diameters of $1$--$2.5$ times the particle diameter; the measured particle-hole ratios of \citet{Murray2023} for their tested velocity of up to 8\,km\,s$^{-1}$, and extrapolating their power-law scaling to 20--30\,km\,s$^{-1}$; and the Kapton law of \citet{Neish2001}. To estimate the impact rate on one side of a surface of unit area, for particles above a minimum mass, we used the model of \citet{Gruen1985} in the analytic form of \citet{yahalomi2025}. We assumed spherical particles with a density of 2\,g\,cm$^{-3}$; more details are given in \cref{sec:expected_impact_rate} about individual meteoroid impacts. 

The computed damage-area growth curves for particles above a minimum mass are shown in the right panel of \cref{fig:timeseries_groups_med}. In \cref{sec:stray_light_estimate}, an order-of-magnitude estimate yields a combined hole size of $A_{\rm h}=3$--30\,mm$^2$ to explain the very high stray-light levels observed at ${\rm AA}=0^\circ$ in August 2023. Scaling the observed 38.5\% relative increase of stray light (over 33 months, see \cref{tab:dark_group_slopes}) to the estimated $A_{\rm h}$ range, yields a hole growth of 0.41--4.1\,mm$^2$\,yr$^{-1}$. This range is well covered by the various models tested, for impactors with a minimum diameter of $\gtrsim40$\,\micron. 

\begin{table}
\centering
\caption{Background evolution in VIS master darks.}
\label{tab:dark_group_slopes}
\small
\setlength{\tabcolsep}{4pt}
\begin{tabular}{lccc}
\hline\hline\noalign{\vskip 1pt}
Group & $B_0$ & Slope & Relative slope \\
 & [$10^{-3}$\,e$^-$\,s$^{-1}$] & [$10^{-4}$\,e$^-$\,s$^{-1}$\,yr$^{-1}$] & [\%\,yr$^{-1}$] \\
\hline\noalign{\vskip 1pt}
A & $2.92$ & $(4.18 \pm 0.13)$ & $(14.3 \pm 0.5)$ \\
B & $7.46$ & $(10.44 \pm 0.33)$ & $(14.0 \pm 0.5)$ \\
C & $8.94$ & $(12.54 \pm 0.44)$ & $(14.0 \pm 0.6)$ \\
D & $12.00$ & $(15.86 \pm 0.94)$ & $(13.2 \pm 1.0)$ \\
\hline
\end{tabular}
\tablefoot{Groups are shown in \cref{fig:dark_evolution_maps}. The dark current is considered negligible, so the whole signal is treated as straylight. $B_0$: fitted level at the first dark (2023-08-12); Relative slope: absolute slope divided by $B_0$. Errors are formal fit errors.}
\end{table}

\begin{figure}[t]
\centering
\includegraphics[angle=0,width=1.0\hsize]{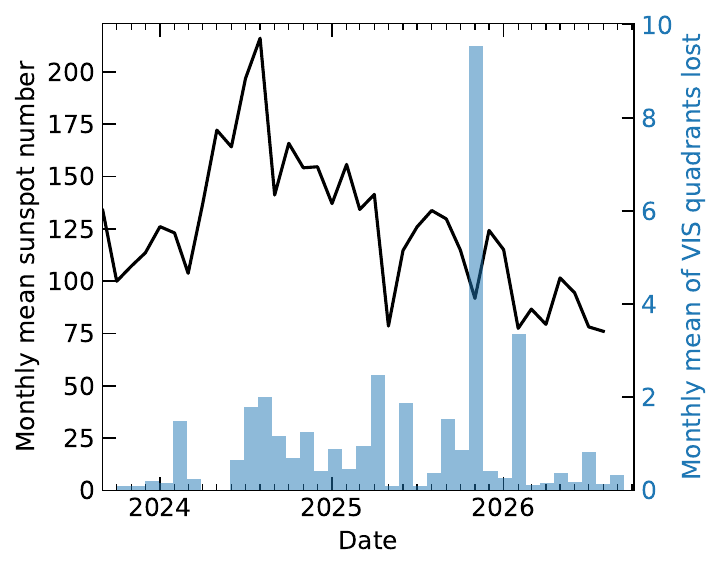}
\caption{Mean monthly sunspot numbers of the 25th solar cycle. The activity has dropped considerably since mid-2024, reflected in a decreasing number of lost detector quadrants in VIS data. In November 2025, a rapid sequence of X-class flares launched fast \acp{CME}, two of which interacted and caused a G4 geomagnetic storm, an S3 radiation storm, and the strongest ground-level enhancement (GLE) of neutron counts since 2005, indicative of an unusually hard particle spectrum \citep{Yasyukevich2026Sensors,mishev2026}. In January 2026, a single X1.9 flare and associated \ac{CME} produced the most intense 
$>10$\,MeV radiation storm (S4) since 2003, but with a soft spectrum and no GLE \citep{chernyshov2026}.}
\label{fig:ssn_2020_present}
\end{figure}

\subsection{Interference by solar X-rays\label{sec:xraypatterns}}
X-rays from solar flares have affected \Euclid since launch \citep[section 5.3 in][]{EuclidSkyOverview}, resulting in a data loss of about 2\% and increased complexity of our data processing routines \citep[section 3.7 in][]{Q1-TP002}. X-rays penetrate the sunshield and thermal insulation, directly reaching the VIS detectors, located further behind in the instrument cavity. The geometry of the resulting patterns on the detectors is highly dependent on \Euclid's attitude, which changes the line of sight through the sunshield. This results in a range of tomographic images of the sunshield for each X-ray flare. Attenuation factors range from $10^2$ to at least $10^8$ at 6\,keV, with an additional strong dependence on photon energy. The most powerful X-class flares, for example on May 15, 2025, saturate VIS detectors in areas of low attenuation. The proton flux received by \Euclid can be considerably enhanced 2--3 days after a major flare.

Solar activity -- parametrised by sunspot numbers\footnote{Source: \url{https://www.spaceweather.gov/products/solar-cycle-progression}} -- has dropped by almost 50\% by June 2026 compared to its peak in the middle of 2024. The data loss scales accordingly (\cref{fig:ssn_2020_present}), and we expect it to decline further toward solar minimum. Additional details about the VIS X-ray sensitivity, including its scientific and technical use, will be presented elsewhere. 

\section{Thermal stability\label{sec:thermalstability}}

During the \ac{ROS}, internal heat-generating processes -- such as instrument operations and data processing -- occur on timescales of a few seconds to a few hours in a continuous, repetitive manner \citep[see also section 2.1 in][]{DR1-TP027}. Residual heat that is not directly managed by radiators is locally injected into the \ac{PLM}, made of silicon carbide (SiC), whose high thermal conductivity keeps temperature gradients small (\cref{fig:cavity_temperature}). Since the thermal-response timescales of the \ac{PLM} (\cref{tab:thermal_fits}) are much longer than the heat-injection timescales, the system settles into a stable quasi-equilibrium. These internal factors are discussed further in \cite{EU-Anselmi}. The main external factors, \Euclid's spatial attitude and heliocentric distance, are discussed below.

\subsection{Attitude and orbital dependence\label{sec:thermal_attitude}}
The internal temperatures, and with them the telescope defocus $z$, depend on the solar heat absorbed by the sunshield, a fraction of which leaks into the interior of the spacecraft. This absorbed heat varies with the heliocentric distance, $d_{\rm h}(t)$, and with the pointing of the spacecraft relative to the Sun (\cref{fig:PLM_cropped}). The optical bench -- hereafter `baseplate' -- is a monolithic structure of $2.5\,{\rm m}\times2.1\,{\rm m}$ that carries the instruments and folding optics on one side, and the telescope truss with the primary and secondary mirrors on the other. All these components are made of SiC \citep{bougoin2016,bougoin2019} and are thermally well coupled. With a mass of 187\,kg, the baseplate dominates the heat capacity of the structure. Its temperature therefore traces the overall thermal state of the telescope. Of the several thermistors mounted on the baseplate, we chose SSSD0170 because of its higher temperature resolution. 

\subsubsection{Data}
We sampled \ac{AA}, \ac{SAA}, and temperature $T$ every 10\,min from 2023-09-01 to 2026-10-31; $d_{\rm h}(t)$ was taken as the nearest neighbour from the orbital ephemeris, which is available at one-hour intervals. The defocus $z$ is defined as the shift of the image plane along the optical axis and was measured for each VIS exposure -- six for every \ac{ROS} that lasts 74\,min -- using diffraction-spike metrology \citep{Neumann26}. We note that this defocus value should not be considered as a pure longitudinal translation of the secondary mirror. This is because the diffraction-spike metrics is sensitive to higher-order wavefront-error terms (Whittam et al., in prep.), and also responds to combinations of movements. 
For the analysis, data points were masked if one of these conditions was met: (i) ${\rm SAA}\geq107^\circ$; (ii) data point within two days following a ${\rm SAA}\geq2^\circ$ step; (iii) epochs with a temperature gradient ${\rm d}T/{\rm d}t>60$\,mK over a period of 4\,h.

\begin{figure}[t]
\centering
\includegraphics[angle=0,width=1.0\hsize]{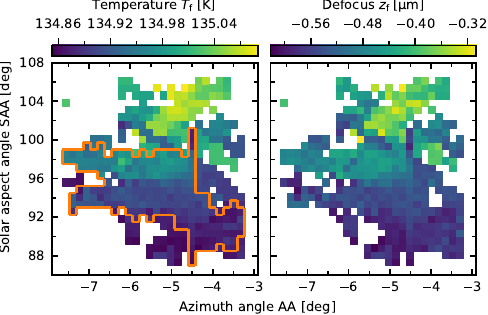}
\caption{Dependence of the modelled stationary baseplate temperature (\textit{left}) and telescope defocus (\textit{right}) on spacecraft attitude. Both patterns are similar, with the attitude driving the temperature and the defocus following. The orange boundary encompasses the \ac{EWS}; attitudes ${\rm SAA}\gtrsim99^\circ$ are mostly from visits to the \ac{EDF-S}. For ${\rm SAA}>90^\circ$, the telescope is tilted in the anti-Sun direction, leading to increased illumination of the spacecraft's bottom for ${\rm SAA}\gtrsim96^\circ$. Larger angles allow for more structural parts to be illuminated, increasing the complexity of the temperature pattern. More details are given in \cref{sec:thermal_attitude}.}
\label{fig:temperature_defocus_attitude}
\end{figure}

\begin{figure}[t]
\centering
\includegraphics[angle=0,width=1.0\hsize]{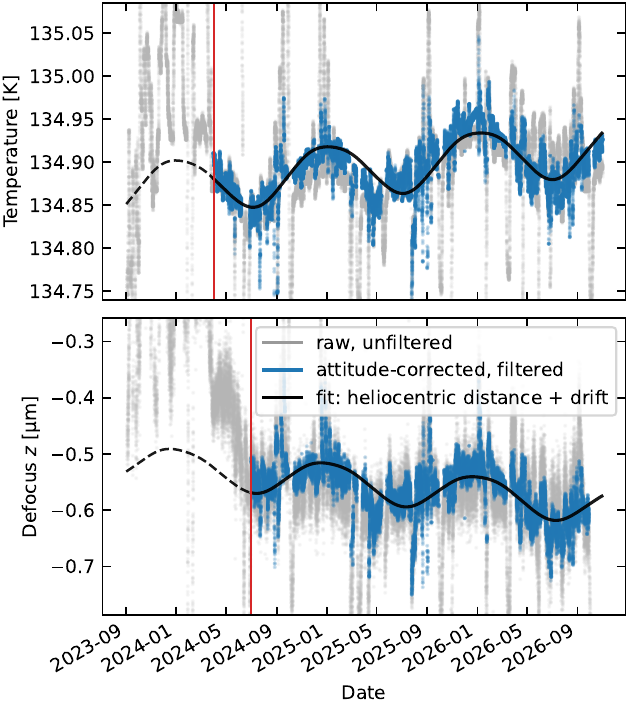}
\caption{Baseplate temperature (\textit{top}) and defocus (\textit{bottom}) as a function of time. The grey and blue dots show the respective measurements before and after correction for spacecraft attitude 
(\cref{fig:temperature_defocus_attitude}). The black lines show the fit allowing for heliocentric distance variation and a linear drift term, evaluated for the periods after specific cut-off dates (red lines). Details are given in \cref{sec:thermal_attitude}.}
\label{fig:T170_timeseries}
\end{figure}

\subsubsection{Modelling}
We consider $T$ and $z$ on a regular (AA, SAA) grid. Some grid cells are visited in only one season, or at times with similar $d_{\rm h}(t)$, thus absorbing part or all of the orbital time dependence. We define $M$ as separate maps of $T_{\rm f}$ and $z_{\rm f}$, the asymptotic values which $T$ and $z$ would attain on the grid in the stationary case. To separate the time evolution from the stationary map, we use the model
\begin{align}
  y_{ci}(t) &= M_c + A\,x_{ci}(t) + C\,(t_{ci}-t_0) + \epsilon_{ci}\,,\quad {\rm with}\label{eq:modeldecomp}\\
  \qquad x_{ci}(t) &= \left[\frac{\langle d_{{\rm h},ci}\rangle}{d_{{\rm h},ci}(t)}\right]^{2} - 1\,.
\end{align}
Here, $c$ labels the grid cell, $i=1,\dots,n_c$ the samples in that cell, and $y_{ci}(t)$ is the measured value, $T$ or $z$. The time dependence is absorbed by $x_{ci}(t)$, the fractional change of the solar flux due to variable heliocentric distance, and by $t_{ci}-t_0$, a linear drift term where $t_0$ refers to the mean epoch of the selected data. The distance and drift coefficients, $A$ and $C$, are shared by all cells. The residual term, $\epsilon_{ci}$, captures measurement noise and temporal variations not described by the model, such as thermal transients following a large step in attitude. Hereafter, we omit indices $c$ and $i$ for simplicity.

Values $y$, $x$, $t$, and $t_0$ are known for each cell, and $M_c$ is unknown. We use an orthogonal decomposition, by subtracting from each of these quantities, $X$, its mean over the samples of the respective cell (hereafter ‘demeaning’), 
\begin{equation}
    \tilde X = X - \langle X\rangle_{\rm cell}\,.
\end{equation}
This yields the reduced model
\begin{equation}
    \tilde y = A\,\tilde x + C\,\tilde t + \tilde\epsilon\,,
\end{equation}
where demeaning removed $M_c$ and $t_0$ because they are constant within a cell. Coefficients $A$ and $C$ are obtained by ordinary least-squares minimisation of $\sum\tilde\epsilon^{2}$. In this reduced model, $A$ and $C$ are constrained only by variations within a cell, that is, by the same attitude observed at different heliocentric distances and times. Samples with $\tilde x\approx0$ carry almost no weight for $A$, that is cells observed in a single season cannot leak the annual signal into the stationary map. The map itself follows as 
\begin{equation}
  M_c = \big\langle\, y - A\,x - C\,(t-t_0) \,\big\rangle_{\rm cell}\,,
\end{equation}
because $\langle\epsilon\rangle_{\rm cell}=0$ by construction.

The fit was iteratively clipped at $3\sigma$, with $\sigma = 1.4826\,\mathrm{MAD}$. For the temperature map we used data taken after 2024-04-01, when the spacecraft has settled into a quasi-stable equilibrium with routine observations; earlier times were affected by \ac{PV}, the phase-diversity campaign \citep{DR1-TP026,DR1-TP027}, and the first thermal decontamination (\cref{sec:decon1st}). For the defocus map we used data taken after 2024-07-01, once the wavefront-altering ice film on \ac{FoM3} was finally cleared by the second decontamination (\cref{sec:iceage}). Only cells with at least three surviving samples were used in the fit; fit parameters are summarised in \cref{tab:fit_parameters}. 

\subsubsection{Results}
The maps in \cref{fig:temperature_defocus_attitude} show that stationary $T_{\rm f}$ and $z_{\rm f}$ have very similar dependence on spacecraft attitude, with $z$ increasing (becoming numerically larger) for increasing temperature. In general, the dependence on \ac{SAA} is considerably larger than on \ac{AA}, unless ${\rm SAA}\gtrsim100^\circ$. The small effect of \ac{AA} is also reported in section 5.2 of \cite{EU-Anselmi}, where a much more comprehensive analysis of \Euclid's thermo-optical performance is presented. Their attitude-related analysis, however, is mostly based on \ac{PV} data taken in October and November 2023, limited to a smaller \ac{AA} and \ac{SAA} range. The notable effect of \ac{AA} on temperature for ${\rm SAA}\gtrsim100^\circ$ was beyond reach in these data, which were also affected by higher thermal instability due to non-standard, early operations (\cref{fig:T170_timeseries}).

\begin{table}[h]
\centering
\footnotesize
\caption{Fit parameters of the attitude-map model from \cref{eq:modeldecomp}, and number $n$ of samples used after masking and clipping.}
\label{tab:fit_parameters}
\begin{tabular}{l|ccc}
\hline\hline
Map $M$ & $A$ & $C$ & $n$ \\
\hline\noalign{\vskip 1pt}
Temperature  & $(931 \pm 96)$\,mK & $(16.2 \pm 1.9)$\,mK\,yr$^{-1}$ & 49219 \\
Defocus & $(969 \pm 161)$\,nm & $(-23.9 \pm 3.9)$\,nm\,yr$^{-1}$ & 39343 \\
\hline
\end{tabular}
\end{table}

\begin{figure}[t]
\centering
\includegraphics[angle=0,width=1.0\hsize]{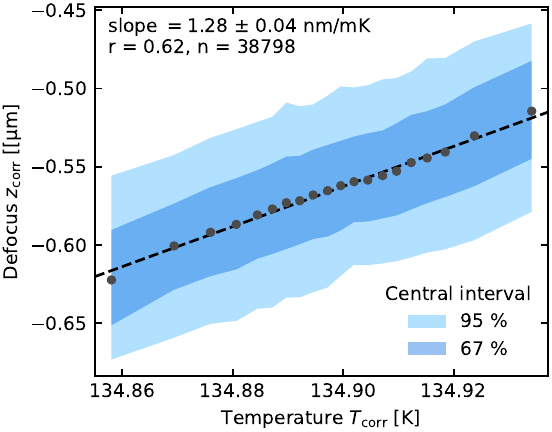}
\caption{Relation between corrected defocus and baseplate temperature, splitting the 38798 data points into 20 equal-count temperature bins. Shown are the attitude-corrected values (blue points in \cref{fig:T170_timeseries}), after removing their fitted temporal evolution (black lines in the same figure).}
\label{fig:temp_defocus_relation}
\end{figure}

The seasonal variations of both baseplate temperature and defocus term are evident in the time-series plots (\cref{fig:T170_timeseries}). The respective amplitudes of 31\,mK and 33\,nm are obtained by scaling fit parameter $A$ with the maximum of the distance regressor, $x_{ci}^{\rm max}=0.0337$.

While the attitude maps show that higher temperatures imply numerically larger $z$, their long-term drift terms have opposite sign (\cref{tab:fit_parameters}). The increase of temperature can be explained, for example, by ageing of the thermal insulation \citep[see][for \textit{Chandra}]{plucinsky2018}. The decreasing (getting more negative) defocus shows that another process must be at play that we have not accounted for. It could be of mechanical nature, such as late-time stress release of the structure and bonds, outgassing, and ageing of adhesives; it could also be an optical change of focal length, since we have not considered curved mirrors, primarily M1 and M3. 

Removing the fitted time evolution from the attitude-corrected data (black lines and blue dots in \cref{fig:T170_timeseries}, respectively) shows a linear dependence of the corrected $z_{\rm corr}$ on $T_{\rm corr}$ (\cref{fig:temp_defocus_relation}). This means that over short time scales of up to a few months, attitudes, and short transients, $z$ follows $T$ with a ratio of $1.28$\,nm\,mK$^{-1}$. The ratio of the long-term drift slopes, $C$, however, is of the same magnitude but opposite direction, $-1.48$\,nm\,mK$^{-1}$, meaning that the defocus is not simply driven by the baseplate temperature alone. The diffraction-spike method \citep{Neumann26} translates the defocus

\begin{figure}[t]
\centering
\includegraphics[angle=0,width=1.0\hsize]{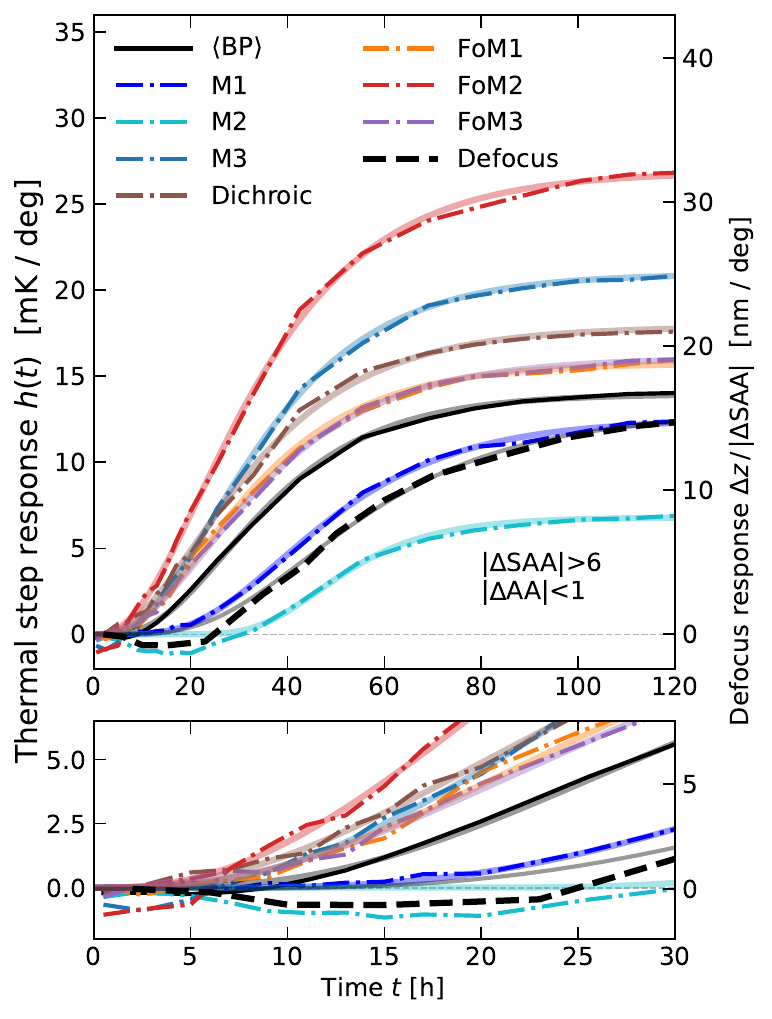}
\caption{Thermal response of optical components, the baseplate, and the defocus term to a large \ac{SAA} step. The dash-dotted curves show the data (dashed for the defocus), and the solid lines their best-fit models. Details are given in \cref{sec:thermal_response_model,sec:thermal_response_interpretation}.}
\label{fig:PLM_thermalresponse_to_SAAstep}
\end{figure}

\subsection{Thermodynamical response: modelling\label{sec:thermal_response_model}}
The thermal response to attitude changes (\cref{fig:temperature_defocus_attitude}) is governed by: (i) external structures moving in and out of the spacecraft's shadow; and (ii) the conductive and radiative couplings of the external and internal structures through layers of thermal insulation. From the mission timeline, we selected five large, undisturbed, and well-defined steps in \ac{SAA} as follows.
\begin{itemize}
    \item{$|{\rm \Delta SAA}|_{\,\rm step}>6^\circ$ and $|{\rm \Delta AA}|_{\,\rm step}<1^\circ$, delivering a strong signal from offsets along the dominant \ac{SAA} direction (\cref{fig:temperature_defocus_attitude}). Typically, the spacecraft moves from the cooler \ac{EWS} to warmer areas at ${\rm SAA}>100^\circ$ -- such as the \ac{EDF-S} -- and back.}
    \item{$|{\rm \Delta SAA}|<6^\circ$ in the preceding 150\,h.}
    \item{$|{\rm \Delta SAA}|<1.5^\circ$ and ${\rm d}T/{\rm d}t<0.02$\,K\,day$^{-1}$ in the preceding 48\,h.}
    \item{$|{\rm \Delta SAA}|<1.5^\circ$ and $|{\rm \Delta AA}|<1^\circ$ in the analysis period following the step.}
\end{itemize}
We then computed the mean stacked impulse response as
\begin{equation}
h(t) = \left\langle\frac{T(t_k + t) - T_{0,\,k}}{|\Delta\mathrm{SAA}_k|} \right\rangle,
\end{equation}
where $t_k$ and $\Delta\mathrm{SAA}_k$ are the time and amplitude of step $k$, and $T_{0,\,k}$ is the mean temperature in the four hours preceding that step. The result is shown in \cref{fig:PLM_thermalresponse_to_SAAstep} for the seven optical components, the baseplate, and the defocus term.

The optomechanical elements are end points of thermal chains that differ in physical length and number of components, and that couple to the baseplate at different locations. We approximated their step response with a special case of the sigmoidal growth curve of \cite{richards1959},
\begin{equation}
h(t)=h_{\infty}\left(1-{\rm e}^{\displaystyle -t/\tau}\right)^{\text{\footnotesize $n$}}\,,
\label{eq:thermalseries}
\end{equation}
where $h_{\infty}$ is the asymptotic temperature change, $\tau$ the characteristic timescale at late times, and $n$ controls the initial delay and the curvature of the response. 

For $n>1$, the derivative at the time of the step vanishes, $h^\prime(0)=0$, which a single exponential cannot reproduce; the response instead sets in with a delay that grows with $n$. The rate of change peaks at the inflection point $t_{\,\rm infl}=\tau\,\ln n$, where $h\approx h_{\infty}/{\rm e}$, after which the remaining difference to the equilibrium temperature decays approximately exponentially on the timescale $\tau$. Higher values of $n$ indicate a longer thermal chain, but in general $n$ does not correspond to the actual number of components because of multiple connectivity and back-reactions. Fit coefficients are listed in \cref{tab:thermal_fits}.

\begin{table}[htbp]
\centering
\footnotesize
\caption{Parameters for the thermal-chain model in 
         \cref{eq:thermalseries}.}
\label{tab:thermal_fits}
\setlength{\tabcolsep}{6pt}
\begin{tabular}{lccccccc}
\hline
\hline\noalign{\vskip 1pt}
Element & $h_{\infty}$ & $\tau$ & $n$
          & $t_{\rm infl}$ & $t_{90}$ & $t_{95}$ & m\\
          & [mK/deg] & [h] & & [h] & [h] & [h] & [kg]\\
\hline\noalign{\vskip 1pt}
BP\_VIS                    &  13.4 &  16.3 &   5.35 &  27.3 &  64 &  76  & \dots\\
BP\_NISP                   &  14.0 &  17.5 &   4.48 &  26.2 &  66 &  78  & \dots\\
BP\_RSU                    &  13.8 &  17.5 &   4.58 &  26.7 &  66 &  79  & \dots\\
BP\_M3                     &  14.5 &  19.0 &   3.85 &  25.6 &  69 &  82  & \dots\\
$\langle{\rm BP}\rangle$   &  13.9 &  17.6 &   4.56 &  26.7 &  66 &  79  & 187\\
\hline\noalign{\vskip 1pt}
M1                         &  12.5 &  20.5 &   6.45 &  38.2 &  85 &  99  & 52\\
M2                         &   6.7 &  12.8 &  36.44 &  45.9 &  75 &  84  & 9.2\\
M3                         &  20.9 &  19.2 &   3.49 &  24.0 &  67 &  81  & 5.4\\
FoM1                       &  15.7 &  18.7 &   3.25 &  22.0 &  64 &  78  & 5.1\\
FoM2                       &  26.9 &  20.5 &   2.85 &  21.5 &  68 &  83  & \multirow{2}{*}{\scalebox{1}[1.3]{$\Big\}$}\,11}\\
Dichroic                   &  17.9 &  20.1 &   2.82 &  20.8 &  66 &  81  & \\
FoM3                       &  16.1 &  20.7 &   2.99 &  22.7 &  70 &  84  & 5.1\\
\hline
\end{tabular}
\tablefoot{The response changes fastest at $t_{\rm infl}$ with $h\approx h_\infty/{\rm e}$, and it reaches 90\% and 95\% of $h_\infty$ at $t_{90}$ and $t_{95}$, respectively. The baseplate has four groups of sensors: near VIS, NISP, the \ac{RSU}, and M3. The parameters of the average baseplate response, $\langle{\rm BP}\rangle$, are the means of the four groups; the mass of the baseplate is arbitrarily listed there. FoM2 and the dichroic element share the same mount and are reported with their total mass. All masses include holders and brackets.}
\end{table}

\subsection{Thermodynamical response: interpretation\label{sec:thermal_response_interpretation}}

Overall, the inflection times increase from the rear cavity (22\,h) through the baseplate (27\,h) and M1 (38\,h) to M2 (46\,h), following the distance of the elements from the heat input -- increased Sun exposure -- at the bottom of the spacecraft when going to higher \ac{SAA} values. The folding mirrors, M3, and the dichroic are located in the rear cavity and all behave similarly, with $\langle t_{\rm infl}\rangle=(22.2\pm1.2)$\,h, $\langle\tau\rangle=(19.8\pm0.9)$\,h, and $\langle n\rangle=(3.08\pm0.29)$ (mean and standard deviation). They start to respond 1--5\,h earlier than the baseplate thermistors and warm more, suggesting that in addition to conductive heating through the baseplate, radiative exchange with the cavity is also relevant.

The instruments as internal heat sources maintain a steady-state temperature gradient across the baseplate (\cref{fig:cavity_temperature}). The four baseplate thermistors show similar step responses, indicating that the step heats the baseplate fairly uniformly. Compared to the cavity optics, the baseplate has similar $t_{90,\,95}$ equilibration times, but settles at a smaller temperature change, $\langle h_\infty\rangle=13.9$\,mK\,deg$^{-1}$, presumably because its front side faces M1 that is 4--9\,K colder.

M1 is the largest and most massive optical element. Its uncoated rear side faces the baseplate, whereas its silver-coated front, with low infrared emissivity, sees mostly the telescope baffle at about 100\,K. It responds later and warms less than the baseplate, which is the expected behaviour when heated mainly through the baseplate. While its $\tau=20.5$\,h matches those of the optical elements in the instrument cavity, the high $n=6.5$ delays the fastest temperature change to $t_{\rm infl}=38.2$\,h.

M2 is the notable exception. At 109\,K, it is 18--25\,K colder than the other optics -- which are all passively cooled -- and is kept at this temperature by survival heaters operating at constant power \citep{Racca2016}. Its connection to the baseplate through the truss, spider arms, and mounting structure contributes to a high value of $n$. In response to the step, M2 initially cools, reaching its minimum temperature after about 15\,h and returning to its initial temperature only after 32\,h. Since \cref{eq:thermalseries} cannot become negative, the fit artificially increases $n$ to delay the temperature rise. 

The initial cooling is unexpected, because the \ac{SAA} step increases the temperature at the bottom of the spacecraft, from where heat is then conducted upwards through the \ac{PLM}. The sunshield, however, cools down because the step tilts it further away from the Sun, reducing the amount of heat leaking through the insulation into the \ac{SVM} and \ac{PLM}. We speculate that the telescope baffle has a better conductive coupling to the sunshield than the much more critical telescope optics. The baffle would then cool before receiving the heat from the warmer bottom; and the quick response of M2 implies an efficient radiative coupling of its holding structure to the baffle (the reflective surface itself couples very weakly). Once warming, M2 equilibrates fastest with $\tau=12.8$\,h, reaching $t_{90}$ and $t_{95}$ 10\,h and 15\,h earlier than M1, presumably because the rear of its holding cell has a large view factor to cold space.

The time constants reported in Whittam et al. (in prep.) differ from ours, because the measurements are not directly comparable. Their table 1 is based on the response to the thermal decontamination of M3, when heat is injected directly and locally into the \ac{PLM}, as compared to the tilts in \ac{SAA} space that we consider. Their table 2 lists time delays between temperatures of the baseplate and M1 to M3, and is based on a selection of six large temperature-change events; these include instrument shutdowns, thermal decontamination, and other disturbances causing high temperature spikes. In contrast, our analysis is not based on selected temperature anomalies, but on selected attitude changes following a defined set of criteria to increase homogeneity.

\section{Water contamination and decontamination\label{sec:contaminationanddecontamination}}

\subsection{On-ground contamination control\label{sec:contaminationcontrol}}
Contamination management was paramount and extensive throughout \Euclid's entire design and development phase. Organic volatile contaminants, particulates, and the environment were controlled during the \ac{AIT} of every element and at each subsequent integration step toward the fully integrated spacecraft, meeting all cleanliness specifications \citep{Portaluppi2023}. The deliberate absence of an aperture cover on the telescope made this particularly critical. The launch campaign and ascent were designed to minimise contaminants, and very clean preparation facilities, commissioned by \ac{ESA}, included an ISO-5 filter wall. Strict control of the Falcon~9 fairing surfaces and ventilation systems was implemented, including an analysis of cross-contamination during ascent and the various separation events. These challenges were all successfully achieved. 

Thanks to these extensive preparations, water is the only molecular contaminant of relevance for \Euclid \citep[][hereafter \citetalias{Schirmer-EP29}]{Schirmer-EP29}. Clean rooms are kept at about 50\% relative humidity for numerous reasons, so water molecules are omnipresent. Water is easily absorbed by various materials, in particular by organic \ac{MLI}, which absorbs approximately 1\% of its mass in water vapour under clean-room conditions within 24\,h after baking \citep{scialdone1993}. Due to this fast reabsorption during \ac{AIT} and the open telescope, measures were taken to limit or remove water immediately before and after launch. First, the \Euclid payload fairing was purged with dry nitrogen for 4\,h on the launchpad right up to lift-off. Second, a full thermal decontamination (\cref{sec:decontamination_full}) began immediately after payload separation from the launcher, approximately 2 hours after launch. Venting holes in the \ac{PLM} permitted rapid decompression during launch. The combined area of all the holes, however,is too small\footnote{Openings to outer space are potential paths for parasitic stray light.} for efficient evacuation of the gas in the molecular flow regime. As a result, water is predominantly redistributed within the cavity and ultimately accumulates on the coldest surfaces (\cref{fig:cavity_temperature}), with measurable consequences (\cref{sec:decon1st,sec:iceage}). An extensive and -- to date -- unique test campaign was performed at the European Space Research and Technology Centre (ESTEC) to measure the optical effects of ice-contaminated optical surfaces \citep{szmolka2023}.

Once in space, the total water content that could eventually be released into the instrument cavity has been estimated at 100\,g, mostly from the 8\,kg of \ac{MLI} that shields the instruments and optics in the rear cavity from outer space. This estimate accounted for all organic absorbing materials used in the cavity, their tabulated outgassing rates, and the amount of re-absorbed water after vacuum baking on the ground. 

\subsection{Decontamination criteria\label{sec:decontamination_criteria}}

Ice on optical surfaces causes scatter loss, optical interference, absorption in the (near-)infrared, mild polarisation, and changes in wavefront error. It therefore affects photometry and also the \ac{PSF} -- \Euclid's core measurements. A decontamination procedure, which is a major thermal disturbance to \Euclid (\cref{fig:decontamination_thermal_response}), is critical only in two cases. Firstly, if the ice itself poses a physical risk to optical coatings and detectors (delamination), which remains largely uncharacterised to date; instead, we rely on experience gained with similar technologies in space. Secondly, if the effects due to ice accumulation cannot be calibrated any more to the accuracy and precision necessary for Stage IV cosmology surveys \citep{Albrecht2006}.

The amplitude and characteristics of the effects that nanometre-thin ice films -- with unknown surface properties -- have on spectro-photometric data were largely unknown before \Euclid's launch. Combined with the large uncertainties in the actual outgassing rates, our estimates of how often decontamination would be needed during the observing phase ranged from several times per year to none at all. Hence, the survey does not pre-allocate time for decontamination; also, because of the long planning horizon, executing a decontamination procedure requires five weeks of lead time. We therefore adopted a wait-and-see approach, using suitable observables (\cref{sec:icedetection_methods}) to reveal contamination. Close monitoring, together with a conservative look-ahead, is an essential part of our data-validation process.

\subsection{Full thermal decontamination\label{sec:decontamination_full}}
A full decontamination procedure was developed for \Euclid to remove ice from most surfaces. The \ac{PLM} is warmed to between 200\,K and 270\,K using internal heaters, and tilting a maximum ${\rm SAA}\geq45^\circ$ toward the Sun for de-icing the telescope baffle. Instruments are turned off during this time. The full thermal cycle lasts around two weeks, followed by at least two weeks for recalibration and the validation of the VIS \ac{PSF} model (Miller et al., in prep.). In case the \ac{PSF} model is found invalid, parts or all of \ac{PDC} \citep[section 6.3 in][]{DR1-TP026} have to be repeated for a duration of up to two months. To this date, a full decontamination procedure was executed only once, after launch in July 2023 and before spacecraft commissioning. While substantial progress has been made compared to heritage cryogenic missions, more accurate and long-term modelling of water transport inside partially closed payloads are needed.

\begin{figure}[t]
    \centering
    \includegraphics[width=1\linewidth]{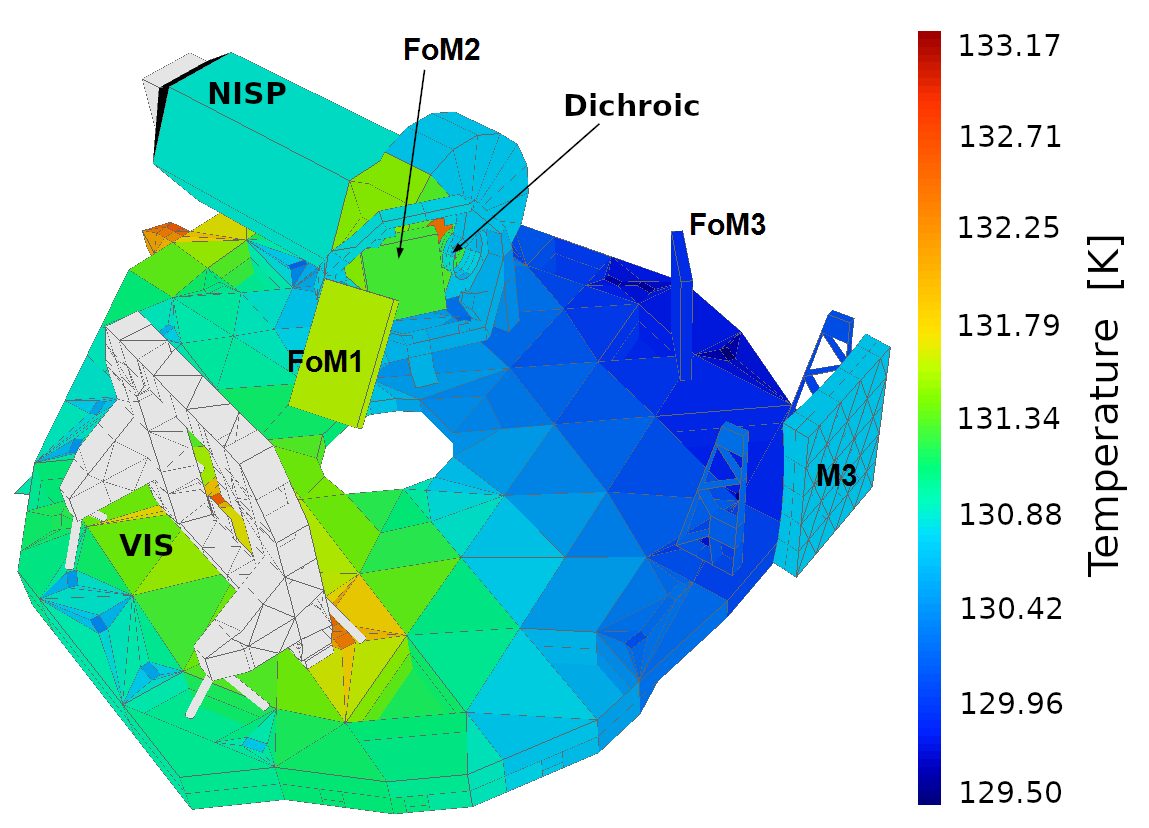}
    \caption{Temperature forecast for the instrument cavity. \ac{FoM3} is the coldest optical element and most susceptible to ice contamination. The baseplate is coldest around and behind \ac{FoM3}, acting as another cold trap. Actual in-flight temperatures are slightly lower than this prediction, with 128.6\,K for \ac{FoM3}. Credit: D. Filleul / Airbus Defence and Space, D. Stramaccioni / ESA.}
    \label{fig:cavity_temperature}
\end{figure}

\subsection{Surgical decontamination\label{sec:decontamination_surgical}}
A few months after launch, contamination of the VIS optical channel became evident (\cref{sec:decon1st}). Only \ac{FoM3} -- the coldest in the instrument cavity 
(\cref{fig:cavity_temperature}) -- appeared affected, which did not justify a full decontamination. Instead, we developed a minimally invasive procedure, targeting individually first \ac{FoM3} and then M3 -- the second-coldest mirror in the cavity -- with their separate heaters only.

First, a full self-calibration visit \citep[see section 5 in][]{DR1-TP026} provides the necessary calibration data to conclude the current survey epoch. The contaminated element is then warmed up in two steps from near 130\,K to 160\,K for two hours, enough to sublime an ice film with a thickness of around $900$\,nm. The instruments are idling during this time, and commence observing the centre of the self-calibration field as soon as the heater is turned off. A diffraction-spike method \citep{Neumann26} is used to determine the defocus wavefront error evolution with a precision of 22\,nm in these images. After three days, the \ac{PLM} is sufficiently stable for the second full self-calibration visit that concludes the procedure and resumes the survey. Full thermo-optical equilibration to within weak lensing requirements, however, occurs only eight days after the beginning of the cool-down (\cref{fig:decontamination_thermal_response}).

\begin{figure*}[t]
    \includegraphics[width=1.0\hsize]{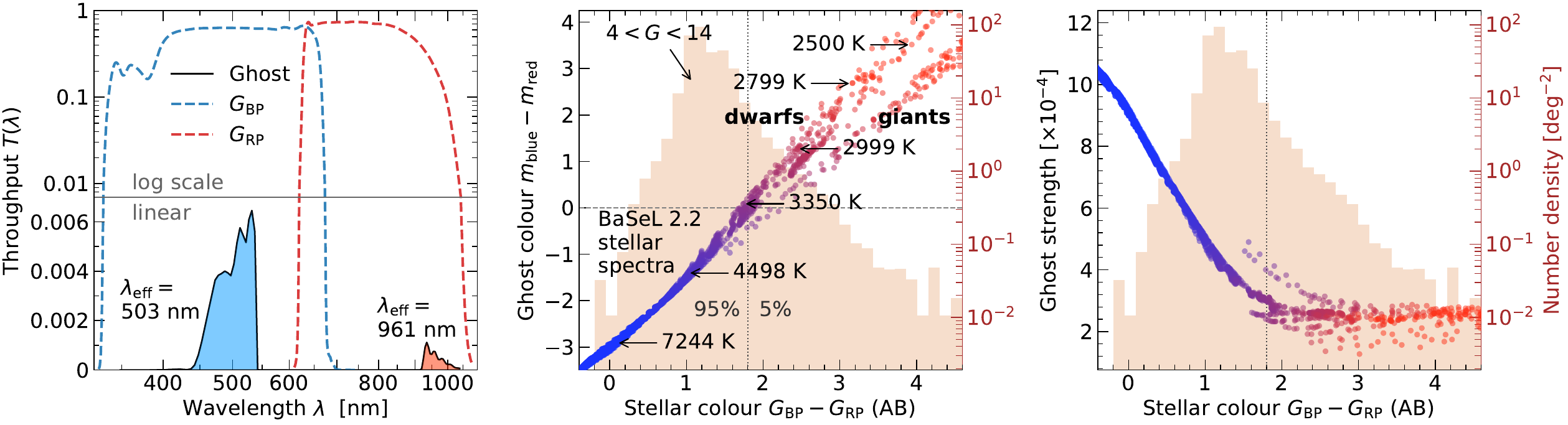}
    \caption{\textit{Left}: the two peaks of the ghost throughput $T_{\rm g}(\lambda)$, straddling the \IE band (530--920\,nm). The peaks fall exclusively into the \textit{Gaia} $G_{\rm BP}$ and $G_{\rm BP}$ bands. \textit{Middle}: Ghost colour as a function of synthetic \textit{Gaia} photometry computed from the BaSeL 2.2 stellar library. The strong response of the ghost colour to the stellar \ac{SED} is a direct consequence of the two peaks falling into separate \textit{Gaia} bands. The data points are arbitrarily coloured from blue to red for visualisation purposes. The histogram shows the number density of stars with $4<G<14$ averaged over the lines of sight toward the \ac{NEP} and the \ac{NGP}. \textit{Right}: \ac{SED}-dependence of the ghost strength. The histogram replicates the stellar density shown in the middle panel. More details are given in \cref{sec:icedetection_dichroicghost}.}
    \label{fig:combined_ghost_panels}
\end{figure*}

The \ac{FGS} continues guiding through the contaminated optics during the entire thermal cycle. Using the improving guide-star flux and temperature readings, the sublimed amount of ice (\cref{sec:iceage}) can be computed and -- for the first time and \Euclid-specific -- ice thickness can be related with optical throughput loss.

The detectors in the Near-Infrared Spectrometer and Photometer \citep[NISP][]{EuclidSkyNISP} also experience notable contamination (\cref{sec:DR1_NISP_contamination}). Our calibration cadences offer sufficiently high temporal sampling to allow for correction at the data level; the methods are currently under development. While a surgical decontamination procedure is under proactive development, it is unclear whether it will be needed during the nominal mission.

\subsection{Ice-detection methods\label{sec:icedetection_methods}}

\subsubsection{Stellar photometry\label{sec:icedetection_stellar}}
The main effects of ice on photometry are scatter losses from rough ice surfaces and interference effects with coating stacks, on mirrors and detectors alike. Both effects are wavelength-dependent, becoming more pronounced at shorter wavelengths. They were not known accurately beforehand, because the surface roughness and thickness of the ice layer are difficult to predict \citepalias{Schirmer-EP29}, and the coating-stack designs that control the interference are proprietary. 

As a first check, the stellar flux ratio between \IE band and \textit{Gaia} $G$ band \citep{vallenari2023} is computed in three \textit{Gaia} $G_{\rm BP}-G_{\rm RP}$ colour bins. This is done for every uncalibrated VIS image as soon as it becomes available, after a static gain correction is applied per detector quadrant. Despite the rudimentary calibration, with this near real-time assessment significant changes are readily detectable, and systematic transmission changes as small as $0.1$\,mmag\,month$^{-1}$ can be followed on longer timescales (\cref{sec:decon1st,sec:iceage}). The assessment with fully calibrated data is part of the monthly self-calibration.

\subsubsection{NISP flat fields\label{sec:icedetection_flats}}
At 95\,K, the NISP detectors are the coldest part in the spacecraft and estimated to accumulate 10\,nm of ice per month \citepalias[table 2 in][]{Schirmer-EP29}. This is detectable with high \ac{SNR} in flat fields, for which the detectors are illuminated directly without intervening optics \citep{EuclidSkyNISP}. 

Indeed we find signal modulations in the flats that can only be explained by interference fringes. Initial modelling with plausible \ac{AR} stacks shows that both the fringe amplitude and frequency decrease with increasing wavelength, matching the observed effects. The fringes are detector-dependent because of the individual \ac{AR} coating stacks and differential ice-deposition rates. The interference models and the code to obtain them are under development for DR2 and DR3 \citep[see also section 7.7 of][]{DR1-TP026}. We note that the VIS detectors with a temperature of 153\,K are too warm for water contamination.

\subsubsection{VIS dichroic ghost strength\label{sec:icedetection_dichroicghost}}
The dichroic beamsplitter reflects the \IE band off its front surface toward VIS. A small fraction of the photons transmitted to NISP, however, is reflected back off the  rear surface. In VIS, they appear as ghosts offset by \ang{;;70} from the bright stars ($G\lesssim14)$ responsible for them. During \Euclid's heavy contamination phase (\cref{sec:iceage}), ghosts were far more suppressed than the stars causing them, albeit with considerable scatter. 

The ghost throughput $T_{\rm g}(\lambda)$ -- which includes the detector \ac{QE} -- is characterised by two widely separated peaks at effective wavelengths 503\,nm and 961\,nm, straddling the \IE band (left panel in \cref{fig:combined_ghost_panels}). The ghost's colour, or ratio of photon fluxes in the blue and red peaks, is expressed in magnitudes as $m_{\rm blue}-m_{\rm red}$ and depends strongly on the stellar \ac{SED}. The colour is not directly measurable in the VIS images, but since the peaks fall separately into the \textit{Gaia} $G_{\rm BP}$ and $G_{\rm RP}$ bands, it is easily predicted using \textit{Gaia} photometry. This is shown in the middle panel of \cref{fig:combined_ghost_panels}, where we computed the colour by integrating $T_{\rm g}(\lambda)$ over the BaSeL 2.2 synthetic stellar spectra \citep{Lejeune1998} provided in {\tt pystellibs} \citep{Fouesneau2025}. In typical survey areas, 95\% of bright stars with $G<14$ have $G_{\rm BP}-G_{\rm RP}<1.8$ and blue ghosts with $m_{\rm blue}-m_{\rm red} < 0$. The remainder comprises stars cooler than $T_{\rm eff}\lesssim3300$\,K that have red ghosts. The ghost colour follows a linear relation,
\begin{equation}
    m_{\rm blue}-m_{\rm red}\approx1.84\,\left(G_{\rm BP}-G_{\rm RP}\right) - 3.27\,,
\end{equation}
except for cool giants and stars hotter than $T_{\rm eff}\gtrsim7200$\,K. The right panel in \cref{fig:combined_ghost_panels} shows the ghost strength, which we define as the ratio of two integrals of the stellar \ac{SED} over the 400--1050\,nm interval: one filtered by $T_{\rm g}(\lambda)$, and the other unfiltered. For $G_{\rm BP}-G_{\rm RP}\lesssim1.8$, that is 95\% of the stars, the ghost strength rises steeply toward bluer colours, whereas for $G_{\rm BP}-G_{\rm RP}\gtrsim1.8$ it flattens to an approximately constant value of $2.6\times10^{-4}$.

\begin{figure}[t]
\centering
\includegraphics[angle=0,width=1.0\hsize]{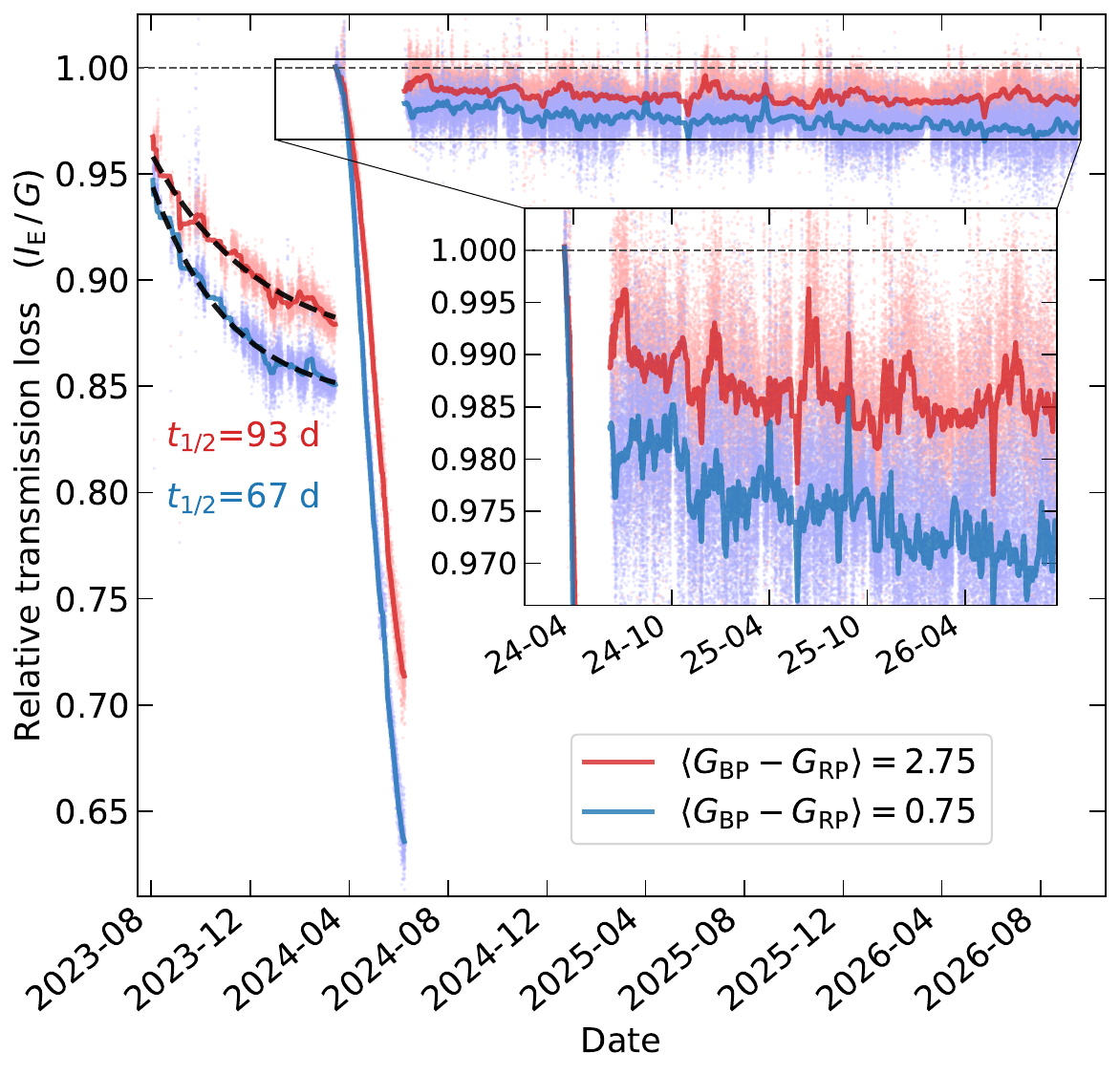}
\caption{Relative VIS transmission loss since 2023-08-03, measured for blue and red \textit{Gaia} sources in uncalibrated data. Shown is the flux ratio between VIS \IE band and \textit{Gaia} $G$ band, normalised to the first 24\,h of observations that were taken after the first decontamination on 2024-03-14 that involved mirrors \ac{FoM3} and M3. In the first six months, the flux losses decayed exponentially with half-times of 67 and 93 days for blue and red sources, respectively. The heavy contamination period that followed the first decontamination was stopped on 2024-06-08, when only \ac{FoM3} was heated, resulting in approximately 1.5\% lower throughput recovery than on 2024-03-14. Since then, the red (blue) sources lost 0.5\% (1.0\%) in the 2.5 years that followed (magnified inset).}
\label{fig:iceplot_smoothed}
\end{figure}

\begin{figure}[t]
    \includegraphics[width=\linewidth]{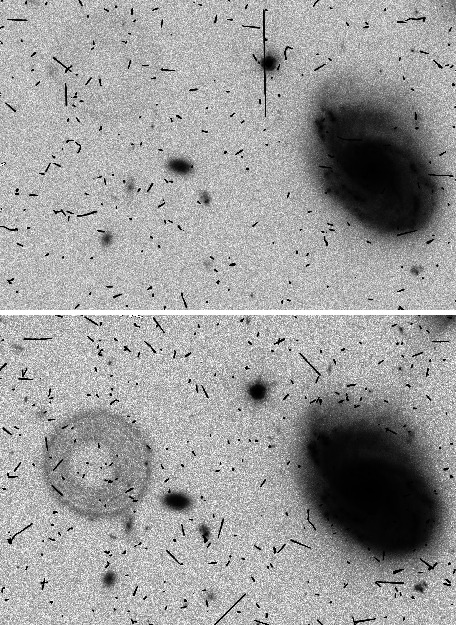}
    \caption{Strength of the VIS dichroic ghost. \textit{Top}: With heavy contamination on \ac{FoM3} on 2024-06-06. Due to its blue colour (\cref{sec:icedetection_dichroicghost}), the ghost is strongly suppressed by surface scattering on ice. \textit{Bottom}: After decontamination on 2024-06-08, the ghost re-appears with nominal strength.}
    \label{fig:visghoststrength}
\end{figure}

\begin{figure}[t]
    \includegraphics[width=\linewidth]{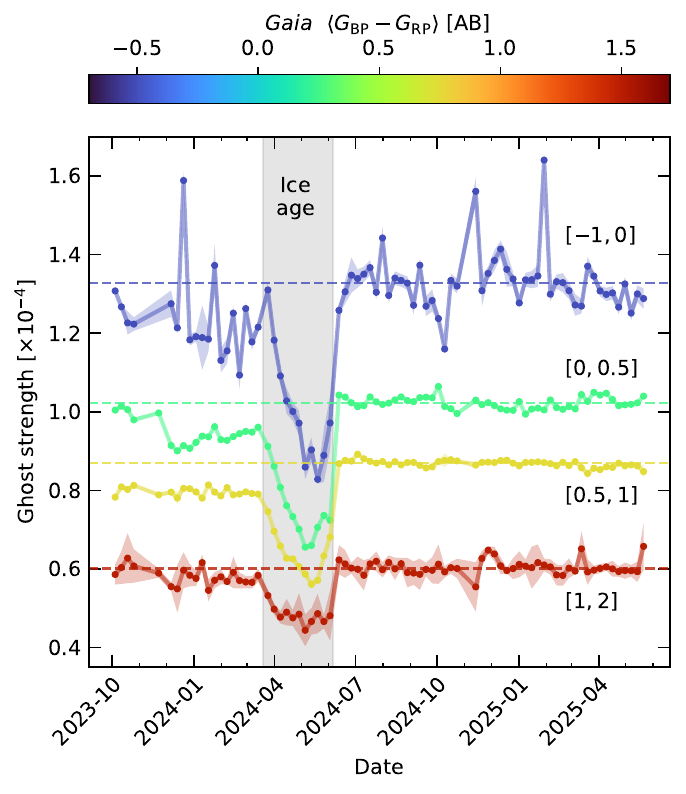}
    \caption{Weekly average of the dichroic ghost strength, for stars in four \textit{Gaia} $G_{\rm BP}-G_{\rm RP}$ bins (brackets). Bluer stars have relatively brighter ghosts (right panel in \cref{fig:combined_ghost_panels}) that respond more strongly to contamination. The dashed lines show the median strength for clean optics, computed from the period after June 2024. This metric will enter the VIS pipeline \citep{DR1-TP002} for DR2.}
    \label{fig:VIS_ghost_gaiacolor}
\end{figure}

\section{In-flight contamination history\label{sec:inflight_contamination_history}}
\subsection{First decontamination on March 14 2024\label{sec:decon1st}}
In October 2023, a cross-match of all VIS science exposures against \textit{Gaia} back to the beginning of \ac{PV} in August 2023 revealed a throughput loss of about 5\% that must had been caused by contamination. Since no changes were recognised in NISP at that time, \ac{FoM3} -- the only mirror not in the NISP optical path -- must have be contaminated. \ac{FoM3} is also the coldest mirror (128.6\,K) in the instrument cavity (\cref{fig:cavity_temperature}) and therefore most susceptible to ice, followed by M3 at 132.0\,K. At this point, the two-month long \ac{PDC} was running and could not be interrupted because it requires extraordinary thermal stability.

The VIS throughput loss reached 12\% in March 2024, following exponential decays with respective half-times of 67 and 93 days for blue and red \textit{Gaia} sources with $G_{\rm BP}-G_{\rm RP}=0.75$ and 2.75, respectively. These half-times are broadly consistent with those of 30--60 days measured for the decay of ambient pressure in the vicinity of the Midcourse Space Experiment \citep[MSX, see][]{uy2003} and \textit{Rosetta} \citep{schlaeppi2010} after their launch. Even though throughput loss and ambient pressure are not directly comparable, outgassing is the physical origin of both and therefore the time-dependence of \Euclid's behaviour was not unexpected. Previous missions had less restrictive contamination management and experienced considerably worse contamination once in flight; for example, the \textit{Gaia} water-contamination anomaly released orders of magnitude more water than \Euclid and for an entire year. \Euclid's strict contamination management was a direct consequence of the \textit{Gaia} anomaly, and resulted in substantially improved water management. In \Euclid, the ice appeared to be localised on \ac{FoM3}, and possibly M3. Since its impact on the wavefront error was uncertain, we used the surgical decontamination method (\cref{sec:decontamination_surgical}) to start nominal science operations with uncontaminated optics. Transmission was fully restored, exceeding the earliest data by about 5\% (\cref{fig:iceplot_smoothed}).

\subsection{Ice age from 14 March to 8 June 2024, second decontamination, and data loss\label{sec:iceage}}
Following the first decontamination, the throughput began a steep decline and lost 5\% just eight days after decontamination (\cref{fig:iceplot_smoothed}). The loss rate accelerated further, reaching a maximum of 0.010\,mag\,day$^{-1}$ in the blue $G_{\rm BP}-G_{\rm RP}$ bin after 34 days, and 0.007\,mag\,day$^{-1}$ in the red bin after 50 days. After 30 days, a large gradient started developing across the \ac{FPA}, indicating that the contamination began to slow down on some part of \ac{FoM3} while continuing unabated elsewhere. The scatter loss observed for blue  dichroic ghosts was very high, near the end by a factor of around five compared to 0.6\,mag loss in the source stars; this is fully explained by the ghost's bimodal throughput curve 
(\cref{sec:icedetection_dichroicghost}).

At this point, a second decontamination was unavoidable. First, the throughput loss meant that \Euclid would not reach its target depths any more. And second, the temporal and spatial evolution was too high to meet the requirement on relative photometric accuracy (1.0\%) using self-calibration alone, thus introducing critical dependencies on external data. At the time of the second decontamination on 2024-06-08, the loss had stabilised at 0.1\,mag in the upper right corner of the \ac{FPA}, and was still deteriorating beyond 0.5\,mag in the lower left corner. The same surgical decontamination method was applied, to \ac{FoM3} only, recovering 99\% and 98\% for the red and blue $G_{\rm BP}-G_{\rm RP}$ bins with respect to the first decontamination, when M3 was included in the process. The throughput has been stable since to within $\sim0.2\,$mmag\,month$^{-1}$, implying that the water reservoirs largely dried up at around the same time.

A retrospective explanation for this behaviour is given in \cref{sec:icethickness_sublimated}. The sky areas observed during this heavy contamination period have been excluded from DR1, since the effect on the wavefront error is still being analysed. The areas have since been re-observed and will be included in DR2. 

\begin{figure*}[t]
    \centering
    \includegraphics[angle=0,width=0.85\hsize]{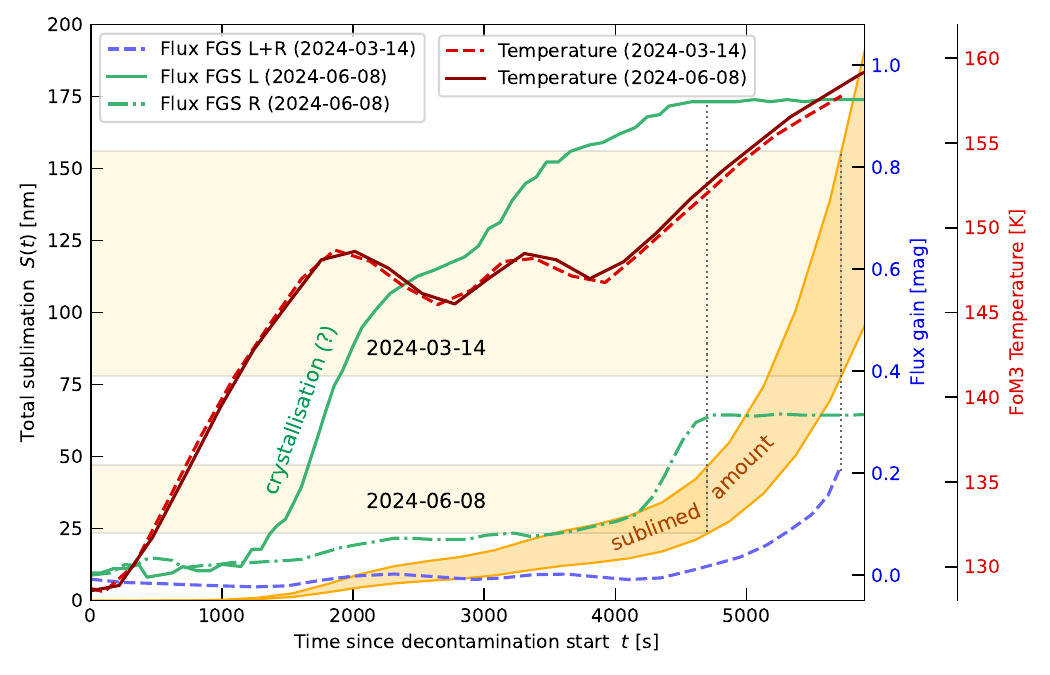}
    \caption{Estimating ice-sublimation amounts. For better understanding, the figure has to be read in a specific order: \textit{Red curves}: Show that both heating profiles were identical. \textit{Dark orange area}: Sublimed amount of ice computed with \cref{eq:murphy} for this heating profile,  accelerating rapidly for $T\gtrsim153$\,K. \textit{Blue dashed line}: Flux gain during the first decontamination, averaged over the \ac{FGS} L and R guide sensors. \textit{Upper yellow rectangle}: The sublimed amount of ice (78--156\,nm) is given by the dark orange area, intersected at the time when the \ac{FGS} flux stabilised (dotted vertical line). During the first decontamination, the \ac{FGS} was interrupted by an antenna repointing, but the gain of $0.20$\,mag was close to final because it matches the recovered VIS flux of 0.19\,mag. \textit{Green lines}: \ac{FGS} flux gain during the second decontamination. While \ac{FGS} L recovered much sooner than \ac{FGS} R, both finished at the same time, 950\,s earlier than during the first decontamination. The ice film was 23--46\,nm thick (\textit{lower yellow rectangle}), yet it caused a much higher flux loss, presumably because of much higher surface roughness  (\cref{sec:icethickness_sublimated}).}
    \label{fig:cumulated_sublimation}
\end{figure*}

\subsection{Computation of ice thickness on FoM3\label{sec:icethickness_sublimated}}
The accelerating flux loss -- and its large amplitude -- during the ice age were unexpected. The temperature change of \ac{FoM3} from 130--160\,K was too small and too localised to plausibly activate a previously dormant, distant water reservoir. Our retrospective analysis of the \ac{FGS} flux evolution during both decontaminations suggests that the ice film present at the time of the second decontamination was about 3 times thinner and amorphous with a rough surface, compared to the thicker and more crystalline and smooth film that had developed before the first decontamination.

\subsubsection{Sublimation mass flux\label{sec:sublimationmassflux}}
Based on kinetic gas theory, the sublimation mass flux of crystalline ice at temperate $T$ in vacuum can be approximated as
\begin{equation}
    j_{\rm s}(T)=\sqrt{\frac{m}{2\pi\, k_{\rm B}\,T}}\;p_0\,\exp\left[c_1+c_2/\mathcal{T}+c_3\,{\rm ln}\,(\mathcal{T}) +c_4\,\mathcal{T}\right]\,,
    \label{eq:murphy}
\end{equation}
where $m=2.99\times10^{-26}$\,kg is the mass of a water molecule and $k_{\rm B}$ Boltzmann's constant. The exponential term with pre-factor $p_0=1$\,Pa and coefficients
$c_1 = 29.3577$, $c_2 = -20.9521$, $c_3 = 3.53068$, and $c_4 = -1.98951$ is the water-ice sublimation pressure for $T>110$\,K given in equation (7) of \cite{murphy2005}, where we renormalised the temperature by the triple-point temperature $T_{\rm t}=273.16$\,K of water, $\mathcal{T}=T/T_{\rm t}$, to obtain a unit-free expression.

\subsubsection{Thermal timescales of SiC and ice films\label{sec:icethermalconductivity}}
The time lag between the actual temperature of the ice film on the front of FoM3 and the readings of the thermistors -- attached to its side or back -- is smaller than the reading cadence of approximately 250\,s. We show this by considering FoM3 and the ice film as slabs of thickness $L$ and density $\rho$, with a temperature of 120\,K. The characteristic timescale for their thermal equilibration is \citep[see e.g.,][]{carslaw1959,incropera2007}
\begin{equation}
    \tau \approx \frac{4\,L^2\,\rho\,c_{\rm p}}{\pi^2\,\kappa}\,,
    \label{eq:thermaltimescale}
\end{equation}
with thermal conductivities $\kappa_{\rm ice}\approx6$\,W\,m$^{-1}$\,K$^{-1}$ for crystalline hexagonal ice \citep{andersson2005} and $\kappa_{\rm SiC}\approx180$\,W\,m$^{-1}$\,K$^{-1}$ \citep{bougoin2016}. The specific heat capacity $c_{\rm p}$ in vacuum is $1021\,\mathrm{J\,kg^{-1}\,K^{-1}}$ for crystalline ice\footnote{$890\,\mathrm{J\,kg^{-1}\,K^{-1}}$ for amorphous ice \citep{handa1988}.} \citep{feistel2006}. For SiC, we interpolated $190\,\mathrm{J\,kg^{-1}\,K^{-1}}$, based on the manufacturer's SiC datasheet \citep{mersen2024}. The densities of ice and SiC are 917\,kg\,m$^{-3}$ and 3150\,kg\,m$^{-3}$, respectively.

Assuming $L=10$\,mm for FoM3 and 100\,nm for the ice, the respective characteristic timescales are approximately 0.1\,s and 0.6\,ns. The ice--substrate interface itself has a thermal resistance because of phonon scattering, adding a few nanoseconds for typical solid--solid interfaces \citep{Swartz1989}. Therefore, FoM3 and the ice are in good thermal equilibrium at our time resolution.

\subsubsection{Film thickness and mass estimates\label{sec:icefilmthickness}}
We computed the thickness $S(t)$ of the ice layer sublimed at a time $t$ after decontamination start by integrating the mass flux \cref{eq:murphy} over the temperature curve,
\begin{equation}
    S(t) = \frac{1}{\rho}\,\displaystyle\int\limits_0^t j_{\rm s}\left[T(t')\right]\,{\rm d}t'\,.
    \label{eq:murphy_integrated}
\end{equation}
The data and results are shown in \cref{fig:cumulated_sublimation}. The dashed red line shows the temperature curve during the 2024-03-14 decontamination, and the blue curve the associated \ac{FGS} flux gain. \Euclid repointed the antenna near the end, and the \ac{FGS} integration was interrupted at $t=5717$\,s, just before the mirror became clean; we know this because the magnitude gain in VIS after decontamination was within a few 0.01\,mag of the gain that could still be recorded with the \ac{FGS}. The total amount sublimed is determined by the intersection of the sublimation curve computed with \cref{eq:murphy_integrated}, shaded orange in \cref{fig:cumulated_sublimation}, at $t=5717$\,s. Accordingly, on 2024-03-14 about 78--156\,nm of ice were present. Given a \ac{FoM3} area of $35.8\times21.5$\,cm$^2$, this implies a mass of 5.5--11.0\,mg.
We note that at 4300\,s, the temperature passed through 152\,K and the flux began recovering rapidly as sublimation became very effective. Furthermore, the \ac{FGS} flux gain (blue line) oscillated 3 times by about 1--2\% during the first 4300\,s after heating began, indicating some restructuring or annealing processes at the lattice level. 

The behaviour during the second decontamination was notably different, despite the same heating profile (red solid line). Consistent with the large zero-point gradient across the \ac{FPA}, \ac{FGS} L (solid green line) recovered about 0.93\,mag of flux, compared to 0.32\,mag for \ac{FGS} R on the opposite side of the \ac{FPA} (dashed green line). Notably, the \ac{FGS} L flux started recovering already after 1200\,s when the mirror reached 140\,K. This is too soon for effective sublimation, but high enough to crystallise amorphous ice phases, and to collapse pores through surface reorganisation \citepalias[for a discussion see][and references therein]{Schirmer-EP29}. Crystallisation and reduction of porosity reduced the surface roughness and therefore scatter losses, before the ice film sublimed once $T$ exceeded 150\,K. For \ac{FGS} R we observe similar oscillations as during the first decontamination, beginning to recover flux rapidly 4100\,s after start, ending at 4700\,s jointly with \ac{FGS} L. This is 950--1000\,s earlier than on 2024-03-14, implying a considerably lower ice thickness of  23--46\,nm, or 1.6--3.2\,mg. This is not surprising because outgassing generally slows down after launch. The much higher flux loss, despite the lower overall amount, can be explained by higher scatter losses from an amorphous ice phase with larger surface roughness.

\subsubsection{Interpretation: two different ice films were present\label{sec:iceinterpretation}}
After the -- very first -- decontamination right after launch, mirror temperatures came down from 270\,K, and the pressure in the gas phase from surface desorption (water accumulated during storage on ground) and outgassing was likely high, implying a high deposition rate. These conditions favour the initial formation of crystalline ice films at already 180\,K or 170\,K, and epitaxy then favours further crystalline growth even at lower temperatures \citepalias{Schirmer-EP29}. This was the ice film that likely grew until 2024-03-14. 

After the decontamination on 2024-03-14, after nine months in vacuum, surface deposits from before launch have evaporated or settled into the thermal sublimation-condensation and outgassing equilibrium in the \ac{PLM}'s instrument cavity. The gas-phase pressure and deposition rate must have been considerably slower than during the first cool-down after launch. An important detail is that \ac{FoM3} cooled down from a lower temperature of 160\,K on 2024-03-14, compared to 270\,K after launch decontamination was over. If the first molecular layers of ice were completed when \ac{FoM3} reached below 140\,K, then the water would have formed increasingly restrained amorphous ice with considerably higher surface roughness and thus higher optical scattering. We estimated 23--46\,nm of ice after 86 days (\cref{sec:icefilmthickness}), that is a mean deposition rate of 0.7--1.4 molecular layers per day (higher at the beginning, and exhausted near the end). \ac{FoM3} cooled from 160\,K to 140\,K (130\,K) in 47\,min (167\,min). Even at a generously estimated contamination rate of 0.4\,\AA\,h$^{-1}$ (projecting the thickness estimate in \cref{sec:icefilmthickness} to March 2024), only a 10--20\% complete monolayer of water molecules could have formed during the cool down before \ac{FoM3} passed below 130\,K, close to its operating temperature. The cool-down was quick and reached temperatures that favour the formation of restrained amorphous ice rather than crystalline ice; at that time, the estimated outgassing rate could have formed only an incomplete monolayer of water molecules, insufficient for a crystalline epitaxy layer that could have seeded further growth of crystalline ice. At the operational temperature of 129\,K the older, deeper layers anneal on timescales of a month \citepalias[with great uncertainties, see the discussion and figure 8 in][]{Schirmer-EP29}. The freshly deposited top layers must have still been amorphous, a few tens of nanometers thick, preventing epitaxial growth and continuously amplifying the roughness of the underlying material. 

This also explains the strong differential response observed during the heavy contamination phase. Short-wavelength photons in the blue peak are more susceptible to surface scattering on ice than those in the red peak. This scatter loss can be approximated as a power-law function of wavelength, with an index that depends strongly on the ice film's surface characteristics, which currently cannot be predicted \citepalias{Schirmer-EP29}. By comparing the observed ghost strength to the expected strength for the uncontaminated state, as a function of $G_{\rm BP}-G_{\rm RP}$, the power-law index can be determined for various contamination levels -- and from it, the spectral change of the \IE throughput curve.

\begin{figure*}[t]
\centering
\includegraphics[angle=0,width=0.8\hsize]{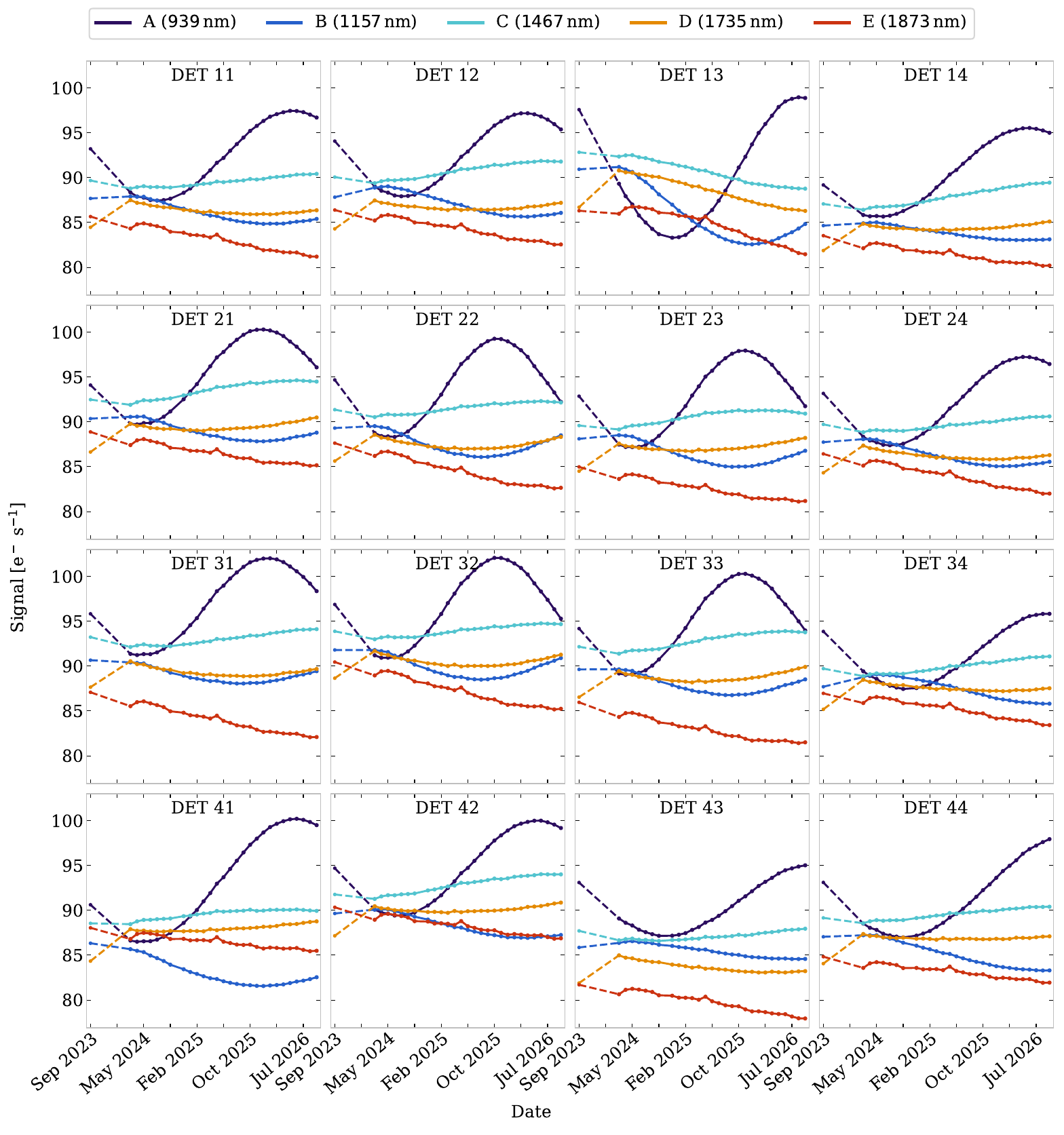}
\caption{Mean flux per NISP detector in the monthly flat fields. The evolution is due to accumulating ice on the detectors' surfaces, leading to optical interference with the \ac{AR} coating stack. Both the temporal change and the amplitude decrease with wavelength.}
\label{fig:nisp_led_response}
\end{figure*}

\subsection{Contamination of NISP detectors\label{sec:DR1_NISP_contamination}} 

The NISP calibration unit \citep[NI-CU,][]{EuclidSkyNISPCU} illuminates the \ac{FPA} directly without intervening optics, using one of five \acp{LED} with near-monochromatic emission from 939\,nm (LED A) to 1873\,nm (LED E). The measured fluxes in the monthly flats have changed continuously (\cref{fig:nisp_led_response}). The emitted fluxes have been stable to within $10^{-3}$ since launch, as judged from driver monitoring; the detector electronics are healthy; and the heavy throughput loss experienced by VIS between March and June 2024 was not detected by NISP. 

A thermodynamic model of the exchange of water molecules between surfaces in the enclosed NISP environment predicts an ice-film growth of 9.9\,nm\,month$^{-1}$ on the detectors \citepalias[table 2 in][]{Schirmer-EP29}. To judge the effects of optical interference, we modelled the \ac{AR} coating stack of detector DET13 that shows the strongest response in \cref{fig:nisp_led_response}. With thin ice films growing at a constant rate included, this model predicts a sinusoidal shape, as well as a higher amplitude and a faster evolution at shorter wavelengths, all of which are seen in all 16 detectors. For DET13 and \acp{LED} A and B, the model predicts fringe minima that are 15.4\% and 9.7\% lower than the maxima, respectively, compared to the measured 15.8\% and 9.4\%. A single fringe -- or $2\,\pi$ phase difference -- is predicted to require 387\,nm (LED A) and 488\,nm (LED B) of ice. \acp{LED} A and B needed 19 and 24 months to complete half a fringe each, implying a deposition rate of 10.2\,nm\,month$^{-1}$ for both (193.5/19, and 244/24). Accordingly, around 378\,nm of ice have accumulated on DET13 by November 2026.

The NISP detectors are located in an enclosed environment, and therefore the water molecules are difficult to evacuate. The continuously evolving phase differences imply that there is currently no change of the outgassing rate. Therefore, a decontamination would clean the detectors for a short period of time only, while considerably disturbing the ongoing survey. Optical models for each detector are in development, and will be used to compute their effective \ac{QE} at any time in the survey to correct for contamination. 

\begin{figure}[t]
\centering
\includegraphics[angle=0,width=1.0\hsize]{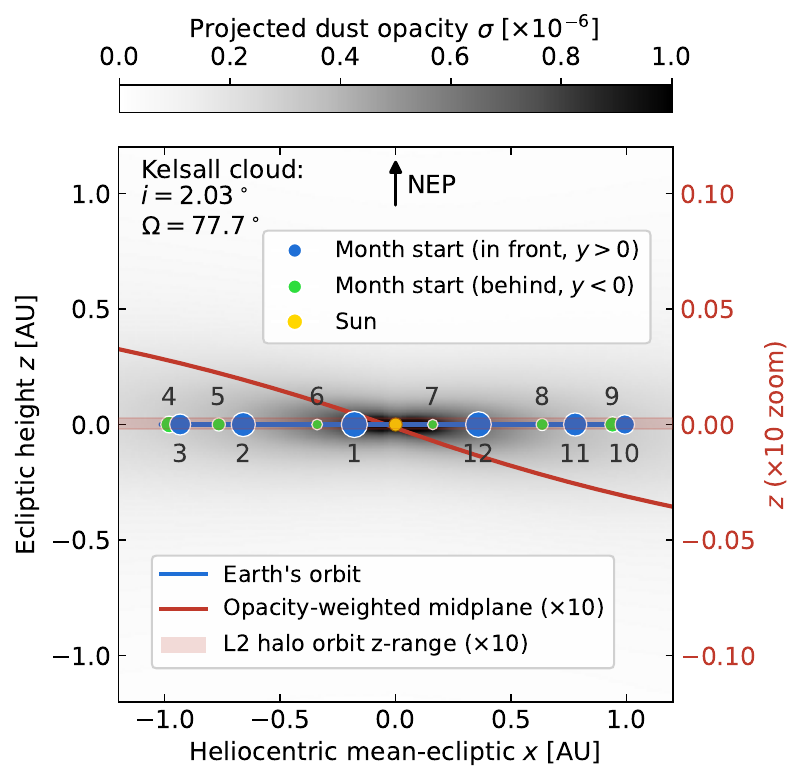}
\caption{Opacity-weighted mid-plane (red line) of the \cite{Kelsall1998} dust cloud with respect to the ecliptic plane (blue line). Its ecliptic height $z$ is amplified by a factor of ten for better visibility of the inclination, $i=2.03^\circ$, whereas the $x$- and $z$-axes are shown to scale. The grey shading shows the dust opacity projected over $y\in[-2,2]$\,AU along the viewer's direction. $\Omega=77.7^\circ$ is the longitude of the ascending node localising the intersection with the ecliptic plane. If $\Omega=90^\circ$, then the mid-plane would be seen fully edge-on.}
\label{fig:earth_orbit_vs_dust_plane_2dproj}
\end{figure}

\section{Zodiacal background\label{sec:zodi}}
Measurements of the zodiacal background light, \ac{ZB}, constrain the 3D dust distribution in our Solar System; accurate zodiacal models are required to remove this foreground emission, for example to constrain of the diffuse \ac{EBL} emitted by faint and unresolved galaxies in the young Universe \citep[see e.g.][]{carleton2022,postman2024,OBrien2026}. The ecliptic plane is inclined by $i=2.03^\circ$ with respect to the mid-plane of the dust distribution (\cref{fig:earth_orbit_vs_dust_plane_2dproj}), and therefore the \ac{ZB} is time-variable for Earth-bound observers. Here, we report high-precision results from \Euclid's self-calibration observations and discuss their accuracy.

\subsection{Measurements\label{sec:zodimeasure}}
The self-calibration field (ecliptic coordinates $\lambda=248.60^\circ, \beta=+88.6^\circ$) near the \ac{NEP} is visited monthly, taking data with all instrument configurations at 76 dither positions \citep[section 5 in][]{DR1-TP026}. The area has some Galactic cirrus because of its lower Galactic latitude ($b=30.3^\circ$), and therefore we used data from only one dither position to maximise background stability. Perfect stability, however, was not achievable because the position angle changes by around
\ang{30;;} every month due to orbit progression.\footnote{For a refined, future analysis we will use a circular area inscribed into the \ac{FoV}.} The background measurement themselves are carried out on the fully calibrated (LE2) single exposures, using iterative outlier rejection. Measurements are done on a per-detector basis, averaged per exposure, and converted to a surface brightness using the known photometric zero points.

\subsection{Background contributors\label{sec:backgroundcontributors}}
The measurements are precise, yet their accuracy is limited by systematic components that must be subtracted. Their contributions are summarised in \cref{tab:skybackground}. 

The VIS instrumental background comprises scattered sunlight (\cref{sec:straylight}) and negligible dark current, estimated from master darks taken at similar spacecraft attitudes. The stray-light patterns are similar but not identical between open-shutter and closed-shutter images. More accurate, attitude-dependent models can eventually be obtained from the large number of survey exposures. The instrumental background for NISP is caused by the charge-persistence signal \citep{Kubik24} from preceding on-sky exposures. A rolling average was determined from all slew darks, using a top-hat filter with a 30-day extent; slew darks are taken every 74 minutes when slewing to the next field, with the same exposure time as the on-sky observations.

\begin{figure}[t]
    \centering
    \includegraphics[angle=0,width=1\hsize]{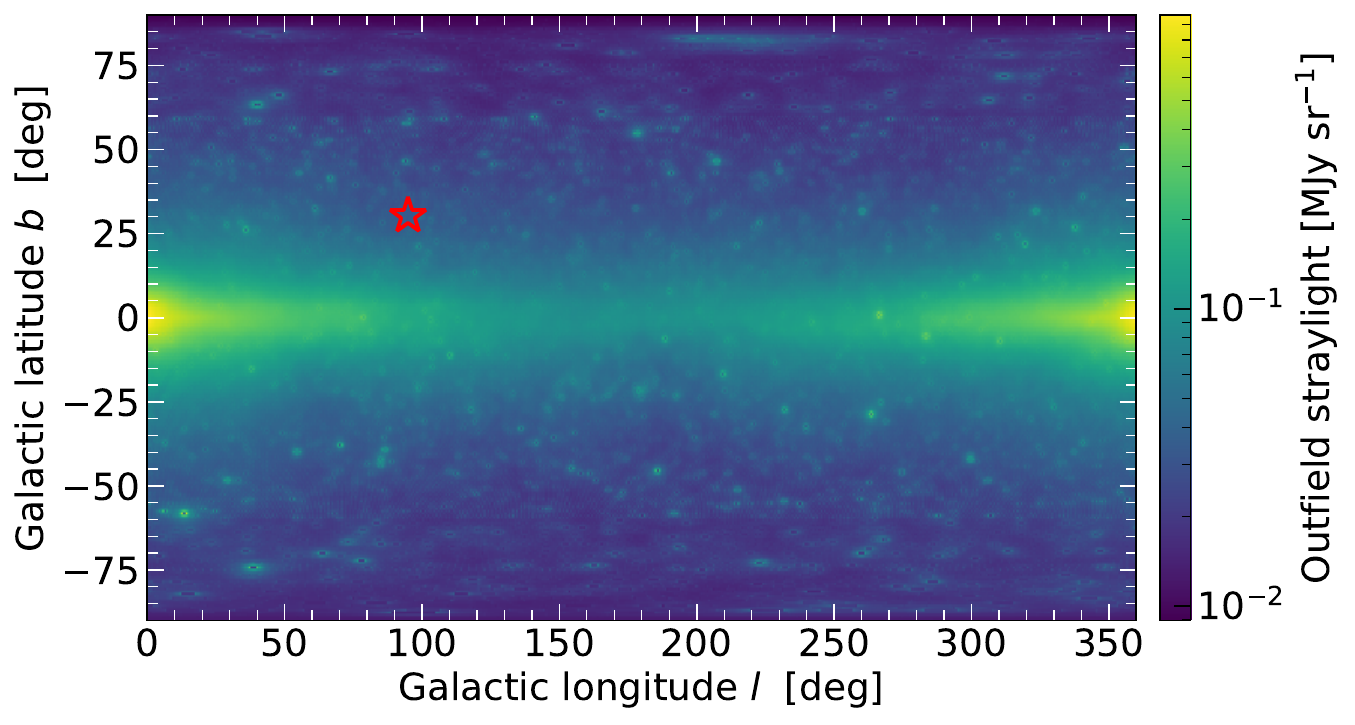}
    \caption{Galactic contribution to the out-of-field stray-light map (OSL) computed for VIS with a very conservative \ac{NDI} model in 2015. The map shows the scattered light received by the detectors when pointing at a position $(l,b)$, integrated over all sources outside the \ac{FoV} within the hemispheric solid angle centred on $(l,b)$. The self-calibration field is shown by the red asterisk. We find that this map overestimates the stellar stray light by a factor $27$ at the self-calibration position (\cref{sec:zodiresults}), and note that it does not include the time-variable contribution from the zodiacal background.}
    \label{fig:outfieldstraylight}
\end{figure}

\Ac{ISL} is caused by sources inside the \ac{FoV}, whose light scatters off optical surfaces; it is spatially variable. \Ac{OSL} is the cumulative flux from all sources outside the \ac{FoV}, illuminating the telescope's input pupil and scattering to the detector plane for angles $\theta\lesssim90^\circ$ between the pointing axis and the source. Dominant contributors are the Galaxy, bright stars that enhance the background locally for $\theta\lesssim2^\circ$, and the \ac{ZB} itself. The resulting \ac{NDI} was conservatively estimated by \cite{venancio2016} to compute \ac{ISL} and \ac{OSL} maps. End-to-end measurements \citep{venancio2023} on the \ac{PLM} show that the true \ac{NDI} is lower by up to an order of magnitude, in particular for $\theta\gtrsim1^\circ$. \cite{EROData} demonstrate that the broad \ac{PSF} is nearly diffraction-limited, implying a very low \ac{ISL}. We consider that, in first order, the \ac{ISL} is constant, because: (i) we used a fixed dither position (apart from field rotation); and (ii) determined the background from the full \ac{FoV} with a small number of bright stars and few, compact optical ghosts \citep[see \cref{sec:icedetection_dichroicghost}, and][]{EuclidSkyNISP}. The \ac{OSL} is constant across the \ac{FoV} at a given time, but varies from field to field. It was computed in 2015 using the conservative \ac{NDI} on a grid with a resolution of $0.7^\circ$ -- the linear \Euclid \ac{FoV} -- across the entire sky. The zodiacal emission was not included in this calculation, which introduces a yet unquantified time variability in \Euclid's \ac{OSL}; we come back to that in \cref{sec:zodiresults}. The resulting map for VIS is shown in \cref{fig:outfieldstraylight}. For the self-calibration field, the \ac{OSL} is dominated by nearby Galactic emission at $\theta\gtrsim10^\circ$, and should be considerably lower than the conservative \ac{NDI} estimates. An updated \ac{OSL} map is not yet available. For the purposes here, we consider the combined \ac{ISL} and \ac{OSL} as an unknown, constant contribution (\cref{sec:zodiresults}).

The \ac{DGL} is caused by scattering of starlight off interstellar dust along the line of sight. \cite{OBrien2026} computed the ratio of the \ac{DGL} to the 100-\micron\ dust emission as a function of wavelength. For the \HE and \JE bands, we extended the ratio beyond $1.54$\,\micron\ using a power law with the local slope at the last data point. The 100-\micron\ emission was taken from the \cite{Schlegel1998} maps and averaged over $1\,{\rm deg}^2$ at the self-calibration position. Integrating the interpolated ratio over the four \Euclid passbands yields a \ac{DGL} surface brightness of $0.006$\,MJy\,sr$^{-1}$. Using the same method, \cite{carleton2022} find 0.001--0.002\,MJy\,sr$^{-1}$ for the darkest areas with high ecliptic and Galactic latitude, consistent with 0.0022\,MJy\,sr$^{-1}$ (\IE) and 0.0011\,MJy\,sr$^{-1}$ (\JE) estimated in \cite{Scaramella-EP1} for typical survey areas. At $b=30.3^\circ$, the self-calibration field is notably closer to the Galactic plane, and the higher \ac{DGL} is not surprising.

Lastly, the \ac{EBL} is a constant contribution from faint sources below the detection threshold. It was estimated using deep galaxy counts in the \ac{EUDF}, covering 3.4\,deg$^2$ centred on the self-calibration field. The galaxy flux was integrated for all sources fainter than the $3\,\sigma$ point-source detection limits, that is $\IE=26.7$ and 24.9 for \YE, \JE, and \HE.
The depth of the \ac{EUDF} exceeds 28 magnitudes in all bands, and will be presented together with many other characteristics elsewhere. Our \ac{EBL} estimates (\cref{tab:skybackground}) are similar to those of \cite{carleton2022}, who report 0.0002--0.0003\,MJy\,sr$^{-1}$ in the 1.2--1.6\,\micron\ range, albeit with a fainter starting point of 26.5. An \ac{EBL} 10 times higher than our value can be considered as an upper bound \citep[see also][]{abeysekara2019,symons2023}. 

\begin{table}[ht]
\centering
\footnotesize
\caption{Background contributions in the self-calibration field [MJy\,sr$^{-1}]$.}
\begin{tabular}{lllll}
\hline
\hline\noalign{\vskip 1pt}
 & \IE & \YE & \JE & \HE \\
\hline\noalign{\vskip 1pt}
EBL & 0.00010 & 0.00032 & 0.00032 & 0.00036 \\
DGL & 0.0065 & 0.0064 & 0.0066 & 0.0070 \\
${\rm ISL}+{\rm OSL}$ (fit) & 0.0013 & 0.0027 & 0.0058 & 0.0115 \\
${\rm ISL}+{\rm OSL}$ (NDI) & 0.0356 & 0.0501 & 0.0431 & 0.0407 \\
$\langle{\rm BG}\rangle$ & 0.0062 & 0.0037 & 0.0030 & 0.0031 \\
\hline
\end{tabular}
\tablefoot{For ${\rm ISL}+{\rm OSL}$ we report both the values found by the model fits, and those predicted by the conservative \ac{NDI}.
$\langle {\rm BG}\rangle$ refers to the time-averaged instrumental background.}
\label{tab:skybackground}
\end{table}

\begin{table}[ht]
\centering
\caption{Best-fit parameters when fitting $\mathcal{M}^{\rm L2\;halo}$ to the data.}
\label{tab:offset_scale_summary}
\begin{tabular}{lcccc}
\hline
\hline\noalign{\vskip 1pt}
Band & $b$ & $s$ & Mean & RMS \\
 &  [MJy\,sr$^{-1}$] & & [MJy\,sr$^{-1}$] &  [MJy\,sr$^{-1}$]\\
\hline\noalign{\vskip 1pt}
\IE & 0.0013 & 1.205 & 0.0000 & 0.0009 \\
\YE & 0.0027 & 1.171 & 0.0000 & 0.0012 \\
\JE & 0.0058 & 1.174 & 0.0000 & 0.0009 \\
\HE & 0.0115 & 1.132 & 0.0000 & 0.0008 \\
\hline
\end{tabular}
\tablefoot{The fitted offset $b$ is treated as an additive systematic, subtracted from the data. The fitted scale $s$ is applied only to $\mathcal{M}^{\rm L2\;halo}$. Mean and RMS refer to the residuals after the fit. Details are provided in \cref{sec:zodiresults}.}
\end{table}

\begin{figure*}[t]
\centering
\includegraphics[angle=0,width=0.9\hsize]{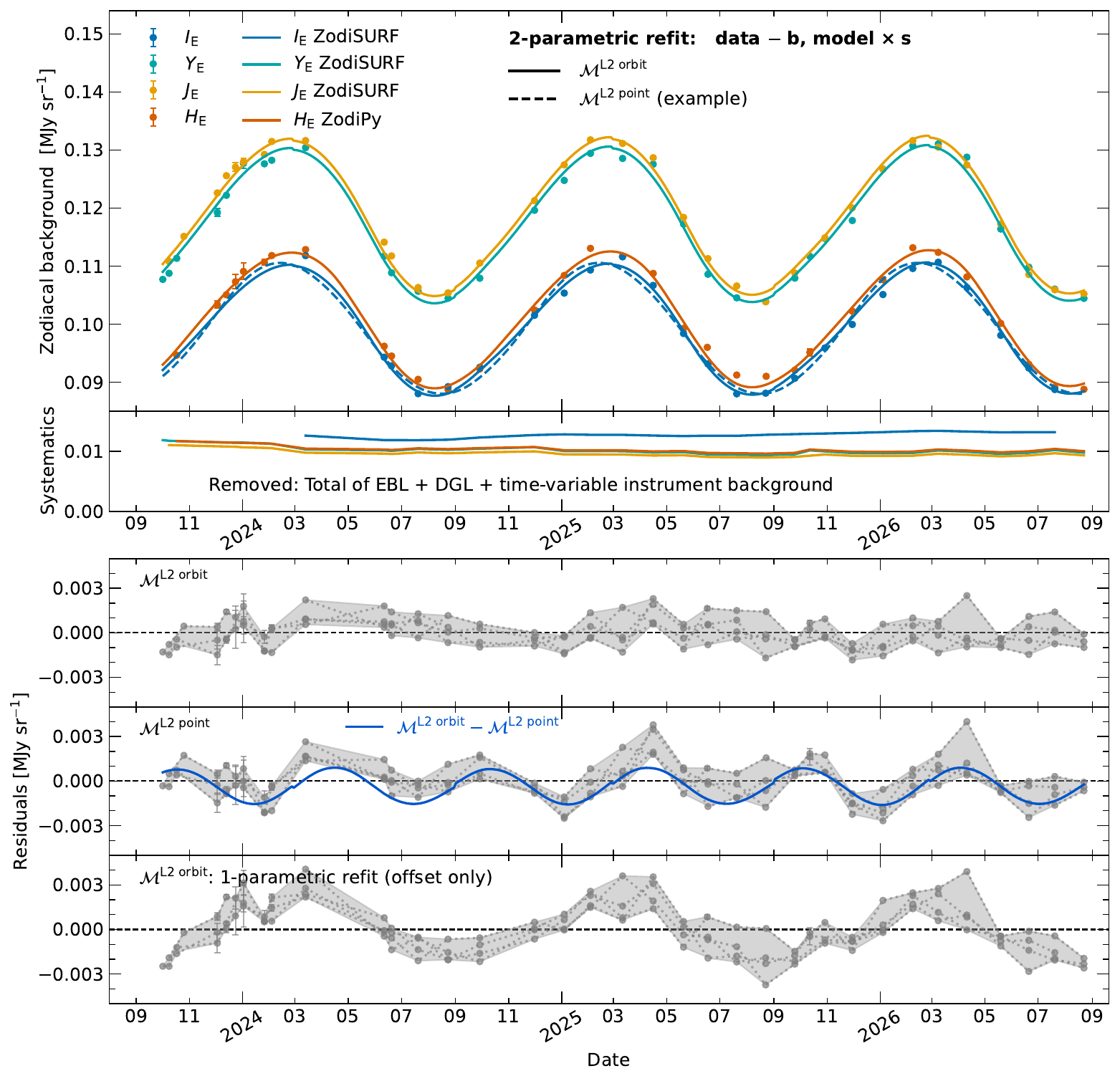}
\caption{Measured \ac{ZB} and modelling. \textit{Top}: Background in the self-calibration field, after subtraction of known systematics as well as a fitted offset $b$ accounting for the \ac{ISL} and \ac{OSL} stray-light components. Error bars are mostly smaller than the symbol size. The solid lines show the \ac{ZB} models evaluated for \Euclid's halo orbit, while the dashed line is evaluated for the L2 point. Models were upscaled by factors $s$ ranging from 1.13--1.20 to match the data (see below). \textit{Bottom}: Residuals with respect to three different refitted model runs. From top to bottom: $\mathcal{M}^{\rm L2\;orbit}$, shows largely uncorrelated residuals with zero mean, which are -- individually -- significantly non-zero because of variable \ac{ISL} due to field rotation; $\mathcal{M}^{\rm L2\;point}$ residuals reveal the halo-orbit libration signal that this simplified model fails to capture, detected at a total significance of $8\,\sigma$; at the bottom, we show $\mathcal{M}^{\rm L2\;orbit again}$ with fixed scaling turned off ($s=1$), revealing a time-variable component in the data that is not captured by the models. This is, probably and to a considerable extent, the \ac{ZB} contribution to \ac{OSL}. Details are discussed in \cref{sec:zodiresults}.}
\label{fig:zodi_timeseries}
\end{figure*}

\subsection{Models\label{sec:zodimodels}}
{\tt ZodiPy} \citep{San2024} implements the \cite{Kelsall1998} model of the interplanetary dust cloud, derived from the Diffuse Infrared Background Experiment \citep[DIRBE;][]{Hauser1998} on board the Cosmic Background Explorer. The model includes a smooth cloud as the dominant extended component, three pairs of asteroidal dust bands, a circumsolar ring, and an Earth-trailing blob. 

There are ten DIRBE photometric bands over 1.25--240\,\micron. The Kelsall model uses one discrete set of coefficients per DIRBE band, for the albedo, the phase function (how much sunlight is scattered toward the observer as a function of scattering angle), and the solar flux \citep[see also][]{OBrien2026}. {\tt ZodiPy}'s passband integration assigns each wavelength the coefficients of the nearest DIRBE band. This is problematic when predicting the \HE background, since a switch to a different set of coefficients occurs at 1.725\,\micron, the mid-point between the 1.25 and 2.2\,\micron\ DIRBE bands and within \HE band. By comparing against a reference of single-wavelength evaluations across the \HE band, we find that this mixing of coefficients underestimates the true \HE band-averaged brightness by 15\%--18\%. We avoid this by splitting the \HE band at 1.725\,\micron\ and evaluating each side separately, recovering the reference value to $0.2$\%.

The \IE, \YE, and \JE passbands fall below or near the 1.25-\micron\ lower wavelength limit of the DIRBE model. We estimate the \ac{ZB} in these bands with {\tt ZodiSURF} \citep{OBrien2026}, which keeps the same dust-cloud geometry but replaces the albedo and phase function with analytic forms fit directly to more than 5000 \ac{HST} sky surface-brightness measurements spanning 0.3--1.6\,\micron.

\subsection{Including the L2 halo orbit geometry\label{sec:zodi_orbit}}
\Euclid's geocentric Cartesian position is known in an inertial, equatorial frame aligned with the International Celestial Reference System, ICRS (see also \cref{sec:orbit}). We added this to Earth's own heliocentric position (from its standard ephemeris) and rotated into ecliptic coordinates, yielding \Euclid's  heliocentric position at any epoch.

Both \texttt{ZodiPy} and \texttt{ZodiSURF} estimate the background for an observer in orbit around L2, by evaluating the background as seen from the L2 point (\cref{fig:euclid_orbit}). While {\tt ZodiPy}'s \texttt{evaluate()} interface accepts an arbitrary heliocentric observer position, \texttt{ZodiSURF} hard-codes the observer into the ecliptic plane internally. We there patched \texttt{ZodiSURF}'s line-of-sight geometry to accept an explicit 3D position; \texttt{ZodiSURF} was recently updated to include this possibility.

\subsection{Results\label{sec:zodiresults}}
We subtracted the \ac{EBL}, \ac{DGL}, and instrumental backgrounds from the data, ignoring the two stray-light components due to the large uncertainty in the \ac{NDI}. We compared the data $\mathcal{D}$ with the models $\mathcal{M}$ evaluated at the respective times, by minimising the unweighted ordinary least-squares,
\begin{equation}
    \left(\hat{b},\ \hat{s}\right) \;=\; \arg\min_{b,\, s} \sum_{i} \Big(\mathcal{D}_i - b - s\, \mathcal{M}_i\Big)^2\,.
    \label{eq:refit}
\end{equation}
Here, $b$ is a constant offset to the data, accounting for the unknown stray-light components, and $s$ is a scale parameter applicable to the model. We use $\mathcal{M}^{\rm L2\;halo}$ and $\mathcal{M}^{\rm L2\;point}$ to distinguish between the models evaluated for the halo orbit around L2, and evaluated at the L2 point, respectively.

The results are shown in the top panel of \cref{fig:zodi_timeseries}. The solid lines show $\mathcal{M}^{\rm L2\;halo}$; the single dashed line shows $\mathcal{M}^{\rm L2\;point}$ for one band, as an example. The difference between the two is small, but it enhances an asymmetry in the light curves that is already present because of Earth's elliptical orbit (steeper decline, shallower rise). Accounting for the time-variable instrumental backgrounds considerably reduced the residuals.

The best-fit parameters are given in \cref{tab:offset_scale_summary}. The offset $b$, interpreted as the combined ${\rm ISL}+{\rm OSL}$, is lower than predicted by the \ac{NDI} model by factors of 27, 19, 7, and 4 for \IE, \YE, \JE, and \HE, respectively (\cref{tab:skybackground}). Under the presumption that the \cite{Kelsall1998} dust model is complete, our observations place strong upper limits on the \ac{NDI} that are consistent with the end-to-end measurements by \cite{venancio2023}. The considerable wavelength-dependence could be inherent to the \ac{NDI} model, which may overestimate the short-wavelength scattering of the blackened non-optical surfaces. The scaling factors $s$ range from 1.13--1.20, which we discuss below.

The bottom panel of \cref{fig:zodi_timeseries} shows the residuals for three different runs. For $\mathcal{M}^{\rm L2\;halo}$, the residuals have zero mean, but they are still significantly different from zero given their very small statistical errors. We attribute this to field rotation between visits, which changes the \ac{ISL} because of the changing Galactic scene (stars and cirrus) in the corners of the \ac{FoV}. VIS data earlier than March 2024 are not shown, because they were already affected by ice on \ac{FoM3} (\cref{fig:iceplot_smoothed}). 

For $\mathcal{M}^{\rm L2\;point}$, the \ac{RMS} of the residuals is 24\% higher than for $\mathcal{M}^{\rm L2\;halo}$. The halo-orbit libration with respect to the dust distribution is detected with a total significance of $8\,\sigma$, and well described by the difference $\mathcal{M}^{\rm L2\;halo}-\mathcal{M}^{\rm L2\;point}$ (blue line in \cref{fig:zodi_timeseries}). 

For $\mathcal{M}^{\rm L2\;halo}$ without rescaling (fixed $s=1$), the residuals follow the libration of the ecliptic plane with respect to the dust cloud, and $\mathcal{M}^{\rm L2\;halo}$ produces 80--87\% of the observed orbit-related signal amplitude. One reason for this is that we let the refitting determine the \ac{OSL} as part of the constant offset $b$. However, the \ac{OSL} is not constant over time, because the time-variable \ac{ZB} also contributes to it; this shortcoming is also evident from \cref{fig:outfieldstraylight}, which does not show the ecliptic emission band. We expect that an updated \ac{OSL} map will resolve most of this discrepancy; the 4.5\% uncertainty in the absolute flux scale of the {\tt ZodiSURF} cannot account for this. Another possibility is that the shape and extent of the components in the \cite{Kelsall1998} dust model are inaccurate. There could also be another extended dust component present in the Solar System that is not described by this model. \cite{OBrien2026} observe an unexplained diffuse signal of $(0.013\pm0.006)$\,MJy\,sr$^{-1}$ and suggest the presence of another dust component. In this case, not all of our offset $b$ should have been subtracted from the \Euclid data; part of it should instead have been added to the model. \ac{OSL} was not considered by \cite{OBrien2026} as an instrumental background effect, but could nevertheless be relevant because of \ac{HST}'s low Earth orbit \citep[O'Brien, priv.comm.; and][]{caddy2022}.

The \Euclid\ zodiacal background model \citep{maris2019} will be improved using the DR1 data set, which has much larger sky coverage, and the results presented here. The current development follows a semi-empirical approach that combines time-dependent all-sky templates with an elongation- and wavelength-dependent characterisation of the zodiacal-light spectrum relative to the sunlight spectrum.

\begin{figure}[t]
    \centering
    \includegraphics[angle=0,width=1\hsize]{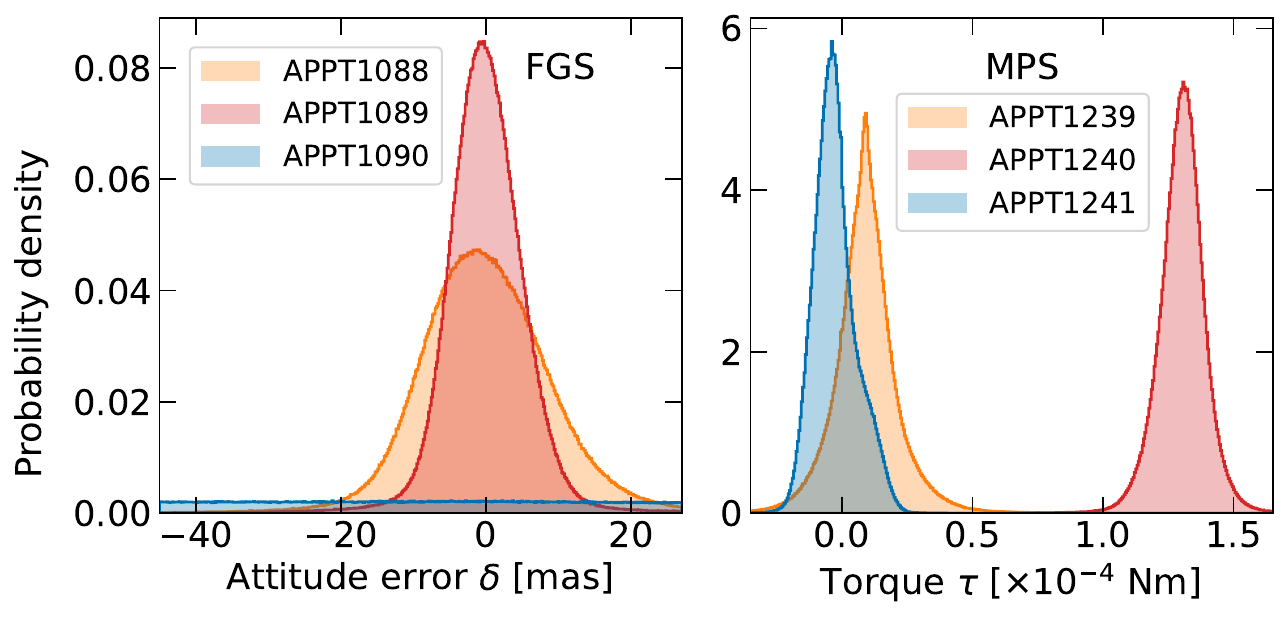}
    \caption{\Acp{RPE} and compensation torques; the colours identify the matching error-torque pairs (\cref{tab:fgs_mps_pairing}). \textit{Left}: The orange and red histograms represent the critical tip-tilt of the telescope, and the blue histogram the much less relevant rotation around the telescope's bore sight. \textit{Right}: Torques applied by the \ac{MPS} cold-gas thrusters. Their means are non-zero, accounting for non-gravitational forces. More details are provided in \cref{sec:guiding_review}.}
    \label{fig:apt_histograms}
\end{figure}

\section{Attitude stabilisation during exposures\label{sec:guiding_review}}

Actuations that disturb spacecraft stability, such as antenna and wheel movements, are prohibited during an exposure, with the exception of the \ac{RSU}, whose shutter blade takes 2.74\,s to open and close 
(\cref{sec:vis_shutter_disturbance}).
In the absence of atmospheric drag around L2, \Euclid still experiences continuous, non-gravitational accelerations from \ac{SRP} and \ac{TER}, which we discuss in this section and which degrade the \ac{PSF} if uncompensated. Molecular outgassing (\cref{sec:orbitalperturbations}) also contributes, but is not considered further here. Additionally, sporadic accelerations are caused by meteoroid impacts (\cref{sec:impacts}).

\begin{table}[h]
\centering
\caption{Pairing of \ac{RPE} and torques (telemetry parameter names).}
\begin{tabular}{ll}
\hline
\hline\noalign{\vskip 1pt}
\ac{FGS}: \ac{RPE} $\bm{\delta}$ & \ac{MPS}: torques $\bm{\tau}$ \\
\hline\noalign{\vskip 1pt}
APPT1088 (AX\_ER, $\delta_X$) & APPT1239 (MPS\_TX, $\tau_X$) \\
APPT1089 (AY\_ER, $\delta_Y$) & APPT1240 (MPS\_TY, $\tau_Y$) \\
APPT1090 (AZ\_ER, $\delta_Z$) & APPT1241 (MPS\_TZ, $\tau_Z$) \\
\hline
\end{tabular}
\label{tab:fgs_mps_pairing}
\end{table}

\subsection{Guide-loop control\label{sec:guideloop_principles}}
During science observations, the \ac{AOCS} control loop continuously calculates the torque $\bm{\tau}$ required on the spacecraft axes (\cref{fig:PLM_cropped}) to drive the \ac{RPE} $\bm{\delta}$ towards zero. The \ac{FGS} measures $\bm{\delta}$ from the guide stars on its detectors. The quantities $\bm{\delta}$ and $\bm{\tau}$ are expressed in the spacecraft coordinate system and are parallel, that is, a positive $\delta_i$ is compensated by a positive $\tau_i$. \Cref{tab:fgs_mps_pairing} gives the pairing of the corresponding telemetry parameters.

\begin{table*}[ht]
\centering
\caption{Properties of \acp{RPE} $\delta_i$ [mas] and torques $\tau_i$ [$\times10^{-4}$\,Nm] in the guiding data set comprising 2611\,h (\cref{sec:FGSMPS_prerequisites}).}
\footnotesize
\begin{tabular}{lccccc}
\hline
\hline\noalign{\vskip 1pt}
\ac{RPE} & $\langle x \rangle$ & $\sigma$ & 67\% & 95\% & 99.7\% \\
\hline\noalign{\vskip 1pt}
APPT1088 $(\delta_X)$ & $-0.1011(50)$ & $8.40$ & $[-8.14,\,8.25]$ & $[-17.3,\,19.9]$ & $[-42.3,\,47.5]$ \\
APPT1089 $(\delta_Y)$ & $0.0028(28)$ & $4.72$ & $[-4.57,\,4.71]$ & $[-10.3,\,10.8]$ & $[-27.7,\,29.7]$ \\
APPT1090 $(\delta_Z)$ & $-1.93(11)$ & $193$ & $[-177,\,183]$ & $[-293,\,317]$ & $[-414,\,471]$ \\
\hline\noalign{\vskip 1pt}
Torque & & & & &\\
\hline\noalign{\vskip 1pt}
APPT1239 $(\tau_X)$ & $0.090005(58)$ & $0.0947$ & $[-0.00727,\,0.189]$ & $[-0.160,\,0.361]$ & $[-0.477,\,0.7]$ \\
APPT1240 $(\tau_Y)$ & $1.303462(47)$ & $0.0772$ & $[1.23,\,1.38]$ & $[1.12,\,1.48]$ & $[0.904,\,1.69]$ \\
APPT1241 $(\tau_Z)$ & $-0.026811(45)$ & $0.0744$ & $[-0.101,\,0.0545]$ & $[-0.167,\,0.157]$ & $[-0.269,\,0.244]$  \\
\hline
\end{tabular}
\tablefoot{$\langle x \rangle$ is the mean after clipping to $|x - \mathrm{median}(x)| \le 3\,\sigma$, with the parentheses denoting the $1\,\sigma$ uncertainty in the last digits of the preceding value. $\sigma$ is the robust standard deviation, computed as $1.485\times\mathrm{MAD}$. The percentile containment intervals are centred on the median. Outliers beyond three times the 95\% half-width from the median were excluded, since they otherwise dominate the 99.7\% interval without affecting the rest.}
\label{tab:appt_stats}
\end{table*}

Attitude correction is achieved using \ac{MPS} cold-gas thrusters \citep{marie2019} as the finest-resolution actuators, after each target slew and once \ac{FGS} tracking has been acquired. The control loop also estimates \ac{SRP}, both at the entry into science observations and during them. The estimate is used as a feed-forward correction by initialising the controller with the previous estimate after each slew, allowing the \ac{RPE} to converge more quickly. A re-initialisation of the controller occurs, for example, if \ac{RPE} exceeds a threshold that is considered a loss of \ac{FGS} tracking (see also \cref{sec:impacts}). This triggers a science-mode abort signal (SC\_ABORT), the \ac{RPE} is set to zero, and star-tracker guiding continues at the new offset position until a new position is acquired; the instruments then gracefully finish their current exposures.

\begin{figure}[t]
    \centering
    \includegraphics[angle=0,width=1\hsize]{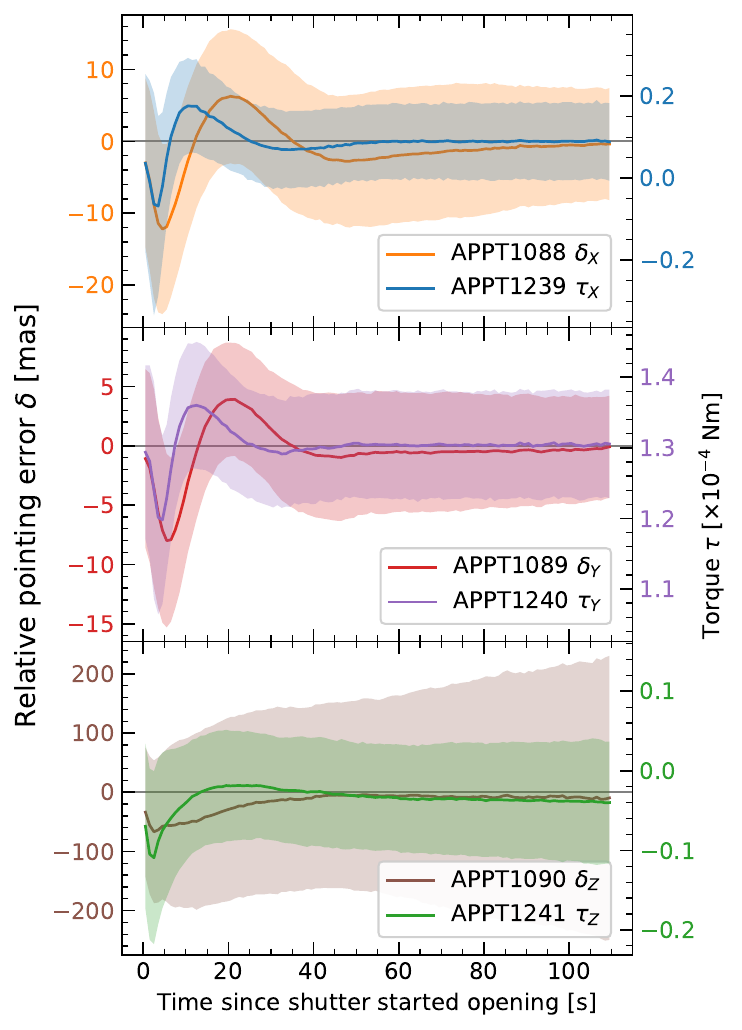}
    \caption{Effect of VIS shutter motion on \ac{RPE}. The disturbance causes a pronounced dip in the first 5--7\,s, followed by a longer overcorrection. The maximum angular deviation is 12\,mas (upper panel). The behaviour resembles that of an damped oscillator that stabilises after 30--40\,s. More details are given in \cref{sec:vis_shutter_disturbance}.}
    \label{fig:vis_shutter_effect}
\end{figure}

\begin{figure}[t]
    \centering
    \includegraphics[angle=0,width=1\hsize]{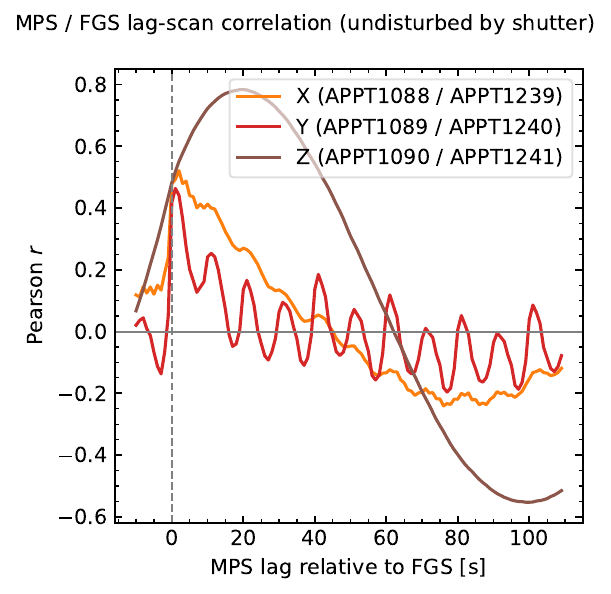}
    \caption{Correlation between \ac{FGS} and \ac{MPS} readings after decay of the shutter disturbance (\cref{fig:vis_shutter_effect}). The \ac{MPS} follows the \ac{FGS} by 1\,s, which is the time resolution of the data. The $X$- and $Y$-components are resonant on a scale of 10\,s. Details are given in \cref{sec:vis_shutter_disturbance}.}
    \label{fig:fgs_mps_correlation}
\end{figure}

\subsection{Tracking quality\label{sec:FGSMPS_prerequisites}}
We extracted 3075 periods of one hour duration between August 2023 and August 2026, about 2.6 per day at a random hour, with a time resolution of one second. From each period, we selected the undisturbed intervals during which science exposures were taken, for a total of 2611\,h, implying 84\% of on-sky time. We refer to this as the `guiding data set', with a mean of 9.4 million data points per sensor (\cref{tab:fgs_mps_pairing}), of which 9.1 million remained after quality masking (noisy guiding periods early in the mission, radiation storms, etc.). The distributions of $\delta_i$ and $\tau_i$ are shown in \cref{fig:apt_histograms} and quantified in \cref{tab:appt_stats}. The \ac{FGS} 99.7\% containment intervals are well within their $3\,\sigma$ allocations \citep[75\,mas for $\delta_X$ and $\delta_Y$; 1500\,mas for $\delta_Z$; see figure 9 in][]{EuclidSkyOverview}.

\subsection{VIS shutter disturbance\label{sec:vis_shutter_disturbance}}
VIS employs a rotating blade shutter \citep{EuclidSkyVIS} with opening and closing times of 2.74\,s. This mechanical disturbance is present during all VIS exposures. To measure its amplitude, we identified the beginning of 2372 nominal (565\,s) VIS exposures\footnote{Telemetry parameter ${\rm VMGT3069}=3$ (opening) and 6 (closing).} and stacked the \ac{FGS} and \ac{MPS} data in the guiding data set. 

The data for the first 100\,s are displayed in \cref{fig:vis_shutter_effect},
showing considerable thrusting during the first 4\,s to control the \ac{RPE}. This puts the closed loop into a temporary, damped oscillatory state where the \ac{MPS} leads the \ac{FGS} by approximately 12\,s (separation between the positive peaks) in the $X$- and $Y$-axes. The $\delta_X$, $\delta_Y$, and $\delta_Z$ components reach a quasi-stable state after approximately 90\,s, 60\,s, and 40\,s, respectively. The pure lag between the \ac{FGS} and the \ac{MPS} without the shutter disturbance is shown in \cref{fig:fgs_mps_correlation}, for which we computed their stacked correlation functions after masking the first 60\,s of all nominal VIS exposures. In this very stable regime, the \ac{MPS} follows the \ac{FGS} by 1\,s in the $X$- and $Y$-axes, the expected behaviour given our time resolution. As a result of the guide-loop control (\cref{sec:guideloop_principles}), a resonance with a period of 10\,s can be seen for the tilt in the anti-Sun direction (rotation around the $Y$ axis), for which radiative effects are most pronounced, and more weakly also for the $X$-axis; these are of no practical relevance.

The shutter disturbance acts on a scale of a few milliarcseconds, or 0.1\,VIS pixel, and is well within the required limits. The full \ac{FGS} time series, as analysed here globally, is available for every exposure to support the \ac{PSF} modelling (Miller et al., in prep.).

\subsection{Radiation pressure and thermal-emission recoil\label{sec:torques}}
Any asymmetry in the mass distribution relative to the spacecraft's geometric centre will turn a uniform \ac{SRP} into a non-zero torque. The same holds for \ac{TER}, which has been identified by \cite{Turyshev2012} as the leading explanation for the Pioneer anomaly.

The torque distributions are shown in the right panel of \cref{fig:apt_histograms}, and their mean and standard deviation are given in \cref{tab:appt_stats}. The largest effect is found for $\tau_Y$, which prevents the bottom-heavy spacecraft from being tipped in the anti-sun direction by both radiative effects. The average 
$\langle\tau_X\rangle>0$ is detected with ${\rm S/N}=1550$, compensating the \ac{TER} that attempts to rotate \Euclid clockwise as seen from the Sun; we note that this force is caused by the thermal emission on \Euclid's unilluminated side. The average 
$\langle\tau_Z\rangle<0$ counters the \ac{SRP} and \ac{TER} that tend to rotate the spacecraft anti-clockwise when looking into the telescope aperture; this is detected with ${\rm S/N}=590$.

To compare \ac{SRP} and \ac{TER}, we consider a greybody with a flat surface. The \ac{SRP} at normal incidence (\crefp{eq:radiationpressure}) is $p_{\rm abs}\approx5.0\times10^{-6}\,{\rm N\,m}^{-2}$. The \ac{TER} for a flat surface with a Lambertian emission profile is given by \cite{Toth2009} as
\begin{equation}
    p_{\rm emit} = \frac{2}{3}\,\epsilon\,\frac{\sigma_{\rm B}\,T^4}{c}\,.
    \label{eq:thermalpressure}
\end{equation}
The numeric prefactor arises from integrating the Lambertian profile over the hemisphere above the surface, and $\epsilon\approx0.9$ is the typical thermal emissivity. For $T=260$\,K (300\,K, 380\,K) we obtain
\begin{equation}
    \frac{p_{\rm abs}}{p_{\rm emit}} \approx 9.5\,\quad(5.3, 2.0)\,.
    \label{eq:pressureratio}
\end{equation}
Depending on heliocentric distance and attitude, the sunshield has a temperature between 361\,K and 385\,K \citep{SpaceTech}. At 380\,K, \ac{TER} contributes 33\% of the force compensated by $\tau_Y$. 
\Euclid's cooler dark sides -- permanently shadowed by the sunshield -- also experience \ac{TER}, as evidenced by $\tau_X>0$. For example, the VIS electronics radiator has an area of 0.95\,m$^2$ and a temperature of 257\,K \citep{EuclidSkyVIS}. And the \ac{SVM}, which is thermally insulated from the \ac{PLM}, is kept at an operating temperature of approximately 290\,K. 

\subsection{Torque effects from heliocentric distance variation\label{sec:fgs_mps_heliocentric}}
\Ac{SRP} and \ac{TER} depend on heliocentric distance, $d_{\rm h}(t)$, well detectable for $\tau_Y$ and $\tau_Z$. A smaller effect is also found for $\tau_X$, even though the corresponding surfaces are not illuminated by the Sun: A power $P_{\rm leak}$, transferred from the insulated sunshield to the \ac{SVM} and \ac{PLM}, is modulated by $d_{\rm h}(t)$ and changes -- for example -- the baseplate temperature by 31\,mK (\cref{fig:T170_timeseries}). The passive radiators compensate efficiently because of the $T^4$ dependence of radiated power, detectable in $\tau_X$.

\subsubsection{Measurements and modelling\label{sec:mam}}
To maximise sensitivity, we filtered the guiding data set by rejecting the first 60\,s after every VIS shutter actuation (\cref{sec:vis_shutter_disturbance}). This removed 4.8\% of the data points. We corrected $\tau_{Y,Z}$ for the sunshield's viewing factor (see also \cref{fig:temperature_defocus_attitude}) using
\begin{equation}
    \tau_{Y,Z}^{\,\rm corr} = \frac{\tau_{Y,Z}}{\sin({\rm{SAA}})\,\cos({{\rm AA}})}\,,
    \label{eq:geometriccorrection}
\end{equation}
where the correction factor $1/[\sin({\rm SAA})\cos({\rm AA})]$ is in the range $[0.932,0.999]$ with a mean of 0.9935. The quantities 
$\tau_X$ and $\tau_Z$ remain strongly correlated with \ac{AA}, and weakly with \ac{SAA}; we removed both trends using linear fits. The attitude-corrected time series are shown in \cref{fig:radiationpressuremodel}, displaying periodicity on a timescale of one year for all components. To decide whether this is indeed caused by $d_{\rm h}(t)$, we need physical models of the joint \ac{SRP} and \ac{TER} for $\tau_Y$ and $\tau_{Z}$, and, for $\tau_X$, an effective \ac{TER} model for the warm surfaces. We derive these models in \cref{sec:derivation_Y} and \cref{sec:derivation_X}, finding that they have the same inverse-distance dependence despite different physical motivation:
\begin{equation}
  \tilde\tau_i = A_i\,\left[\frac{1\,{\rm AU}}{d(t)}\right]^{2} +\, B_i\, +\, C_i\,(t-t_{\rm ref})\quad[\times10^{-4}\,{\rm N\,m}]\,.
  \label{eq:fitting_tau_XYZ}
\end{equation}
Molecular outgassing, after linearisation of a joint outgassing and sublimation model, the same inverse-square dependence on distance arises as in \cref{eq:fitting_tau_XYZ}, so that outgassing and radiative effects cannot be distinguished to first order. The pressure of solar energetic particles can also be neglected, as it is 2--3 orders of magnitude below \ac{SRP} in non-storming conditions around solar maximum. The linear drift term $C_i\,(t-t_{\rm ref})$, where time is in units of years and $t_{\rm ref}$ is the mean epoch of the observations, was introduced to account for a long-term evolution evident in the data. The torque was then fitted to \cref{eq:fitting_tau_XYZ} by joint ordinary least squares, where we fixed $B_{Y,Z}=0$ because this constant term is absent in the sunshield's model (\cref{sec:derivation_Y}). Since $d_{\rm h}(t)$ is taken directly from the spacecraft's ephemeris and is not itself a fit parameter, the regression can only rescale -- and for $\tau_{X}$ also offset -- the predicted curves vertically, but not shift them along the time axis. Therefore, a phase shift between model and data would invalidate the model and our hypotheses. Fit parameters are listed in \cref{tab:radiationpressure_fitparams}.

\begin{table}
\centering
\caption{Parameters for the torque fit in \cref{eq:fitting_tau_XYZ}.}
\begin{tabular}{lccc}
\hline
\hline\noalign{\vskip 1pt}
Torque & $A$ & $B$ & $C$\,[yr$^{-1}$]\\
\hline\noalign{\vskip 1pt}
$\tau_X^{\,\rm corr}$ & $0.1181(43)$ & $-0.0246(42)$ & $-5.5(1.3) \times 10^{-4}$\\
$\tau_Y^{\,\rm corr}$ & $1.33995(20)$ & $\equiv 0$ & $0.01084(24)$\\
$\tau_Z^{\,\rm corr}$ & $-0.03714(12)$ & $\equiv 0$ & $8.3(1.5) \times 10^{-4}$\\
\noalign{\vskip 1pt}
\hline
\end{tabular}
\label{tab:radiationpressure_fitparams}
\end{table}

\subsubsection{Results\label{sec:timeseries_fittingresults}}
An inverse-square dependence on heliocentric distance is evident for all three components, detected even for the two weaker ones, with ${\rm S/N}=26$ ($\tau_X^{\rm corr}$) and 15 ($\tau_Z^{\rm corr}$). A linear drift in time is found for $\tau_Y^{\rm corr}$ and $\tau_Z^{\rm corr}$, with respective ${\rm S/N}=36$ and 19; both components are governed by \ac{SRP} and \ac{TER} acting on the sunshield. 

A physical cause could be radiation-induced darkening of the protective coverglasses of the solar cells, that is, the formation of colour centres in the substrate \citep{Messenger2018,Aihara2026}. While the total absorptivity $\alpha_{\rm tot}=\alpha_{\rm glass} + \alpha_{\rm cell}$, reflectance $\rho$, and therefore also \ac{SRP} (\crefp{eq:radiationpressure}) remain constant, the generated electric power decreases because a fraction of the photons is converted to heat in the coverglass. In this scenario, \ac{TER} would increase and be compensated for by an increase of $\langle\tau_Y\rangle$ by 0.81\%\,yr$^{-1}$ (\cref{tab:appt_stats,tab:radiationpressure_fitparams}). For a sunshield temperature of 380\,K, \ac{TER} contributes 33\% of the total radiative force, and would thus have to increase by $0.81\%/0.33 = 2.4\%$ per year to account for the observed drift.

\begin{figure*}[t]
    \centering
    \includegraphics[angle=0,width=0.45\hsize]{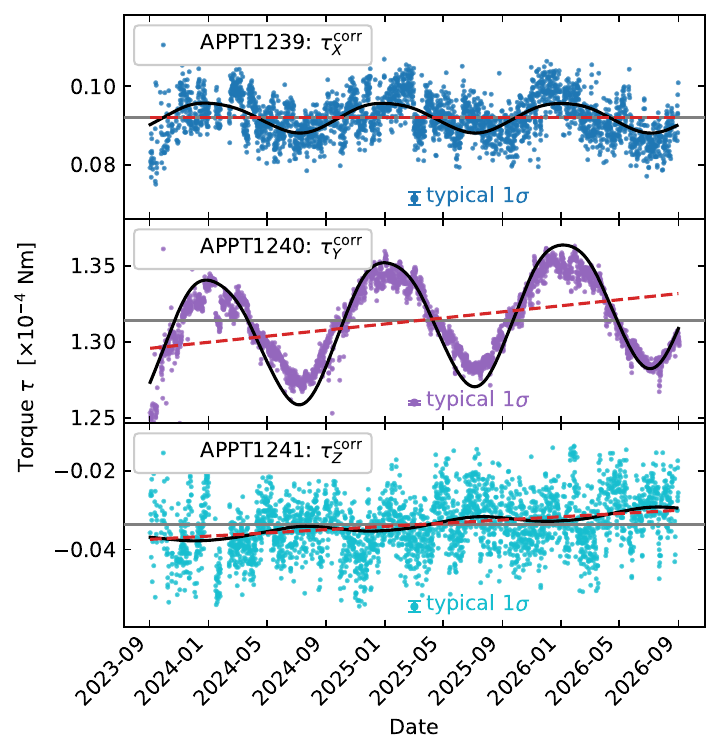}
    \includegraphics[angle=0,width=0.45\hsize]{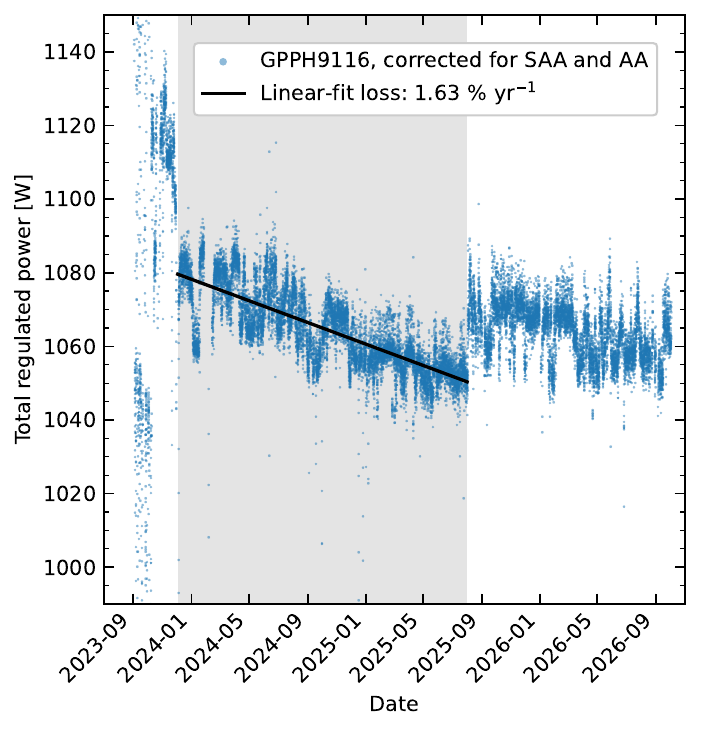}
    \caption{\textit{Left}: Time series of the attitude-corrected torques compensating non-gravitational accelerations. The black lines show fits of joint \ac{SRP}--\ac{TER} models accounting for heliocentric-distance variation. The dashed red lines are a linear fit that shows a systematic increase of $\tau_{Y,Z}$, attributed to radiation-darkening of the solar array (\cref{sec:timeseries_fittingresults}). All model and fit parameters are given in \cref{tab:radiationpressure_fitparams}. \textit{Right}: Regulated power provided to the spacecraft after shunting, stabilising after the end of \ac{PV} in November 2023. The decrease of regulated power is attributed to radiation-darkening of the solar-cell coverglass (\cref{sec:timeseries_fittingresults}). The higher absorption in the coverglass increases the sunshield's temperature and \ac{TER}, compensated by higher torques (red dashed lines in the left panel).}
    \label{fig:radiationpressuremodel}
\end{figure*}

A corresponding relative loss should then be seen in the electric power generated. The solar array was designed with margins, producing more power than needed under normal circumstances. The total power provided to the spacecraft (GPPH9116 telemetry parameter) is regulated by shunting: when production exceeds demand, the current from individual array sections is diverted back into those same sections, where it is dissipated as heat, while the remaining sections meet the demand. The regulated power is shown in the right panel of \cref{fig:radiationpressuremodel}, indicating long-term declines of 1.1--1.6\%\,yr$^{-1}$; the effects of active shunting are also evident, by periodic readjustments and -- above all -- the complete absence of a heliocentric-distance modulation. The only correction we applied was the geometric term of \cref{eq:geometriccorrection}, which reduced the scatter by about 20\%.

The decrease of 1.1\%--1.6\%\,yr$^{-1}$ in regulated power is below the 2.4\%\,yr$^{-1}$ increase in torque. Some of the difference is due to the calibration bias caused by the drift anomaly of the offset voltage in the mass-flow sensors, which was also observed in \textit{Gaia} and the Laser Interferometer Space Antenna (LISA) Pathfinder \citep{marie2019}. Other uncertainties arise because of: (i) the active shunting of array sections, which can bias the slope estimate; (ii) the simplistic modelling with ad-hoc estimates for $\alpha$; and (iii) uncertainties in the modelled sunshield temperature. The drift anomaly linked to $\tau_Y$ appears to have stabilised since January 2026, and further monitoring will better discriminate between physical and instrumental effects.

\begin{table}
\centering
\footnotesize
\caption{List of 34 high-confidence meteoroid impacts.}
\label{tab:mm_events}
\begin{tabular}{llrrr}
\hline
\hline\noalign{\vskip 1pt}
Impact (UTC) & Type & $S_{\rm VIS}$ & $S_{\rm NISP}$ & $t_{\rm opt}$ \\
 & & [ADU] & [ADU] & [s] \\
\hline\noalign{\vskip 1pt}
2023-11-25T05:49:26 & flash & \dots & 732 & $<1$ \\
2024-01-31T12:50:09 & flash & 1 & 217 & $<1$ \\
2024-03-26T16:48:34 & flash & 2.5 & 663 & $<1$ \\
2024-06-21T12:16:48$^{\star\star}$ & debris & 21 & 662 & 27 \\
2024-07-03T17:18:23$^{\star\star}$ & debris & 8 & 531 & 20 \\
2024-08-16T03:06:06 & debris & 85 & 1405 & 22 \\
2024-11-13T05:09:19 & flash & \dots & 270 & $<1$ \\
2024-12-05T17:06:23$^\star$ & dark & \dots & \dots & \dots \\
2024-12-12T10:05:52$^\star$ & dark & \dots & \dots & \dots \\
2025-04-04T10:31:39 & \dots & 1.8 & \dots & \dots \\
2025-06-09T21:12:16 &  dark & \dots & \dots & \dots \\
2025-07-16T10:20:19$^\star$ &  dark & \dots & \dots & \dots \\
2025-08-18T03:41:52 & \dots & 0.5 & \dots & \dots \\
2025-09-13T06:40:05 & \dots & 0.5 & \dots & \dots \\
2025-11-15T17:47:45$^{\star\star}$ & debris & 36 & 466 & 95 \\
2025-11-19T04:59:23 & debris & 0.8 & 427 & 30 \\
2025-11-24T17:04:13 & dark & \dots & \dots & \dots \\
2025-12-13T21:35:28 & flash & \dots & 27 & $<1$ \\
2025-12-21T03:53:02 & debris & 2.5 & 32 & 40 \\
2026-01-03T06:26:20 & debris & 1.7 & 344 & 40 \\
2026-01-03T23:28:01 & debris & 3 & 35 & 29 \\
2026-03-03T15:00:55 & debris & 8 & 642 & 22 \\
2026-03-06T14:34:30 & dark & \dots & \dots & \dots \\
2026-03-06T14:59:12 & \dots & 1 & \dots & \dots \\
2026-03-08T02:44:18 & debris & 18 & 81 & 30 \\
2026-03-16T08:28:59$^\star$ & dark & \dots & \dots & \dots \\
2026-04-05T03:49:43 & debris & \dots & 161 & 17 \\
2026-04-07T03:26:50 & dark & \dots & \dots & \dots \\
2026-05-09T22:04:13 & flash & \dots & 136 & $<1$ \\
2026-07-08T11:46:46$^\star$ & flash+debris & 22 & 8555 & 55 \\
2026-07-31T01:38:33$^{\star\star}$ & debris & 16 & 711 & 22 \\
2026-09-15T01:27:46 & flash & 5 & 570 & $<1$ \\
2026-09-17T17:34:10 & debris &  1.7 & 325 & 36 \\
2026-09-28T22:12:32$^\star$ & dark & \dots & \dots & \dots \\
\hline
\end{tabular}
\tablefoot{Single asterisks ($^\star$) denote particularly strong events that triggered SC\_ABORT mode; ($^{\star\star}$) denotes debris signals without \ac{FGS} detection. `Type' characterises the optical signal as seen in the NISP data (\cref{sec:impact_detections}). The quantity $S$ refers to the peak brightness measured. The duration $t_{\rm opt}$ of the optical counterpart is available only if NISP data were taken. Optical counterparts that were detected only in VIS do not have a type classification. An impact is `dark' if no optical counterpart is available, but the mechanical counterpart is unmistakable. Affected exposures are flagged as invalid and excluded from further processing.}
\end{table}

\section{Dust and meteoroid impacts\label{sec:impacts}}

\subsection{\textit{Gaia} and \textit{JWST} impact recordings\label{sec:gaia_jwst_impacts}}
\textit{Gaia} sensed meteoroid\footnote{The IAU discouraged the use of the term `micrometeoroid' in 2017. Dust particles are smaller than 30\,\micron, and meteoroids are larger.} hits at L2 via attitude-rate disturbances, with a mean rate of around $5$\,month$^{-1}$ \citep{Lindegren2018}. Most impacts left no lasting trace once the \ac{AOCS} absorbed the disturbance. Five particularly strong events are discussed in \cite{Altmann2026}, with notable attitude rate excursions. The strongest event occurred on 2 April 2024, causing a considerable stray-light leak that was mitigated by an onboard detection-threshold adjustment.

The \ac{JWST} detected 25 meteoroid-induced surface deformations between March 2022 and January 2023, a rate of $2.5$\,month$^{-1}$; all but one produced no measurable degradation in \ac{WFE} after correction \citep{Rigby2023}. The exception was an impact between 22 and 24 May 2022, for which the \ac{WFE} of the affected mirror segment could only be partially compensated \citep{McElwain2023,Gialluca2025}. This event exceeded prelaunch damage expectations, and \ac{JWST} operations have since reduced the fraction of time spent on viewing within a $75^\circ$-wide cone in the direction of orbital motion, where high-velocity meteoroids are more common \citep{jwsttech2026}.

\begin{figure*}[t]
    \includegraphics[width=0.50\hsize]{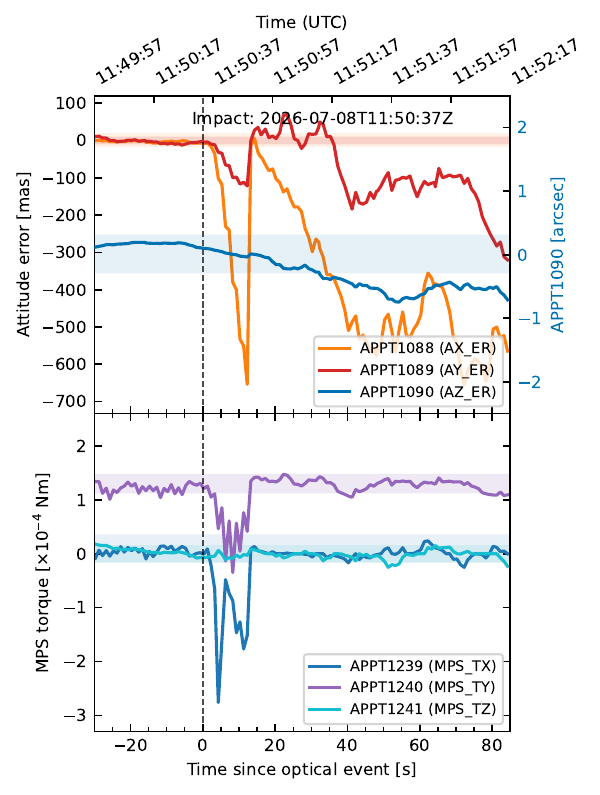}
    \includegraphics[width=0.50\hsize]{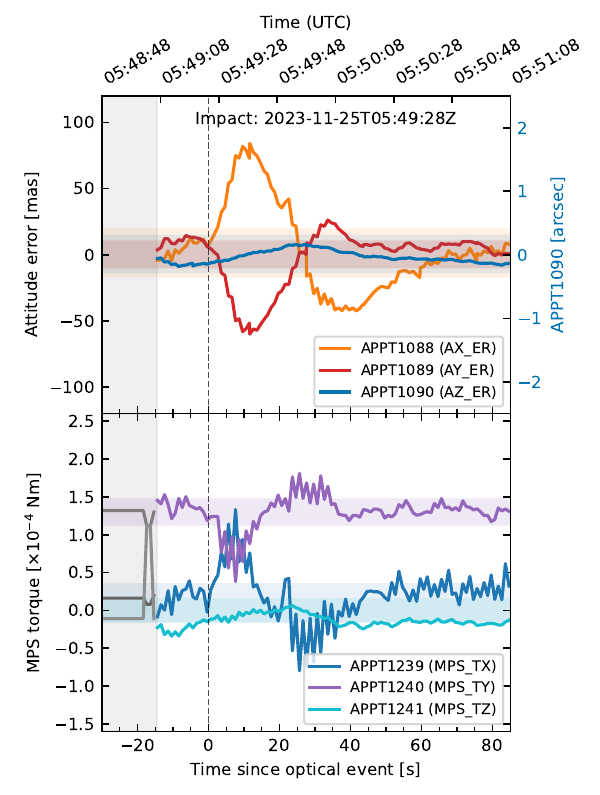}
    \caption{\textit{Left}: Spacecraft response to the strong impact on 2026-07-08 (\cref{sec:impact_major}). Twelve seconds after the event, $\delta_X$ (AX\_ER) exceeded the maximum allowed \ac{RPE} and triggered SC\_ABORT mode (\cref{sec:euclid_impact_detection}); the \ac{RPE} was reset to zero and star trackers guided at the offset position until the next acquisition. The horizontal shaded areas depict the central 95\% intervals of the respective parameters. \textit{Right}: A medium-strong impact. Stable tracking was recovered after 60\,s. The grey vertical area denotes the period before the exposure started.}
    \label{fig:euclid_appt1088_appt1089}
\end{figure*}

\subsection{Impact detection with \Euclid\label{sec:euclid_impact_detection}}
Our list of impacts is shown in \cref{tab:mm_events}. Events were selected based on mechanical and optical detection criteria as follows.

Impacts with sufficient angular-momentum transfer leave a characteristic, detectable signal in the closed \ac{FGS}-\ac{MPS} guide loop (see also \cref{sec:guiding_review}). For a given particle, this signal decreases as the impact parameter with respect to the spacecraft's centre-of-mass becomes smaller. Some impacts were large enough to exceed \ac{RPE} thresholds, opening the guide loop and causing an SC\_ABORT (\cref{sec:guideloop_principles}); we note that SC\_ABORT can have numerous other causes unrelated to impacts. A blind search in \ac{FGS} data for minor, unflagged impact events has not yet been performed, requiring filtering of the time-series data with suitable search patterns. The characteristic \ac{FGS} signature of impacts that open the guide loop are shown in the left panel of \cref{fig:euclid_appt1088_appt1089}, and those of less strong events in the right column.

Impacts also cause optical flashes. The microscopic particles considered here carry the same kinetic-energy-to-mass ratios of several $10^8$\,J\,kg$^{-1}$ as macroscopic impactors, far above the vaporisation and ionisation energy of hard-surface materials. Their plumes of hot and partially ionised material expand quickly and cool adiabatically, setting short timescales on their optical emission; see also \cite{Eichhorn1976}, \cite{Burchell1996ASR}, and \cite{Burchell1996Icarus}. Flashes can be detected if the impact occurs inside the telescope baffle, around the secondary mirror and its holding mechanism, or inside the instrument cavity if the impactor survives penetration of the \ac{MLI}. Longer optical signals are generated if ejected and sun-lit debris clouds drift across the viewing direction; this was demonstrated by a camera with near-field focus onboard the Midcourse Space Experiment \citep[MSX; see figure 9 in][]{uy2003}. Both types are initially flagged by screening the $\chi^2$ parameter of the on-board signal estimator \citep{kubik2016}, which computes the slope fit for the ramp. An example for an impact that caused both flash and debris is shown in \cref{fig:nisp_rawlines}.

The optical signals are generated in the near-field and are strongly defocused, typically illuminating the entire \ac{FPA}. \Euclid's telescope has an aperture $D=1.2$\,m and a focal length $f=12.5$\,m for NISP and $24.5$\,m for VIS. The respective diagonal \ac{FPA} diameters are $s=0.21$\,m and $0.42$\,m. For a point-like debris particle with the telescope focused at infinity, the thin-lens approximation gives the particle's distance $d$ at which its defocused image has diameter $s$ as
\begin{equation}
    d=\frac{D\,f}{s}\,.
    \label{eq:thinlens}
\end{equation}
Setting $s$ to the \ac{FPA} diagonal, the image becomes smaller than the \ac{FPA} for $d\gtrsim70$\,m. 

Flashes and debris clouds are easily detected in the NISP raw line (one per detector), for which the signal accumulating during the exposure is available for each frame \citep[details about the raw lines are given in][]{EU-Cogato,kubik2016}. The frame duration is 1.45048\,s, setting the precision by which the flash can be timed if it occurs within a group of frames; for an event between two frame groups the accuracy is limited to several seconds. If a flash occurs during frame readout, its light profile can be sampled with millisecond resolution (\cref{sec:flashduration}). Similar considerations apply for debris clouds that appear in front of the telescope, with measured delays ranging from 0--140\,s after the impact. Their optical decay is approximately exponential while they disperse in front of the telescope, with discontinuities in the light curve when bright particles enter or leave the viewing cone. 

In VIS, optical signals are diluted by the long exposure time accumulating the zodiacal background, and because the pixels have a 2.25 times smaller collecting area than those in NISP. The highest sensitivity is achieved by comparison with a preceding or subsequent exposure.

\subsection{Flash duration\label{sec:flashduration}}
\begin{figure}[t]
    \includegraphics[width=1.0\hsize]{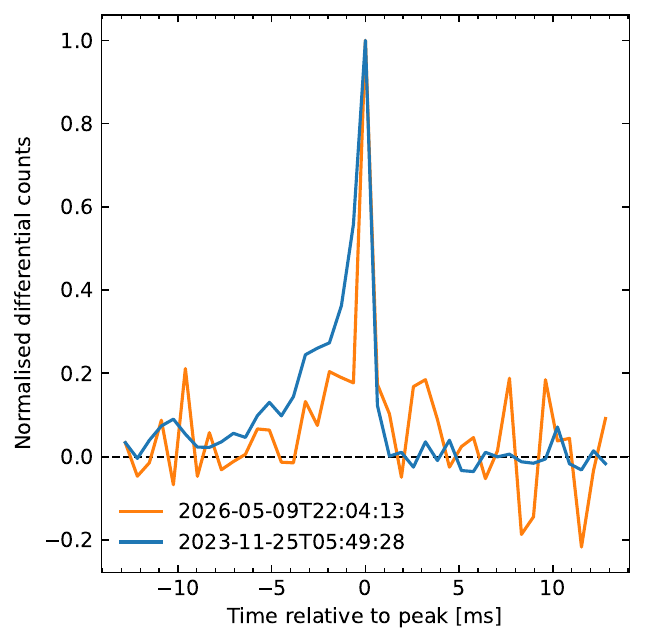}
    \caption{Light curves for two impacts that occurred during NISP frame readout, offering millisecond time resolution.}
    \label{fig:flashprofile}
\end{figure}

The impacts on 2023-11-25T05:49:28 and 2026-05-09T22:04:13 occurred during the NISP frame readout, causing a discontinuous step along the image columns. The time resolution along a column is the row readout time of 640\,\ms. For each of the 16 detectors, we collapsed the image into a single column: for every row, pixels deviating from the row median by more than $2\times1.486\,\mathrm{MAD}$ were rejected, and the mean of the remaining pixels was taken. The step signal in the average column was differentiated, yielding one light curve per detector. The curves from all detectors were aligned on their peaks using integer-row shifts, averaged, and normalised to unit peak height. The \ac{FWHM} was measured on this mean profile, using linear interpolation to locate the half-maximum crossings. The light curves are shown in \cref{fig:flashprofile}.

The 2023 event (also in the right panel of \cref{fig:euclid_appt1088_appt1089}) showed a rise starting 5--8\,ms before peak brightness, followed by an unresolved drop-off, and had an \ac{FWHM} of $(1.2\pm0.1)$\,ms. The 2026 event was 5.3 times fainter, with an \ac{FWHM} of $(0.8\pm0.1)$\,ms and a rise detectable about 3\,ms before peak brightness. These \ac{FWHM} values are comparable to the row sampling time of 640\,\ms\ and should be regarded as upper limits on the flash durations. For comparison, the flashes of lunar impactors with masses of several grams or more last from a few tens to about a hundred milliseconds \citep{liakos2024,yanagisawa2025}.

\subsection{The 2026-07-08 impact\label{sec:impact_major}}
Out of the six strongest events that triggered an SC\_ABORT (\cref{sec:guideloop_principles}), the impact on 2026-07-08T11:50:39 was the only one accompanied by an optical counterpart. NISP recorded the event 218\,s after the beginning of a spectroscopic exposure, between frame groups six and seven in the ramp. The flash was 10--100 times brighter than typical, and the \ac{RPE} limits were exceeded 12\,s later. \Cref{fig:euclid_appt1088_appt1089} compares the impact to a more typical one that did not open the guide loop. 

\begin{figure}[t]
    \includegraphics[width=1.0\hsize]{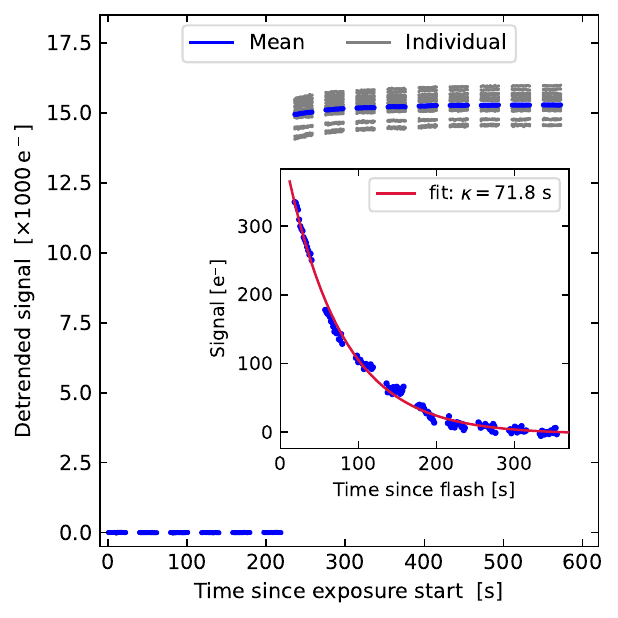}
    \caption{Flash and debris signal for the 2026-07-08 impact. Shown in grey are the detector raw values for every frame taken during the exposure, comprising 15 groups of 16 frames, with the background signal subtracted. Their mean is shown in blue. The impact occurred between groups six and seven at $t\approx218$\,s, and its flash added 15\,000\,e$^{-}$\,pixel$^{-1}$. The signal from the dissipating debris followed, decaying with a time constant of $\kappa\approx72$\,s and detectable until the end of the exposure (inset).}
    \label{fig:nisp_rawlines}
\end{figure}

To isolate the flash from the -- virtually negligible -- astrophysical signal that accumulates at a constant rate, we subtracted a linear fit to groups 1--6 from all data points. The flash deposited approximately 15\,000\,e$^{-}$\,pixel$^{-1}$ in the red-grism exposure. Assuming it lasted 1\,ms (\cref{sec:flashduration}), this would be a surface brightness of 5.2\,AB\,mag\,arcsec$^{-2}$, or equivalent to the flux of a $-11.9$\,AB\,mag star distributed evenly over the \ac{FPA}. The flash was followed immediately by the characteristic decay from the Sun-lit dissipating debris cloud, which was observed for almost 6\,min until the end of the exposure (\cref{fig:nisp_rawlines}).

\begin{figure*}[t]
    \centering
    \includegraphics[angle=0,width=1.0\hsize]{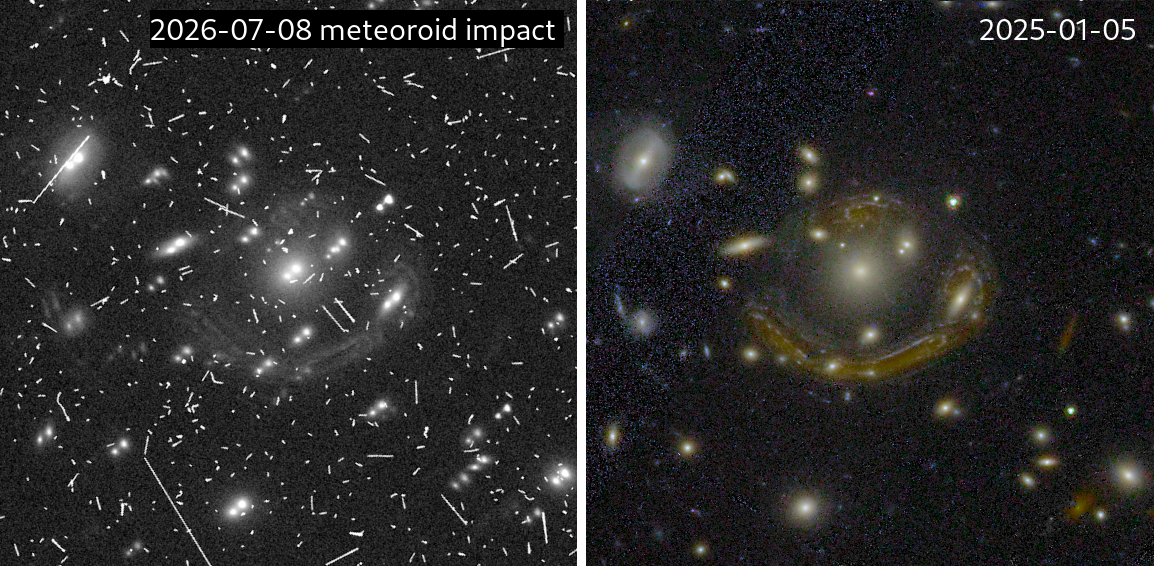}
    \caption{Strong meteoroid impact on 2026-07-08. \textit{Left}: The impact transferred enough angular momentum to offset the telescope pointing by \ang{;;1.3}, exceeding the limits for recovery by the \ac{FGS}. At the time of the impact, the Molten Ring strong-lens system was in the \ac{FoV}. \textit{Right}: The lens is in the overlapping area of two adjacent survey fields, and was observed previously on 2025-01-05.}
    \label{fig:moltenring}
\end{figure*}

The impact resulted in double sources in the images with an angular separation of \ang{;;1.3} (\cref{fig:moltenring}), caused by the impact itself and the zero-setting of the \ac{RPE} with continued star-tracker guiding at the new offset position. Coincidentally, the impact occurred during an exposure of an area overlapping with a previously observed survey patch, containing the Molten Ring strong-lens system \citep[see \cref{fig:moltenring}, and][]{diazsanchez2021}.

\subsection{Observed impact rate and completeness\label{sec:impact_detections}}
The impacts listed in \cref{tab:mm_events} were selected heterogeneously. 21 events have both a mechanical and an optical counterpart (flash or debris, or both), making then unambiguous. Smaller impact parameters with respect to the spacecraft's centre-of-mass mean less transferred angular-momentum, weakening the signal in the \ac{FGS} data. Without an \ac{FGS} signature, a strong impact is still unmistakeable if a debris cloud is detected, for which no other plausible explanation exists. On the other extreme, `dark impacts' have a strong, characteristic \ac{FGS} signature, but without an optical counterpart, either because of unfavourable impact locations, or because no image was taken at that moment. Outside science exposures, the signal must be large enough to not be masked by onboard mechanical activities. In total, nine dark impacts were detected (\cref{tab:mm_events}). 

Before June 2025, \ac{FGS} sensitivity to impacts was reduced by a sporadic vibration-noise anomaly of the VIS shutter. In the period that followed, we detected 24 high-confidence impacts or 1.5\,month$^{-1}$ until 2026-09-30. Based on the size and geometry of the spacecraft, telescope aperture, random particle trajectories and impact parameters, we estimate that the detected impacts comprise perhaps 10--20\% of all impacts that could, in principle, be detected under the right circumstances.

\begin{figure*}[t]
    \includegraphics[width=\linewidth]{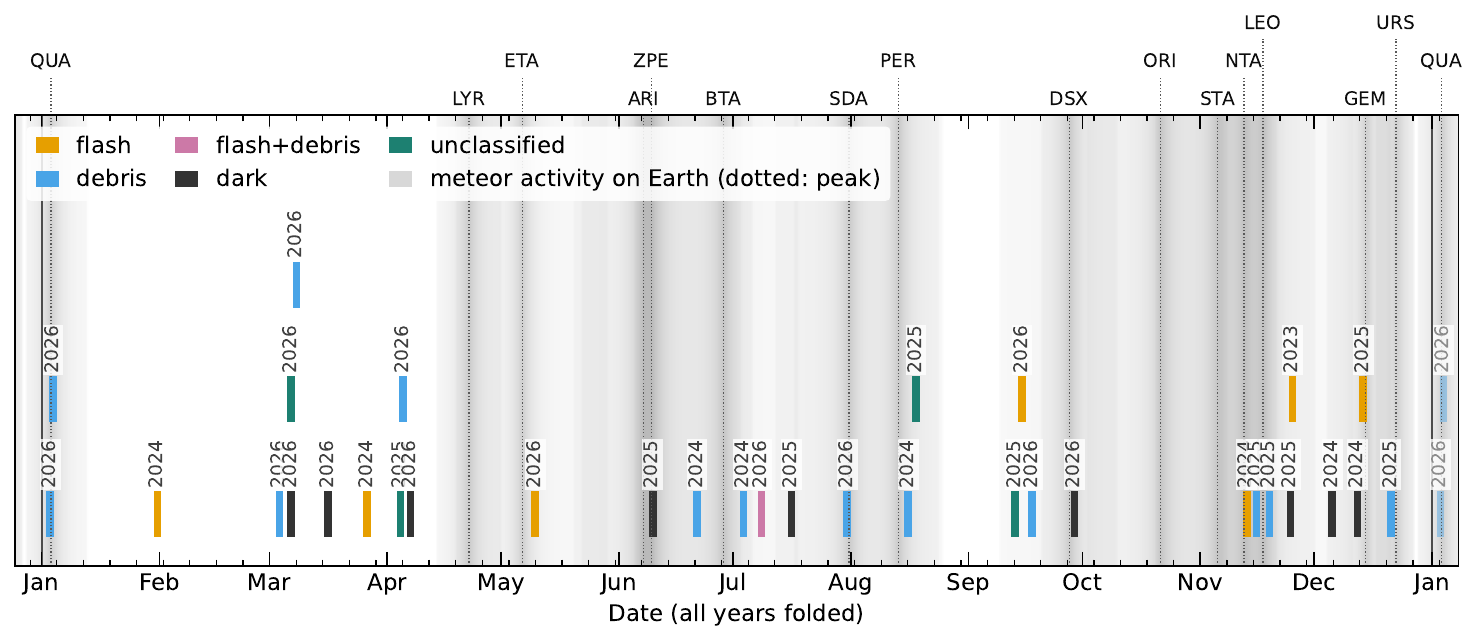}
    \caption{Phase-folded impact time series, with major meteor streams shown in grey. The impacts appear sporadic, with a likely origin in the interplanetary particle background field and not associated with any of the known streams.}
    \label{fig:impact_timeline_phase}
\end{figure*}

\subsection{Expected impact rate\label{sec:expected_impact_rate}}

\cite{Gruen1985} have built an empirical model of the interplanetary, cumulative, omnidirectional meteoroid flux at 1\,AU, hitting on one side of a randomly tumbling flat plate \citep{moorhead2020}. The analytical approximation of \cite{yahalomi2025} describes the flux above mass $m$ as
\begin{equation}
    F(>m) = 3.156 \times 10^{7}
    \left[ F_1(m) + F_2(m) + F_3(m) \right]\,{\rm m}^{-2}\,{\rm yr}^{-1}\,.
    \label{eq:yahalomi}
\end{equation}
The numeric prefactor is the number of seconds per year, the mass is measured in grams, and $F_1$ to $F_3$ represent three particle populations dominated by large ($m \gtrsim 10^{-9}$\,g), intermediate ($10^{-14}\,{\rm g} \lesssim m \lesssim 10^{-9}$\,g), and small ($m \lesssim 10^{-14}$\,g) particles, respectively,
\begin{align}
    F_1(m) &= \left(2.2\times10^{3}\,m^{0.306} + 15\right)^{-4.38}, \\
    F_2(m) &= 1.3\times10^{-9}
              \left(m + 10^{11}\,m^2 + 10^{27}\,m^4\right)^{-0.36}, \\
    F_3(m) &= 1.3\times10^{-16}
              \left(m + 10^{6}\,m^2\right)^{-0.85}.
\end{align}
To estimate particle diameters from particle masses, we used a bulk density of $\rho=2$\,g\,cm$^{-3}$ and spherical grains, noting that the true density can deviate considerably, depending on porosity and chemical composition.

We modelled \Euclid's shape as a cuboid with the sunshield as an attached, larger rectangle with area $A_{\rm sun}=4.8\,\mathrm{m} \times 3.1\,\mathrm{m}$ (shown yellow in \cref{fig:euclid3d_meteoroid_model}). This rectangle accounts for impacts from the Sun- and anti-Sun-facing directions. The \ac{PLM} has lateral faces (green) with  $A_{\rm lat}=2.8\,\mathrm{m} \times 4.0\,\mathrm{m}$ that -- when observing an ecliptic pole -- point into the leading and trailing direction of the orbital track; the anti-sun facing side of the \ac{PLM} is already accounted for by the sunshield rectangle. The top and bottom faces (blue) measure $A_{\rm top}=2.8\,\mathrm{m}\times2.8\,\mathrm{m}$. The total surface area of the model is 67.8\,m$^2$; shadowing effects are ignored. 

\begin{figure}[t]
    \includegraphics[width=\linewidth]{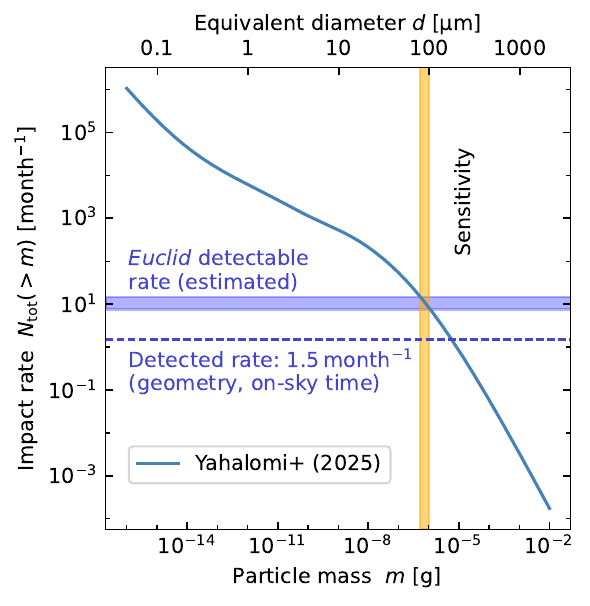}
    \caption{Estimated impact rate on \Euclid. The dashed line shows the observed rate, the horizontal shaded region shows what could be observed with 100\% on-sky time and if no geometric factors reduced the detectability. The intersection with the model flux then yields the minimum mass (vertical yellow area).}
    \label{fig:yahalomi2025_impacts}
\end{figure}

The expected cumulative impact rate per month,
\begin{equation}
    N_{\rm tot}(>m) = 12\,F(>m)\,\left[2\,(A_{\rm sun} + A_{\rm lat} + A_{\rm top})\right]\,,
\end{equation}
is shown in \cref{fig:yahalomi2025_impacts} and quantified in \cref{tab:yahalomi} for selected masses. 

\subsection{Comparing observed and expected impact rates\label{sec:meteoroid_rate_comparison}}
For $m \geq 10^{-9}$\,g ($d\gtrsim10$\,\micron), we expect 537 impacts per month across all six faces, but most would be below our detection sensitivity. The minimum mass that is consistent with our rate of 1.5\,month$^{-1}$ is $m\gtrsim5\times10^{-7}$\,g, corrected for the events missed because of the considerations discussed in \cref{sec:impact_detections}. The associated minimum diameter for a density of $\rho=2$\,g\,cm$^{-3}$ is 78\,\micron.

For macroscopic impactors with diameters of 1\,mm and above ($m \gtrsim 10^{-3}$\,g), the rate drops to $3.4\times10^{-3}$\,month$^{-1}$, or 0.25 over the nominal 6-year mission duration. These estimates apply to the interplanetary background at 1\,AU at L2, where neither Earth shielding nor gravitational focusing is relevant.

\subsection{Optical degradation of the primary mirror\label{sec:degradation}}
Continuous exposure to dust and meteoroid pitting will slowly degrade the performance of the primary mirror: cratering, cracking, and delamination of the coating layers increase the internal stray light, and -- if the particle breaches the reflective layer -- area damage occurs. Since the crater depth can be up to 3.5 times the impactor's diameter \citep{watts1993}, even sub-micron particles can breach thin optical coatings. The details of \Euclid's protected silver coating are proprietary; assuming a total coating thickness of 1\,\micron, a particle with a diameter of 280\,nm ($m=2.3\times10^{-14}$\,g) and a velocity of 20\,km\,s$^{-1}$ suffices to breach it.

Particles with $m\lesssim10^{-11}$\,g -- referred to as $\beta$-particles -- are dominated by \ac{SRP}. They mostly move in the anti-sun direction, are blocked by the sunshield and telescope baffle, and can be ignored. Therefore, all surface-degrading impactors on \Euclid's primary mirror breach the coating and increase area damage. The effects on optical performance are complicated \citep{watts1993}, and also relevant for \ac{JWST} and the future Habitable Worlds Observatory \citep[see e.g.][]{Heaney2010,Gialluca2025,shepherd2026}. 

The VIS end-to-end flux loss -- averaged over blue and red sources -- is approximately 0.23\,mmag\,month$^{-1}$ (\cref{fig:iceplot_smoothed}) since the last decontamination on 2026-06-08. It is the result of contamination and area damage with unknown fractions, and currently most likely due to contamination. At this rate, another $\approx9$\,mmag would be added until the end of the nominal mission, which is well below the budgeted pre-launch estimate of 5\% throughput loss.

\begin{figure}[t]
    \includegraphics[width=\linewidth]{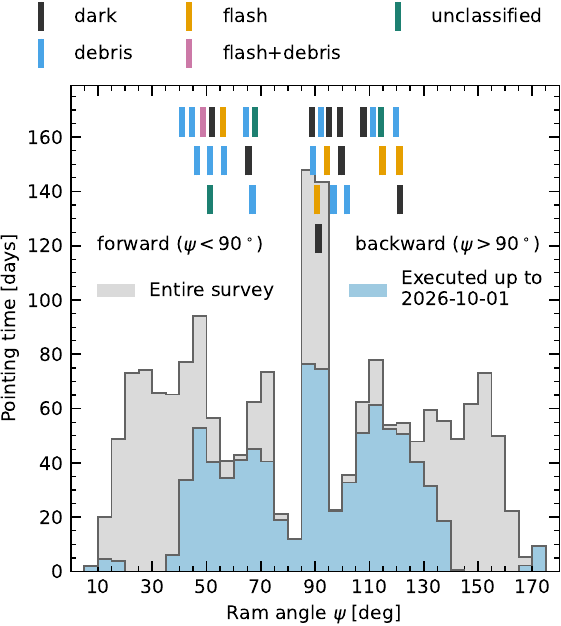}
    \caption{Ram angle between pointing direction and orbital motion. Meteoroid impacts do not show a directional preference. 12\% of the time is spent at $\psi\lesssim30^\circ$forward-looking into the particle stream, where a particle confined to the ecliptic plane can realise a high-velocity, direct impact on the primary mirror.}
    \label{fig:ram_angle_histogram}
\end{figure}

This could change, however, because initially \Euclid observed the best areas with low Galactic and zodiacal background, and now continues at increasingly lower ecliptic latitudes. We define the ram angle $\psi$ as the angle between the orbital motion vector and the telescope's pointing direction. The distribution of $\psi$ is shown in \cref{fig:ram_angle_histogram}, for the entire survey (grey) and the already completed area by 2026-10-01 (blue), reflecting this optimised survey strategy. The remainder of the nominal mission time is characterised by $\psi\lesssim50^\circ$, with higher chances for direct impacts on the primary mirror. Increasing the fraction of observations in the trailing or backward direction -- as done for \ac{JWST}, see \cref{sec:gaia_jwst_impacts} -- would reduce this risk. However, the survey is already highly constrained by numerous factors \citep{Scaramella-EP1,EuclidSkyOverview,DR1-TP027}, and such a severe restriction would not allow for its timely completion.

\begin{table*}[th]
\centering
\caption{Cumulative expected impact rates per month, for particles of mass $\geq m$, computed using \cref{eq:yahalomi} based on \cite{yahalomi2025}. }
\label{tab:yahalomi}
\begin{tabular}{llllll}
\hline
\hline\noalign{\vskip 1pt}
$m$ [g] & $d$ [\micron] &
Sunshield ($14.9\,\mathrm{m}^2$) &
Lateral ($11.2\,\mathrm{m}^2$) &
Top/bottom ($7.8\,\mathrm{m}^2$) &
Total \\
\hline\noalign{\vskip 1pt}
$10^{-15}$ & 0.10  & $4.2\times10^{4}$ & $3.1\times10^{4}$ & $2.2\times10^{4}$ & $1.9\times10^{5}$ \\
$10^{-12}$ & 0.98  & $1.3\times10^{3}$ & $1.0\times10^{3}$ & $7.1\times10^{2}$ & $6.2\times10^{3}$ \\
$10^{-9}$  & 9.9   & $118$             & $89$              & $62$              & $537$             \\
$10^{-7}$  & 46    & 12              &  8.9              &    6.2            &    54\\
$10^{-6}$  & 124   & $1.8$             & $1.4$             & $0.97$            & $8.4$             \\
$10^{-5}$  & 267   & $0.18$            & $0.14$            & 0.096  &    $0.83$\\
$10^{-4}$  & 576   & $1.3\times10^{-2}$  &   $9.7\times10^{-3}$   &  $6.8\times10^{-3}$  &   $5.9\times10^{-2}$\\
$10^{-3}$  & 1240  & $7.4\times10^{-4}$& $5.6\times10^{-4}$& $3.9\times10^{-4}$& $3.4\times10^{-3}$\\
$10^{0}$   & 15\,600 & $8.7\times10^{-8}$& $6.6\times10^{-8}$& $4.6\times10^{-8}$& $4.0\times10^{-7}$\\
\hline
\end{tabular}
\tablefoot{Particle diameters assume a bulk density of $\rho=2$\,g\,cm$^{-3}$. The columns with the areas correspond to the sunshield (sun/anti-sun), the lateral \ac{PLM} faces (along track), and the top (telescope aperture) and bottom faces. The last column gives the total over all six faces. See also \cref{fig:euclid3d_meteoroid_model}.}
\end{table*}

\section{Summary and conclusions\label{sec:conclusions}}
\Euclid has performed extremely well in the first three years of its flight, as evidenced by the numerous technical and scientific papers accompanying DR1. However, the time was not without challenges. 

The optical stray-light leaks in the thermal insulation were mitigated by reducing and shifting the allowed roll-angle range for calibration and survey observations, notably reducing the survey efficiency. Fortunately, the cosmological science is not affected by the remaining residual stray light, which is removed with nominal background subtraction \citep[section 2.2 in][]{DR1-TP002}. For low-surface brightness studies, dedicated background models successfully remove correct for remaining stray-light residuals \citep{EROData,Borlaff-EP16,EP-Urbano,EROPerseusICL}. The stray-light level increases by 14\% per year, which is broadly consistent with meteoroid puncturing, albeit with considerable modelling uncertainties (\cref{sec:straylight}).

Significant transmittance of X-rays -- emitted by Solar flares -- through the sunshield was unexpected. Complex and variable tomographic patterns of the sunshield are projected onto the VIS detectors. This required considerable changes to numerous pipeline processing tools to detect and to reject affected image areas in the DR1 data. The pipeline has reached sufficient maturity to handle these events largely automatically for future data releases. 

Heavy water-ice contamination on \ac{FoM3} between March and June 2024 resulted in the loss of 300\,deg$^2$ of VIS data. While contamination was anticipated and accounted for by the calibration plan \citep[see section 5 in][]{DR1-TP026}, the amount of surface scattering on rough ice films could not be predicted before flight. Using surgical decontamination methods developed on-the-fly (\cref{sec:inflight_contamination_history}), the VIS throughput could be restored and has been stable since. The NISP detectors experience continuous ice deposition, modulating their effective \ac{QE}, which is well understood and can be calibrated. The estimated accumulation rates of a few nanometer per month on \ac{FoM3} and NISP are low compared to that of other missions that use \ac{MLI} as part of their thermal design. This is consistent with the effectiveness of the measures taken to minimise ice contamination during \ac{AIT} and prelaunch operations, including the dry-nitrogen purge of the fairing, and of the temperature profile maintained during the in-orbit cool-down of the payload. For missions like \Euclid, even nanometer-scale ice films are detectable and need to be accounted for, through operations, calibration, and data processing. 

Analysis of the guiding errors demonstrates excellent tracking well within allocated margins (\cref{sec:guiding_review}). The statistical power of the very large guiding data set, in combination with the cold-gas correction torques, enabled highly significant detections of solar-radiation pressure and thermal-emission recoil, the effect of heliocentric distance variation on them, and -- possibly -- radiation-darkening of the solar-cell coverglass. 

The complexities of the thermal behaviour of \Euclid were expected from detailed modelling before launch. Yet, the actual behaviour offered surprises in flight. For example, we expected that attitude changes around the \ac{AA} roll angle would dominate the internal temperatures, when actually it is the \ac{SAA} tilt angle that dominates (\cref{fig:temperature_defocus_attitude}). The onboard thermal control through radiators and high thermal conductance telescope design reduces modulation of the optical bench temperature to 31\,mK from \Euclid's heliocentric distance variation over a full orbit. Analysis of in-orbit data and telemetry show thermal response times of key telescope elements of 13--20\,hours (\cref{tab:thermal_fits}), and the spacecraft needs several days to reach a steady state. A unique spacecraft attitude sequence was enforced to minimise thermal drifts for \ac{PSF} calibration purposes \citep[see sections 6.3 and 6.4 in][]{DR1-TP026}. Temperature variations during nominal survey operations are well within requirements. 

The rigorous calibration programme, using repeated observations of the self-calibration field, has enabled high-precision measurements of the zodiacal background, including effects caused by the relative perspective change from the halo orbit. Comparison with background models allowed us to tighten \Euclid's out-of-field stray-light sensitivity by factors of 4--27 across the four \Euclid bands, compared to pre-launch estimates.

\Euclid has weathered numerous severe radiation storms successfully during the maximum of the 25th Solar cycle. Another risk factor is the meteoroid environment at L2 (\cref{sec:impacts}). Impacts are detected \Euclid's spacecraft attitude control system, and -- given suitable geometric conditions and trajectories -- the flashes and Sun-lit debris generated by the impacts. To date, 34 unambiguous events were identified with an average impact rate of around 1.5\,month$^{-1}$, similar to those seen by \textit{Gaia} and \ac{JWST}. They comprise only 10--20\% of all impacts of similar strength that occurred during that time, but were not detectable in the data for geometric reasons. So far, the impacts had no consequences other than the loss of the images taken at the time of the hit.

\Euclid's orbital station-keeping is efficient, with low consumption of propellant. The step-and-stare process to observe the tessellated sky requires frequent spacecraft stabilisation using a cold-gas micro-propulsion system numerous times per day. The current survey scheme has proven very efficient in terms of cold-gas consumption that is predicted to deplete in May 2040 with the current observing technique (\cref{sec:consumables}).

\begin{acknowledgements}
\AckEC
\AckDRone
\end{acknowledgements}

\bibliography{main,Euclid}

\begin{appendix}
\crefalias{section}{appendix}

\onecolumn 
\nolinenumbers
\section{Figures\label{sec:apdx_figs}}

\begin{figure}[h]
\centering
\includegraphics[width=0.9\hsize]{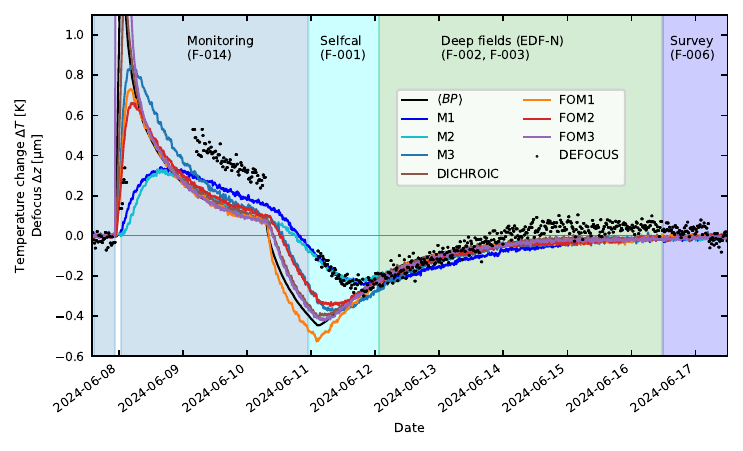}
\caption{Thermal response to the second decontamination procedure. \ac{FoM3} was warmed to 160\,K for 2\,h. Throughput and focus monitoring commenced as soon as the heating was turned off, and \ac{FoM3} came to within 1.5\,K of its long-term equilibrium temperature of 128.6\,K after 2.8\,h. Despite the surgical and local procedure, it took three days to reach sufficient stability for self-calibration and observations of the deep fields. Equilibration was disturbed on 2024-06-10 by a crash of the VIS command and data processing unit (CDPU), in response to which VIS was turned off. The immediate drop of temperature -- and lack of focus measurements -- are evident. Observations resumed 20\,h later. Around 2024-06-17T05:00:00, a jump in focus occurred between two exposures, requiring an update of the VIS distortion model. Arguably, this was a mechanical relaxation event in the telescope's SiC structure.}
\label{fig:decontamination_thermal_response}
\end{figure}

\section{Estimating stray-light levels and effective hole size in the multi-layer thermal insulation\label{sec:stray_light_estimate}}

This is an order-of-magnitude estimate of the photon-number flux density reaching a small hole, with area $A_{\rm h}$, in the thermal insulation surrounding the instrument cavity. Two successive diffuse reflections enter: sunlight $\rightarrow$ thruster nozzle ($A_1$) $\rightarrow$ \ac{MLI} surface ($A_2$) $\rightarrow$ hole. We made the following simplifying assumptions.
\begin{itemize}
    \item Both reflecting surfaces are Lambertian scatterers with reflectance $\rho_1=0.2$ (dark surface) and $\rho_2=0.8$ (aluminised \ac{MLI}).
  \item The solar photon flux density $\Phi_1$ received by the first scattering surface is obtained from a simple AB magnitude scaling. 
  \item The nozzle tip has an illuminated flat surface area of $A_1 = 10\,{\rm cm}^2$ at normal incidence, and is located at a distance $d_1 = 1.5\,$m from the second surface on the backside of the sunshield. Since $d_1 \gg \sqrt{A_1}$, the nozzle is treated as a point source for the first reflection, diluting the photon stream in the inverse-square sense.
  \item The area on the backside of the sunshield that can scatter photons onto the hole has $A_2 = 1\,{\rm m}^2$ with angle of incidence $\theta= 45^{\circ}$.
  \item The hole is located $d_2 = 0.5\,$m in front of the second scatter surface. The point-source approximation is not valid, since $d_2\approx \sqrt{A_2}$. We use the on-axis irradiance of a uniform Lambertian disc of the same area, with the hole located along the surface normal.
\end{itemize}

\paragraph{Solar photon flux density:} We estimate $\Phi_1$ directly from the Sun's $m_{\rm AB}\approx-27$, and assume a frequency-flat spectrum across the relevant passband range.  The Sun's spectral flux density is then
\begin{equation}
  f_\nu = 10^{\displaystyle -0.4\,(m_{\rm AB}+48.60)}\;{\rm erg\,s^{-1}\,cm^{-2}\,Hz^{-1}}
        \approx 2.3\times10^{-9}\;{\rm erg\,s^{-1}\,cm^{-2}\,Hz^{-1}} .
\end{equation}
We consider photons in the 400--900\,nm wavelength range detected by VIS. Integrating $f_\nu$ over the corresponding frequency range, we obtain $\Phi_1 \approx 2.8\times10^{17}$\,photons\,s$^{-1}$\,cm$^{-2}$.

\paragraph{First scattering event (nozzle to sunshield):}
The brightness or radiance of the illuminated nozzle is $L_1 = \rho_1\,\Phi_1/\pi$, from any viewing direction. The normalisation with $\pi$ comes from integrating the Lambertian $\cos(\theta)$ dependence over the hemisphere above the scatterer. The photon flux received by the second scattering surface is 
\begin{equation}
  \Phi_2 = \frac{L_1\,A_1\,\cos\theta}{{d_1}^{2}}
           = \frac{\rho_1\,\Phi_1\,A_1\,\cos\theta}{\pi\,{d_1}^{2}} .
  \label{eq:phi2}
\end{equation}

\paragraph{Second scattering event (sunshield to hole):} On the backside of the sunshield we consider an illuminated disc with $A_2=\pi R^2$ and radiance 
\begin{equation}
    L_2 = \rho_2\,\Phi_2/\pi\,
    \label{eq:L2}
\end{equation}
into the hemisphere above. The emitting surface element is ${\rm d}A_{\rm s} = r\,{\rm d}r\,{\rm d}\varphi$ at a distance $r$ from the centre of the disc, in polar coordinates. The receiving element is ${\rm d}A_{\rm r}$ at the hole's squared distance $s^2 = r^2 + d_2^2$. The transferred photon flux (in s$^{-1}$\,cm$^{-2}$) per ${\rm d}A_{\rm r}$ is then
\begin{equation}
  \frac{{\rm d}\Phi_{\rm h}}{{\rm d}A_{\rm r}} = L_2\,\frac{\cos\theta_s\cos\theta_r}{s^2}\,{\rm d}A_{\rm s}\,, 
  \qquad {\rm with}\quad \cos\theta_s=\cos\theta_r=\frac{d_2}{s}\,.
\end{equation}
The total photon flux density at the hole is obtained by integrating the transferred flux over the disc,
\begin{equation}
  \Phi_{\rm h}
  = \int_0^{2\pi}\!\!\int_0^{R} L_2\,\frac{d_2^2}{\left(r^2+d_2^2\right)^2}\,r\,{\rm d}r\,{\rm d}\varphi
  = 2\pi\,L_2\,d_2^2 \int_0^{R}\frac{r\,{\rm d}r}{\left(r^2+d_2^2\right)^2}
  = \pi L_2\left(1-\frac{d_2^2}{R^2+d_2^2}\right)
  = \frac{\pi\,L_2\,A_2}{A_2+\pi\,d_2^2}\,.
\end{equation}
Substituting with \cref{eq:phi2} and \cref{eq:L2},
\begin{equation}
  \Phi_{\rm h} = \frac{\rho_1\,\rho_2\,\Phi_1\,A_1\,A_2\,\cos\theta}
             {\pi\,d_1^{2}\,\left(A_2+\pi\,d_2^{2}\right)}\;.
  \label{eq:phihole}
\end{equation}
The photon rate through the hole into the instrument cavity is then $\dot{N} = \Phi_{\rm h}\,A_{\rm h}$, which can be used to estimate $A_{\rm h}$ from the observed stray-light levels, which we do in the following.

\paragraph{Estimating the hole size:}
 With $\rho_1=0.2$, $\rho_2=0.8$,
$A_1=10\,{\rm cm}^2$, $d_1=150\,{\rm cm}$, $A_2=10^4\,{\rm cm}^2$, $d_2=50\,{\rm cm}$, $\theta=45^{\circ}$,  \cref{eq:phihole} gives
\begin{equation}
  \Phi_{\rm h} \;\approx\; 2.5\times10^{12}\;{\rm photons\,cm^{-2}\,s^{-1}}\;.
\end{equation}

The observed mean stray-light flux in VIS is approximately $2.4\times10^6$\,photons\,s$^{-1}$\,cm$^{-2}$ when the \Euclid's sunshield points squarely at the Sun and the thruster nozzle is fully illuminated, as in the example above. Assuming an instrument-cavity surface area of $4$\,m$^2$, a mean quantum efficiency of ${\rm QE}=0.6$ across the 400--900\,nm range, and that all photons are uniformly distributed over the inner cavity surface, the observed total photon rate entering the cavity becomes $\dot{N}_{\rm obs} = 1.6\times10^{11}$\,photons\,s$^{-1}$. The effective hole area needed to explain these stray-light levels becomes
\begin{equation}
    A_{\rm h} = \frac{\dot{N}_{\rm obs}}{\Phi_{\rm h}}=0.064\,{\rm cm}^2=6.4\,{\rm mm}^2\,.
\end{equation} 
The uncertainties in the assumptions made are considerable, though. For example, the \ac{MLI} at the back of the sunshield is not a Lambertian scatterer, but a crinkled shiny surface that may send multiple glints -- or only very few -- through the same hole. A range of
$A_{\rm h}=3$--30\,mm$^2$ is entirely plausible.

\begin{figure}[t]
    \centering
    \includegraphics[angle=0,width=0.5\hsize]{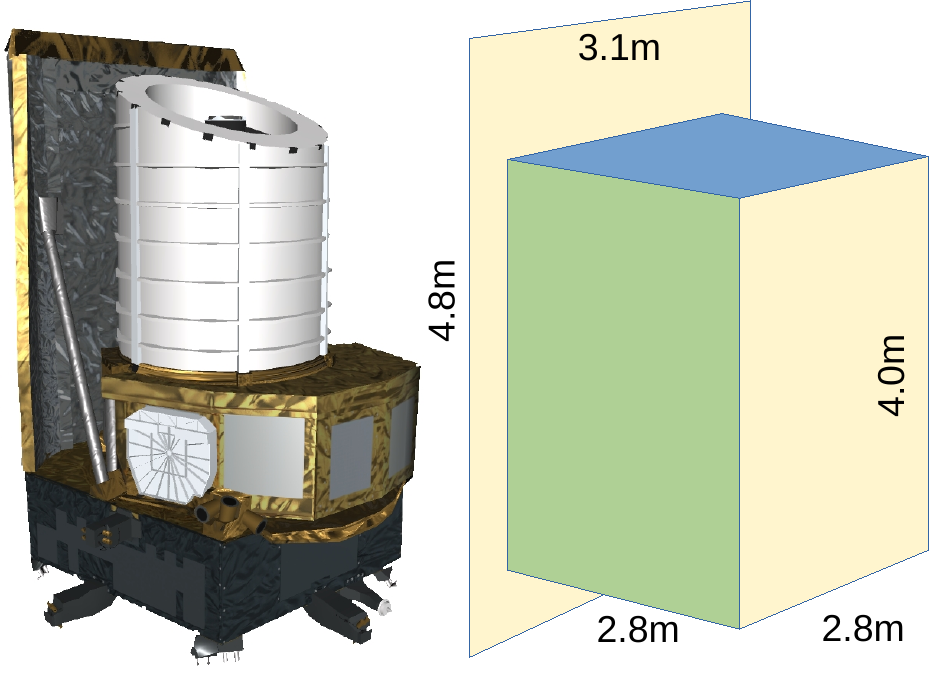}
    \caption{Sketch of the 3D polygonic model to compute the meteoroid impact rate on the spacecraft (\cref{sec:expected_impact_rate}).}
    \label{fig:euclid3d_meteoroid_model}
\end{figure}

\section{Modelling the torque time series}
\subsection{Tilt in the anti-Sun direction (torque \texorpdfstring{$\tau_Y$}{Y-torque}) and rotation around telescope axis (torque \texorpdfstring{$\tau_Z$}{Z-torque})\label{sec:derivation_Y}}
The quantities $\tau_{Y,Z}$ counter the action of \ac{SRP} and \ac{TER} on the the sunshield. The radiation-pressure force on the sunshield is 
\begin{equation}
    F_{\rm abs} = \frac{A\,S}{c}\,(\alpha + 2\,\rho) = \frac{A\,S}{c}\,(2-\alpha)\,,
    \label{eq:force_abs}
\end{equation}
where $A$ is the sunshield's area, $\alpha$ the absorptivity and $\rho$ the reflectivity, and we postulate conservation of energy, so $1=\alpha+\rho$. In this framework, $\rho$ is the fraction of incident energy that leaves the surface without exciting the atomic lattice, that is, reflected photons transfer momentum and no energy. $\alpha$ is the fraction of incident energy that gets transferred to the sunshield in an inelastic process. $S$ is the distance-dependent solar irradiation,
\begin{equation}
    S(d_{\rm h}) = S_0\,\left(\frac{1\,{\rm AU}}{d_{\rm h}}\right)^2\,,
    \label{eq:solarconstant}
\end{equation}
where $S_0$ is the solar constant and $d_{\rm h}(t)$ \Euclid's the heliocentric distance. The sunshield is in thermal equilibrium because the geometric correction factor $[\sin({\rm SAA})\,\cos({\rm AA})]^{-1}$ in \cref{eq:geometriccorrection} is rather stable with a long-term mean of $0.993\approx1$, and because the orbital distance variation is very slow in comparison to the sunshield's thermal timescale. Balancing absorbed and emitted power, we obtain
\begin{equation}
    P_{\rm abs} = P_{\rm emit} \longrightarrow\alpha\,A\,S = \epsilon\,A\,\sigma_{\rm B}\,T^4 \longrightarrow T^4 = \frac{\alpha\,S}{\epsilon\,\sigma_{\rm B}}\,.
    \label{eq:t4dependence}
\end{equation}
The reflectance $\rho$ does not appear in this equation because it does not transfer energy. Using \cref{eq:force_abs} for \ac{SRP} and \cref{eq:thermalpressure} for \ac{TER}, the total force acting on the sunshield is
\begin{equation}
    F_{\rm tot} = F_{\rm abs} + F_{\rm emit} = \frac{A\,S}{c}\,(2-\alpha) + \frac{2}{3}\,\frac{\epsilon\,A\,\sigma_{\rm B}}{c}\,T^4\,
    =\frac{A}{c}\left(2-\frac{\alpha}{3}\right)\,S 
    =k_1\,\left(\frac{1\,{\rm AU}}{d}\right)^2\,\label{eq:firstinvsquarelaw}\,.
\end{equation}
Here, we substituted \cref{eq:t4dependence} for $T^4$, used \cref{eq:solarconstant}, and $k_1$ is a constant. Because of the linear relation between force and torque, our joint pressure and recoil model for $\tau_Y$ thus has a simple one-parameter inverse-square dependence on heliocentric distance,
\begin{equation}
  \tilde\tau_i = A_i\,\left[\frac{1\,{\rm AU}}{d(t)}\right]^{2} \quad[\times10^{-4}\,{\rm Nm}]\,,\,\quad{\rm with}\;i\in\{Y,Z\}\,,
  \label{eq:fitting_tau_Y}
\end{equation}
where $A_i$ is a fitting parameter.

\subsection{Rotation around the Sun--\Euclid axis (torque \texorpdfstring{$\tau_X$}{X-torque})\label{sec:derivation_X}}
The case for $\tau_X$ is different from $\tau_{Y,Z}$, but results in a similar model. The periodicity in their time series (top and bottom panels in \cref{fig:radiationpressuremodel}) implicates $d_{\rm h}(t)$, yet these sides of the spacecraft are in the shadow. The modulation of the baseplate temperature shown in \cref{fig:T170_timeseries} is the key signature that a small fraction of the incident power, 
\begin{equation}
    P_{\rm leak}(d) = \eta\,A\,S\,,\quad {\rm with}\; \eta\ll1\,,
    \label{eq:pleak}
\end{equation}
is conducted from the sunshield to the \ac{SVM} and \ac{PLM}. The spacecraft continuously generates $P_{\rm internal}\gg P_{\rm leak}$ as part of its routine operations, with the radiators keeping the \ac{SVM} at $T_0\approx290$\,K. When the \ac{SVM} warms because of the heliocentric distance change, then the radiators warm as well. The associated temperature change is small, $\Delta T(d)\lesssim1\,{\rm K}\ll T_0$, also because radiators are highly efficient because of the $T^4$ dependence of the power radiated away. We linearise
\begin{equation}
    T^4\approx T_0^4\left(1+4\,\frac{\Delta T}{T_0}\right)\,.
    \label{eq:t4_linearised}
\end{equation}
The \ac{MPS} torque balances the \ac{TER} given by \cref{eq:thermalpressure}, and we have 
\begin{equation}
    \tau = k_2\,T^4 = k_2\,T_0^4 \left[1+4\,\frac{\Delta T}{T_0}\right] = \tau_0 + \Delta \tau\,,
    \label{eq:taulinearised}
\end{equation}
where $k_2$ and $\tau_0$ are constants. The thermal timescales of the \ac{PLM} (\cref{sec:thermal_response_model,tab:thermal_fits}) and the \ac{SVM} are much shorter than the changes in $P_{\rm leak}$. In the equilibrium state, where the total generated power equals the power radiated away by an area $A_{\rm r}$,
\begin{equation}
    P_{\rm internal} + P_{\rm leak} = \epsilon\,A_{\rm r}\,\sigma_{\rm B}\,T^4
    \approx \epsilon\,A_{\rm r}\,\sigma_{\rm B}\,T_0^4 + 4\,\epsilon\,A_{\rm r}\,\sigma_{\rm B}\,T_0^3\,\Delta T\,,
\end{equation}
where we have substituted \cref{eq:t4_linearised} for $T^4$. Using \cref{eq:pleak},
\begin{equation}
    P_{\rm leak}(d) = \eta\,A\,S = 4\,\epsilon\,A_{\rm r}\,\sigma_{\rm B}\,T_0^3\,\Delta T(d)\,,
\end{equation}
which we solve for $\Delta T$ and expand $S$ with \cref{eq:solarconstant},
\begin{equation}
    \Delta T(d) = k_3\,\left(\frac{1\,{\rm AU}}{d}\right)^2\,,\quad{\rm with}\,k_3={\rm constant}\,.
\end{equation}
Using this result in \cref{eq:taulinearised}, we find 
\begin{equation}
    \tau = \tau_0 + k_4\,\left(\frac{1\,{\rm AU}}{d}\right)^2\,,
\end{equation}
where $\tau_0$ represents the constant \ac{TER} generated by $P_{\rm internal}$ in absence of any distance-related effect. Thus, our model for torques $\tau_X$ and $\tau_Z$ have the same inverse-square dependence on $d_{\rm h}(t)$ as for $\tau_Y$ in \cref{eq:fitting_tau_Y}, with an additional additive constant,
\begin{equation}
  \tilde\tau_Z = B + A\,\left[\frac{1\,{\rm AU}}{d(t)}\right]^{2} \quad[\times10^{-4}\,{\rm Nm}]\,.
  \label{eq:fitting_tau_XZ}
\end{equation}
In \cref{sec:mam} we added a linear drift term to the model, which was necessitated by the data.

\end{appendix}

\end{document}